\documentclass[%
superscriptaddress,
nofootinbib,
 amsmath,amssymb,
 aps,
prd,
twocolumn,
floatfix,
]{revtex4-2}

\usepackage{graphicx}%
\usepackage{dcolumn}%
\usepackage{bm}%
\usepackage{enumitem}
\usepackage{amsmath}
\usepackage{mathrsfs}
\usepackage{url}
\usepackage{hyperref}
\usepackage{amssymb}
\usepackage[export]{adjustbox}
\usepackage[english]{babel}
\usepackage{float}
\usepackage{silence}
\usepackage{lipsum}
\usepackage{listings}
\usepackage{tensor}
\usepackage{siunitx}
\usepackage{array}
\usepackage{graphicx}%
\usepackage{dcolumn}%
\usepackage{bm}%
\usepackage{hyperref}%
\usepackage[mathlines]{lineno}%
\usepackage{pifont}
\usepackage{mathtools}
\usepackage{caption}
\usepackage[capitalize]{cleveref}
\usepackage{bigints}
\usepackage[italicdiff]{physics} 
\usepackage{slashed}
\usepackage{cancel}
\newcommand{\nn}{\nonumber\\}
\newcommand{\av}[1]{\langle #1 \rangle }

\newcommand{\ddphi}[1]{\dfrac{d #1}{d\trueanom}}
\newcommand{\ddtheta}[1]{\dfrac{d #1}{d \zeta}}
\newcommand{\ppphi}[1]{\dfrac{\partial #1}{\partial \trueanom}}
\newcommand{\pptheta}[1]{\dfrac{\partial #1}{\partial \zeta }}

\newcommand{\bigO}[1]{\mathcal{O} (#1) }
\newcommand{\xn}[2]{X^{(#1)}_{#2}}

\newcommand{\ddt}[1]{\dfrac{d #1 }{dt}}

\newcommand{\trueanom}{\vartheta}
\newcommand{\mft}{\mathfrak{t}}

\usepackage{xcolor}
\newlist{todolist}{itemize}{2}
\setlist[todolist]{label=$\square$}

\begin{document}

\newcommand{\sissa}{SISSA, Via Bonomea 265, 34136 Trieste, Italy \&  INFN Sezione di Trieste}

\title{Post-adiabatic effects sourced by radiation reaction \\for inspiraling, eccentric binaries in post-Newtonian theory}

\author{Rohit S. Chandramouli}%
\email{rchandra@sissa.it, Corresponding author}%
\affiliation{\sissa}
\affiliation{IFPU - Institute for Fundamental Physics of the Universe, Via Beirut 2, 34014 Trieste, Italy}
\author{Nijaid Arredondo}%
\email{josena2@illinois.edu}%
  \affiliation{Illinois Center for Advanced Studies of the Universe \& Department of Physics\\
University of Illinois Urbana-Champaign, Urbana, Illinois 61801, USA.}%
\author{Lidia J. Gomes Da Silva}%
\email{lidiajoana@pm.me}%
\affiliation{\sissa}%
\affiliation{IFPU - Institute for Fundamental Physics of the Universe, Via Beirut 2, 34014 Trieste, Italy}%
  \author{Nicol\'as Yunes}%
\email{nyunes@illinois.edu}%
  \affiliation{Illinois Center for Advanced Studies of the Universe \& Department of Physics\\
University of Illinois Urbana-Champaign, Urbana, Illinois 61801, USA.}%

\date{\today}

\begin{abstract}

The adiabatic approximation, in which eccentric binaries evolve under orbit-averaged fluxes, is widely used in gravitational-wave and astrophysical modeling. We use resummed multiple-scale analysis, which is closely related to near-identity averaging transformations, to derive post-adiabatic corrections from leading-order radiation reaction and explain how the accuracy of the adiabatic approximation depends on the initial orbital phase. While the secular evolution equations for the eccentricity and semi-major axis receive no explicit first post-adiabatic corrections, their solutions acquire such corrections through the initial conditions. We also derive the change in the functional form of the gravitational-wave polarizations that cancels the gauge dependence of the orbital elements at first post-adiabatic order, ensuring agreement between waveforms evaluated at the same physical event in different coordinates. At a radial turning point, the first post-adiabatic oscillations of the semi-major axis and eccentricity vanish, so the initial osculating energy and angular momentum agree with their orbit-averaged values up to second post-adiabatic corrections. When initialized elsewhere along the orbit, the adiabatic approximation can accumulate significant dephasing, which the first post-adiabatic approximation can substantially reduce. Both dephasings can change sign across parameter space, consistent with competing contributions from successive orders. Accounting for initialization is therefore essential for assessing the accuracy of these approximations, particularly for long-lived, highly eccentric binaries.

\end{abstract}

\maketitle

\section{Introduction}
The emission of gravitational waves (GWs) by a binary system leads to a loss of energy and angular momentum, causing the binary to inspiral and merge~\cite{Peters:1963ux,Peters:1964zz}.
During the early inspiral, when the binary's characteristic velocity is small compared to the speed of light, the orbital evolution can be studied using post-Newtonian (PN) theory~\cite{Blanchet:2013haa}.
Owing to the separation of timescales between the conservative orbital dynamics and the dissipative radiation-reaction dynamics, the energy and angular momentum change \emph{adiabatically}, and their evolution can be computed by averaging the far-zone radiative fluxes over the orbital timescale, as first done by Peters and Mathews~\cite{Peters:1963ux,Peters:1964zz}.
The resulting adiabatic evolution of the orbital elements causes the orbit to shrink and circularize~\cite{Peters:1964zz}.
Since the work of Peters and Mathews, the orbit-averaged fluxes have been computed to high PN orders~\cite{Blanchet:2013haa}, from which the secular changes in the energy, angular momentum, and orbital elements can be obtained~\cite{Arun:2009mc}.
This raises the following natural questions: what are the \emph{corrections} to the adiabatic approximation? At what PN order must they be accounted for? And when does the adiabatic approximation become inaccurate?

Several related, systematic formalisms exist for computing post-adiabatic (PA) corrections, including multiple-scale analysis (MSA)~\cite{BenderOrszag1978,KevorkianCole1996}, the dynamical renormalization group (DRG)~\cite{DRG,Galley:2016zee,GalvezGhersi:2021sxs}, and the recently developed Magnusian formalism~\cite{Kim:2025gis,Blanco:2026prj,DaSilva:2026pdk}, each of which has been applied to binaries driven by radiation reaction~\cite{Lincoln_Will_1990,PoissonWill2014,Will:2019lfe,Galley:2016zee,Blanco:2026prj, DaSilva:2026pdk}. In the DRG formalism~\cite{DRG}, one works with a single time variable and absorbs the secularly growing terms of the na\"ive perturbative solution into slowly varying integration constants, whose evolution follows from requiring the solution to be independent of the arbitrary renormalization time. In the Magnusian formalism, finite-time evolution is generated by a phase-space function through nested Poisson brackets~\cite{Kim:2025gis}. For binaries driven by radiation reaction, the leading-order generator is known in \emph{closed-form}\cite{Galley:2012hx, Tsang_2015} and defines a \emph{canonical} map from one orbital cycle to the next, whose secular content coincides with the effective Hamiltonian of a near-identity transformation \cite{Blanco:2026prj}. Because this generator is an explicit functional of the force, it can also be learned directly from the force law, without trajectory data, as a canonical flow map that outperforms structure-preserving integrators at the same time-step \cite{DaSilva:2026pdk}. The PA corrections studied here coincide with the beyond orbit average content learned in that manuscript. 

In this work, we focus on MSA, as outlined for orbital dynamics, e.g., by Poisson and Will~\cite{PoissonWill2014} (also referred to as two-timescale analysis~\cite{Lincoln_Will_1990,Hinderer:2008dm}).
In MSA, one explicitly separates the fast and slow timescales from the outset, which yields PA corrections with both secular and oscillatory contributions.
MSA is a natural approach for this problem, since the fast orbital timescale and the slow radiation-reaction timescale can be identified \emph{a priori}.

At leading order (LO) in MSA, one obtains the adiabatic evolution of the orbital elements.
Within the PN literature, the next-to-leading-order (NLO) oscillatory 1PA corrections were computed in~\cite{Damour:2004bz} using an equivalent two-timescale (variation-of-constants) approach, where $n$PA refers to PA corrections at $n$th order relative to the LO adiabatic approximation.
These 1PA corrections enter the dissipative dynamics at relative 2.5PN order, where an $n$PN term scales as $\sim v^{2n}$, with $v$ the orbital velocity; this follows from the fact that they are corrections to orbit-averaging, and thus scale with the ratio of the orbital to the radiation-reaction timescale.
Later,~\cite{Boetzel:2019nfw} recomputed these 1PA oscillatory corrections analytically using a small-eccentricity expansion, and propagated them to the gravitational-wave modes of eccentric, non-spinning binaries, in order to maintain relative 2.5PN waveform accuracy.

Separately,~\cite{Loutrel:2018ssg,Loutrel:2018ydu,Will:2019lfe} computed both the oscillatory and secular 1PA corrections to the orbital elements, focusing largely on the small-eccentricity limit.
In particular,~\cite{Loutrel:2018ydu} showed that the secular 1PA corrections to the semi-major axis and eccentricity can be nonzero when the binary is initialized far from pericenter or apocenter.
The accuracy of the 0PA approximation, and the 2.5PN contributions from the 1PA terms, therefore have a subtle but important dependence on the initial true (or mean) anomaly.
The same dependence appears in~\cite{Fumagalli:2025rhc}, where the accuracy of the 0PA approximation was studied by comparing the merger time to that of a direct (non-averaged) evolution.
Furthermore,~\cite{Nee:2025zdy} found that the initial mean anomaly is important in characterizing numerical-relativity simulations of non-spinning eccentric binary mergers.

Given these prior works, it is clearly important to obtain an analytical understanding, within PN theory, of why the accuracy of the 0PA approximation and the importance of the 1PA corrections depend crucially on the initial true (or mean) anomaly.
Obtaining this understanding, and assessing the validity of the 0PA and 1PA approximations across parameter space for a wide range of initial conditions, is the main goal of this work.
This complements recent works that assessed the waveform accuracy of PA approximants built from eccentric effective-one-body (EOB) models~\cite{Lynch:2026ibo,Lynch:2026eyw,Gamba:2026ado}.
These works also highlight that PA approximations can reduce the computational cost of evolving eccentric binaries while retaining waveform accuracy. This is important for current ground-based GW detectors, including LIGO, Virgo, and KAGRA~\cite{LIGOScientific:2014pky,VIRGO:2014yos,KAGRA:2020tym}, future ground-based detectors, including the Einstein Telescope~\cite{Punturo:2010zz,Maggiore:2019uih} and Cosmic Explorer~\cite{Evans:2021gyd}, and space-based detectors, such as the Laser Interferometer Space Antenna (LISA)~\cite{LISA:2017pwj,LISAConsortiumWaveformWorkingGroup:2023arg,LISA:2024hlh}.

In order to assess the validity of the PA approximation in a robust manner, one has to account for a few subtleties in obtaining the solutions at each $n$PA order.
The first subtlety is that the MSA formulation, in particular the MSA ansatz for the secular and oscillatory pieces, is not unique.
The second subtlety, which follows from the first, concerns how initial conditions are provided to the secular and oscillatory pieces at successive PA orders.
The third subtlety is that the radiation-reaction acceleration is gauge-dependent, so that the PA corrections to the orbital elements are also gauge-dependent.
As a secondary goal of our work, we address each of these subtleties. 

Some of these subtleties are related to the specific formulation of MSA that is implemented in the two-body problem. 
In the MSA formulation of~\cite{PoissonWill2014,Loutrel:2018ssg,Loutrel:2018ydu}, the secular and oscillatory pieces are each expanded in a power series in the ratio of the orbital to the radiation-reaction timescale.
On the other hand, in the alternative formulation of~\cite{Mora:2003wt,Will:2019lfe}, one expands only the oscillatory piece in such a series, so that the secular piece is effectively a resummation of the one in~\cite{PoissonWill2014}.
This ``resummed'' MSA formulation is closely related to near-identity averaging transformations (NITs)~\cite{KevorkianCole1996}, as used in extreme-mass-ratio-inspiral (EMRI) modeling
~\cite{VanDeMeent:2018cgn,Lynch:2021ogr} and, more recently, in EOB modeling\footnote{In both cases, the MSA/NIT expansion is in the ratio of the orbital to the radiation-reaction timescale. For EMRIs, this ratio is set by the mass ratio, whereas for comparable-mass binaries in the PN and EOB approaches it scales as $\sim \eta\, v^5$, where $v$ is the orbital velocity and $\eta$ is the symmetric mass ratio, and is small owing to the slow motion of the binary, rather than to the mass ratio. Furthermore, the EOB approach resums the PN dynamics, extending its accuracy to the late inspiral, whereas we use here the PN-truncated dynamics.}~\cite{Lynch:2026eyw,Lynch:2026ibo,Gamba:2026ado}.
For the solutions obtained from these two formulations (that of~\cite{PoissonWill2014} and that of~\cite{Mora:2003wt,Will:2019lfe}) to match, one has to account for how the initial conditions are split between the secular and oscillatory pieces.
Such subtleties were discussed in earlier works~\cite{Pound:2007ti,Pound:2007th,Hinderer:2008dm}, but an explicit demonstration of the equivalence of the two approaches has not been provided.
In our work, starting from the results of~\cite{PoissonWill2014}, we resum the secular piece to 1PA order and show explicitly that the resulting equations are consistent with those of~\cite{Mora:2003wt,Will:2019lfe}.
We also discuss how to choose the initial conditions in each formalism so as to obtain order-consistent solutions, thereby demonstrating the equivalence of the two MSA formulations and addressing the first and second subtleties.

The gauge dependence of the radiation-reaction acceleration arises from the coordinate freedom in defining the near-zone energy and angular momentum at 2.5PN order, within which the instantaneous balance laws are preserved~\cite{Iyer:1995rn}.
The orbital elements are therefore naturally gauge dependent at 2.5PN order.
Recently,~\cite{Fumagalli:2025rhc} proposed using near-identity transformations (without averaging) to eliminate the gauge dependence of the orbital elements.
Physical observables, such as the far-zone radiative fluxes or the gravitational-wave polarizations, must nonetheless be gauge invariant, irrespective of the gauge dependence of the orbital elements.
Following~\cite{Arun:2009mc}, we derive the functional transformation of the polarizations with respect to the orbital elements that cancels, to 1PA order, the contribution from the 2.5PN gauge transformation of the orbital elements.
The gauge dependence of the 1PA corrections to the orbital elements is thus compensated by the corresponding change in the functional form of the polarizations, which allows for meaningful comparisons of the 1PA contributions in different gauges and addresses the third subtlety.

With these subtleties accounted for, we perform numerical experiments comparing the 0PA and 1PA evolutions against a direct (non-averaged) evolution at fixed initial conditions and system parameters.
These comparisons isolate the errors introduced by the PA approximations and quantify the improvement from including 1PA corrections.
This differs from assessing their performance in matched filtering or Bayesian parameter estimation, where the waveform parameters are allowed to vary.
We consider initial eccentricities ($e_0 \in [10^{-4}, 0.9]$), semi-latera recta ($p_0/m \in [20, 75]$), true anomalies ($\vartheta_0 \in [-\pi/2, \pi/2]$), and mass ratios ($q \in [1/10, 1]$).

When initialized at a radial turning point, the 1PA oscillatory terms in the semi-major axis and eccentricity vanish, so that the initial osculating energy and angular momentum equal their orbit-averaged values up to 2PA corrections.
Consequently, the 0PA and 1PA approximations are comparably accurate when initialized at pericenter, with orbital dephasings relative to the direct solution of magnitude $\lesssim$~a few $\times\,10^{-1}$~rad for $e_0 \lesssim 0.8$.
For these configurations, including 1PA corrections offers little improvement in orbital-phase accuracy.

When initialized far from a radial turning point ($|\vartheta_0| > \pi/4$), however, the magnitude of the 0PA dephasing exceeds $1$~rad already at moderate eccentricities, $e_0 \gtrsim 0.1$, independently of the mass ratio.
The magnitude of the 1PA dephasing, by contrast, depends only weakly on $\vartheta_0$ and, for equal masses, remains $\lesssim$~a few $\times\,10^{-1}$~rad for $e_0 \lesssim 0.8$, reaching $\sim 1$~rad only for $e_0 \sim 0.9$.
At small eccentricities, $e_0 \lesssim 10^{-2}$, the 0PA approximation is accurate irrespective of the initial true anomaly, and 1PA offers no significant improvement.

We also find that both the 0PA and 1PA dephasings change sign depending on the initial conditions.
The 0PA dephasing changes sign for $\vartheta_0 < 0$, along a curve that starts at $e_0 \approx 5\times10^{-3}$ for $\vartheta_0 = -\pi/2$ and moves to larger $e_0$ as $\vartheta_0$ approaches $0$.
The 1PA dephasing changes sign at $e_0 \approx 0.6$--$0.75$, with a weak dependence on the other parameters.
We attribute these sign changes to competing contributions from successive PA orders.

The implications for GW data analysis also depend on initialization.
For a highly eccentric case study, re-initializing the 0PA evolution at a radial turning point reduces the GW mismatch from $\approx 0.67$ to $\approx 2\times10^{-5}$, comparable to that of the 1PA waveform.
This suggests that large orbital dephasings at fixed initial conditions need not imply a substantial loss of recoverable SNR when waveform parameters are allowed to vary.

The remainder of the paper is organized as follows. In Sec.~\ref{sec:orbital_MSA}, we first provide an overview of the perturbed Kepler problem, discuss the MSA framework of~\cite{PoissonWill2014} applied to 1PA order, and show how to resum the dynamics, to obtain consistent results with the equivalent MSA formalism of~\cite{Mora:2003wt}.
In Sec.~\ref{sec:PA_RR}, we apply the MSA framework to the 2.5PN radiation-reaction perturbation, and obtain the 1PA results for the osculating orbital elements.
We then show in Sec.~\ref{sec:quasicirc} how to obtain the quasi-circular limit by regularizing the orbital elements, and compare with previous results from the literature.
In Sec.~\ref{sec:waveform}, we show that the orbital elements transform under a 2.5PN gauge transformation, and how to account for the functional change in the gravitational-wave polarizations with respect to the orbital elements, leaving the polarizations gauge-invariant.
We then perform our numerical experiments and assess the validity of the 0PA and 1PA approximations in Sec.~\ref{sec:num_exp}, and present our conclusions in Sec.~\ref{sec:conclusion}.

Henceforth, we use the following conventions and notations throughout the paper: geometric units with $G=1=c$, the Einstein summation convention over repeated dummy indices, $m_1$ and $m_2$ for the primary and secondary masses, $m = m_1 +m_2$ as the total mass, $q = m_2/m_1 \leq 1$ as the mass ratio, and $\eta = q/(1+q)^2$ as the symmetric mass ratio.

\section{Orbital dynamics using multiple-scale analysis} \label{sec:orbital_MSA}
In Sec.~\ref{subsec:Kepler} we review the perturbed Kepler problem to fix notation and conventions. In Sec.~\ref{subsec:expanded-MSA} we summarize the MSA formalism of Poisson and Will \cite{PoissonWill2014}, and in Sec.~\ref{subsec:resum_IC} we show that it is equivalent to the resummed (near-identity) formulation of~\cite{Mora:2003wt,Will:2016}. This equivalence resolves the first subtlety raised in the introduction: the two ans\"atze differ only in how the 1PA secular terms are grouped, and lead to the same evolution equations \emph{once the initial conditions  for the secular and oscillatory pieces are chosen consistently}. 

\subsection{The perturbed Kepler problem} \label{subsec:Kepler}
 We review the essentials of the Kepler problem, following~\cite{PoissonWill2014}, to establish the notations and conventions used in our work. 
 In the center of mass frame, we introduce the instantaneous Cartesian orthonormal triad $(\vec{e}_x, \vec{e}_y, \vec{e}_z)$. The binary dynamics is restricted to the plane orthogonal to the orbital angular momentum. In this orbital plane, the motion is more conveniently described in terms of the $(\vec{n},\vec{\lambda})$ basis,
 \begin{align}
 \vec{n} &= \cos \vartheta \vec{e}_x +\sin \vartheta \vec{e}_y,\\
 \vec{\lambda} &= -\sin \vartheta \vec{e}_x + \cos \vartheta \vec{e}_y,
 \end{align}
 where $\vartheta$ is the true anomaly.
 For bound systems, the orbit is an ellipse and is parameterized by the relations below: 
 \begin{subequations}
 \label{eqn:kepler_solution}
 \begin{align}
 r &= \dfrac{p}{1+e\cos \trueanom} = a (1-e \cos u),\\
 \phi &= \trueanom + \omega, \\
 \sqrt{\frac{m}{a^3}}(t-t_0) &= \ell = u - e \sin u \label{eqn:Kepler_eqn}, \\
 \trueanom &=  u + 2 \tan^{-1} \left( \dfrac{\beta_e \sin u}{1-\beta_e \cos u} \right) ,\\
 \ \beta_e&=(1-\sqrt{1-e^2})/e ,
 \end{align}    
 \end{subequations}
where $p = a (1-e^2)$ is the semi-latus rectum, $e$ the orbital eccentricity, $a$ the semi-major axis, $u$ the eccentric anomaly, $\ell$ the mean anomaly, $\omega$ longitude of the pericenter, and $\phi$ the orbital phase. The time dependence enters through~\cref{eqn:Kepler_eqn}, known as Kepler's equation.

The above description of the Keplerian orbit is given in the orbital plane.
Adopting the fundamental frame of~\cite{PoissonWill2014}, we use an invariant triad $(\vec{e}_X,\vec{e}_Y,\vec{e}_Z)$, which is related to the orbital triad through
\begin{subequations}
\begin{align}
\vec{e}_x &= (\cos\Omega\cos\omega - \cos\iota\sin\Omega\sin\omega)\,\vec{e}_X \nn
&\quad + (\sin\Omega\cos\omega + \cos\iota\cos\Omega\sin\omega)\,\vec{e}_Y \nn
&\quad + (\sin\iota\sin\omega)\,\vec{e}_Z\,, \\
\vec{e}_y &= -(\cos\Omega\sin\omega + \cos\iota\sin\Omega\cos\omega)\,\vec{e}_X \nn
&\quad - (\sin\Omega\sin\omega - \cos\iota\cos\Omega\cos\omega)\,\vec{e}_Y \nn
&\quad + (\sin\iota\cos\omega)\,\vec{e}_Z\,, \\
\vec{e}_z &= \sin\iota\sin\Omega\,\vec{e}_X - \sin\iota\cos\Omega\,\vec{e}_Y + \cos\iota\,\vec{e}_Z\,,
\end{align}
\end{subequations}
where $\Omega$ is the longitude of the ascending node and $\iota$ is the
inclination of the orbital plane with respect to the plane of the fundamental frame.

When a gravitating binary is affected by some additional force, the orbital evolution is typically obtained by either directly integrating the equations of motion (EoM) to evolve the radial coordinate $r(t)$ and orbital phase $\phi(t)$, or by evolving the orbital elements so that the trajectory is constructed through a sequence of osculating Keplerian orbits. The latter is simply the method of variation of constants applied to the orbital elements, which we describe below.

Let us introduce the acceleration three-vector associated with the additional force
\begin{equation}
\delta \Vec{a}= \mathcal{R} \vec{n} + \mathcal{S} \vec{\lambda} + \mathcal{W} \Vec{e_z}, \label{eqn:acc_decomp}
\end{equation}
where $\mathcal{R}$ and $\mathcal{S}$ are the radial and azimuthal components, respectively, in the orbital plane. $\mathcal{W}$ is then the component perpendicular to the orbital plane (instantaneously parallel to the Keplerian orbital angular momentum).
 The method of variation of constants then yields the EoM for the orbital elements, known widely as the Lagrange planetary equations (LPE)s, which are given by~\cite{PoissonWill2014}
 \begin{subequations}\label{eqn:LPE}
     \begin{align}
     \ddt{p} &=2 \sqrt{\frac{p^3}{m}} \dfrac{\mathcal{S}}{1+e \cos \trueanom }, \\
  \ddt{e} &=\sqrt{\dfrac{p}{m}}\left[ \mathcal{R} \sin \trueanom + \mathcal{S} \left( \dfrac{2 \cos \trueanom +e(1+ \cos ^2 \trueanom)}{1+e \cos \trueanom } \right) \right], \\
 \ddt{\varpi} &=\dfrac{1}{e} \sqrt{\dfrac{p}{m}}\left[- \mathcal{R} \cos \trueanom + \mathcal{S} \left( \dfrac{2+e \cos \trueanom}{1+e \cos \trueanom} \right) \sin \trueanom \right], \\
 \ddt{\iota} &=\sqrt{\dfrac{p}{m}} \dfrac{\cos (\omega+\trueanom)}{1+e \cos \trueanom} \mathcal{W}, \\
 \ddt{\Omega} &=\sqrt{\dfrac{p}{m}} \dfrac{\sin (\omega+\trueanom)}{1+e \cos \trueanom} \dfrac{\mathcal{W}}{\sin \iota}, \\
 \ddt{\trueanom} &= \sqrt{\dfrac{m}{p^3}} (1+e \cos \trueanom )^2 \nn
 & + \dfrac{1}{e} \sqrt{\dfrac{p}{m}} \left [ \mathcal{R} \cos \trueanom - \mathcal{S} \left( \dfrac{2+e \cos \trueanom}{1+e \cos \trueanom} \right) \sin \trueanom \right ],
 \label{eq:dthetadt}
 \end{align}
 \end{subequations}
 with $\dot{\varpi}= \dot{\omega} + \dot{\Omega} \cos \iota$.

Although there is no formal expansion involved in deriving the LPEs, in most practical applications, $\delta \vec{a}$ is a perturbation on the Keplerian acceleration $\vec{a}_K = - m/r^2 \vec{n}$, and, thus, we will expand in $\epsilon \sim |\delta \vec{a}| / |\vec{a}_K| \ll 1 $.
When doing so, one can convert derivatives of $t$ to derivatives of $\trueanom$ using the chain rule and re-expand the sourcing functions to a desired order of the perturbation.
At LO, this gives us the perturbed osculating orbit equations
  \begin{subequations}
 \begin{align}
  \dfrac{d p}{d \trueanom} &= 2 \frac{p^3}{m} \dfrac{\mathcal{S}}{(1+e \cos \trueanom)^3 } \left[ 1 + \mathcal{O}(\epsilon) \right ], \\
\begin{split}
  \dfrac{d e}{d \trueanom}&= \dfrac{p^2}{m (1+e \cos \trueanom)^2}\left[ \mathcal{R} \sin \trueanom \right. \\
  &+ \left. \mathcal{S} \left( \dfrac{2 \cos \trueanom +e(1+ \cos ^2 \trueanom)}{1+e \cos \trueanom } \right) \right]  \left[ 1 + \mathcal{O}(\epsilon) \right ],
\end{split} \\
\begin{split}
 \dfrac{d \varpi}{d \trueanom}&= \dfrac{1}{e} \dfrac{p^2}{m (1+ e \cos \trueanom)^2} \left[- \mathcal{R} \cos \trueanom \right. \\
 &+ \left. \mathcal{S} \left( \dfrac{2+e \cos \trueanom}{1+e \cos \trueanom} \right) \sin \trueanom \right] \left[ 1 + \mathcal{O}(\epsilon) \right ],
\end{split} \\
 \dfrac{dt}{d \trueanom} &= \sqrt{\dfrac{p^3}{m}} (1+ e \cos \trueanom)^{-2} \left[ 1 + \mathcal{O}(\epsilon) \right ],
 \end{align}
 \end{subequations}
 where for simplicity, we have restricted to the case of co-planar perturbations ($\mathcal{W} = 0$).

When iterating to NLO, the evolution of the orbital elements is described by the system of ordinary differential equations:
\begin{align}
 \dfrac{d X_{\alpha}}{d \trueanom} &= \epsilon  Q^{(0)}_{\alpha} (X_{\beta}(\trueanom),\trueanom) + \epsilon^2 Q^{(1)}_{\alpha} (X_{\beta}(\trueanom),\trueanom) , \label{eqn:ODE_main}
\end{align}
where $X_{\alpha} = \{ p, e, \varpi, \mft\}$, with $\mft = \epsilon \, t$ and $\epsilon \ll 1$. The smallness of $\epsilon$ gives rise to a separation between the orbital timescale $\tau_{\rm orb}$ and the secular evolution timescale $\tau_{\rm sec} \sim \tau_{\rm orb}/\epsilon$.
We have introduced $ Q^{(0)}_{\alpha} (X_{\beta}(\trueanom),\trueanom) $ and $ Q^{(1)}_{\alpha} (X_{\beta}(\trueanom),\trueanom) $ as the sourcing functions for  $X_{\alpha}$ at LO and NLO respectively, where $Q^{(1)}_\alpha$ accounts for the correction sourced by \cref{eq:dthetadt}.

We exploit this separation of timescales, $\tau_{\rm orb} \ll \tau_{\rm sec}$, using MSA~\cite{BenderOrszag1978,KevorkianCole1996} to seek solutions for $X_{\alpha}$ that are perturbative in $\epsilon$ and consist of the LO evolution of the orbital elements and its PA corrections.
These corrections can be split into secular and oscillatory pieces, but the MSA ansatz for this split is not unique. In the approach of~\cite{PoissonWill2014}, both the secular and the oscillatory pieces are expanded in a series in $\epsilon$. In the approach of~\cite{Mora:2003wt,Will:2016}, only the oscillatory piece is expanded in $\epsilon$, while the secular piece is resummed, which is closely related to performing a near-identity averaging transformation. Although these two formulations appear different, they are equivalent, as we show below: starting from the MSA formulation of~\cite{PoissonWill2014}, we resum the secular pieces and recover the results of~\cite{Mora:2003wt,Will:2016} to the order considered.

\subsection{MSA formalism} \label{subsec:expanded-MSA}
Following~\cite{PoissonWill2014}, we first introduce a slow variable $\zeta= \epsilon\trueanom$, which converts the differential operator into
\begin{align}
    \ddphi{} = \ppphi{} + \epsilon \pptheta{}. \label{eqn:P_W_diff_op}
\end{align}
We next promote $X_{\beta}(\trueanom) \rightarrow X_{\beta} (\zeta,\trueanom)$ and propose an ansatz given by 
\begin{align}
\begin{aligned}
    X_{\alpha} (\zeta,\trueanom) &= X^{(0)}_{\alpha} (\zeta, \trueanom) + \epsilon X^{(1)}_{\alpha} (\zeta, \trueanom) \\
    &+ \epsilon^2 X^{(2)}_{\alpha} (\zeta, \trueanom)+\bigO{\epsilon^3}, \label{eqn:P_W_ansatz}
\end{aligned}
\end{align}
where we assume that each of the $X^{(n)}_{\alpha}$ functions are periodic in $\trueanom$. 

We now iterate~\cref{eqn:ODE_main} using~\cref{eqn:P_W_diff_op,eqn:P_W_ansatz} and obtain the following equations at each order: 
\begin{subequations}\label{eqn:MSA_PoissonWill}
    \begin{align}
 \quad \ppphi{X^{(0)}_{\alpha} } &= 0 , \label{eqn:MSA_LO} \\
\quad \pptheta{X^{(0)}_{\alpha}}+\ppphi{X^{(1)}_{\alpha}} &= Q^{(0)}_{\alpha} (X^{(0)}_{\gamma},\trueanom) , \label{eqn:MSA_NLO}\\
 \quad \pptheta{X^{(1)}_{\alpha}}+\ppphi{X^{(2)}_{\alpha}} &= X^{(1)}_{\beta} \partial_{\beta}  Q^{(0)}_{\alpha} (X^{(0)}_{\gamma},\trueanom) + Q^{(1)}_{\alpha}(X^{(0)}_{\gamma},\vartheta),
 \label{eqn:MSA_NNLO}
    \end{align}
\end{subequations}
where we have introduced $\partial_{\beta} \equiv \partial / \partial X_{\beta}$; note that the derivative with respect to $X_{\beta}$ has to be taken first and then evaluated at $X_{\beta} = X_{\beta}^{(0)}$. At $\bigO{\epsilon^{0}}$, \cref{eqn:MSA_LO} tells us that the quantities $X^{(0)}_{\alpha}$ do not vary over the short timescale $\trueanom$, and so $X^{(0)}_{\alpha} = X^{(0)}_{\alpha} (\zeta)$. 
In other words, at LO in MSA, the solution does not depend on the fast timescale, but rather, it depends only on the slow, secular timescale. At $\bigO{\epsilon^1}$, we 
average~\cref{eqn:MSA_NLO} over $\trueanom$, where the average of a function $A (\zeta, \trueanom)$ is defined as
\begin{align}
    \av{A} (\zeta) \equiv \dfrac{1}{2 \pi} \int \limits_{0}^{2\pi} A (\zeta, \trueanom) d\trueanom.
\end{align}
By~\cref{eqn:MSA_LO}, $X_{\alpha}^{(0)}$ depends only on $\zeta$, so the first term of~\cref{eqn:MSA_NLO} is unaffected by the average and is a total derivative. The average of $\partial X^{(1)}_{\alpha}/\partial\trueanom$ vanishes by periodicity in $\trueanom$. We thus find,
\begin{align}
 \ddtheta{X_{\alpha}^{(0)}}  = \av{Q^{(0)}_{\alpha} (X_{\beta}^{(0)},\trueanom)}, \label{eqn:MSA_LO_adiabatic}   
\end{align}
which governs the 0PA evolution.
Now, inserting~\cref{eqn:MSA_LO_adiabatic} back into~\cref{eqn:MSA_NLO}, we obtain
\begin{align}
    \ppphi{X^{(1)}_{\alpha}} &= Q^{(0)}_{\alpha} (X^{(0)}_{\beta},\trueanom) - \av{Q^{(0)}_{\alpha} (X_{\beta}^{(0)},\trueanom)} ,\label{eqn:MSA_post_adiabatic}
\end{align}
which integrates to
\begin{align}
    \xn{1}{\alpha} = \tilde{X}^{(1)}_{\alpha}(\zeta) +  Y_{\alpha}^{(1)}(\zeta,\trueanom). \label{eqn:osc_sec_split}
\end{align}
The oscillatory piece is
\begin{align}
    Y_{\alpha}^{(1)}(\zeta,\trueanom) = \int \limits^{\trueanom} [Q^{(0)}_{\alpha} (\tilde{X}^{(0)}_{\gamma},\trueanom') - \av{Q^{(0)}_{\alpha} (\tilde{X}_{\gamma}^{(0)} ,\trueanom')} ] d \trueanom', \label{eqn:Y_osc}
\end{align}
where we introduced $\tilde{X}^{(0)}_{\alpha} \equiv X^{(0)}_{\alpha}$ as the purely secular LO solution, and the constant in the indefinite integral is chosen such that $\av{Y_\alpha^{(1)}}=0$.
The secular piece $\tilde{X}_{\alpha}^{(1)}(\zeta)$ arises as an ``integration constant'' (constant with respect to $\trueanom$)~\cite{BenderOrszag1978,PoissonWill2014}.

At ${\cal{O}}(\epsilon^2)$, we proceed similarly and average~\cref{eqn:MSA_NNLO} over $\trueanom$. 
Periodicity again removes $\av{\partial X_\alpha^{(2)} /\partial\trueanom} =0$, and 
$\av{Y^{(1)}_{\alpha}} = 0$ gives $\av{\partial X_{\alpha}^{(1)}/\partial\zeta} = d\tilde{X}_{\alpha}^{(1)}/d\zeta $. Since $\tilde{X}^{(1)}_{\beta}$ does not depend on $\trueanom$, it factors out from the average integral, and we find
\begin{align}\label{eqn:MSA_NLO_adiabatic_full}
\begin{split}
    \ddtheta{\tilde{X}^{(1)}_{\alpha}(\zeta)} &= \tilde{X}^{(1)}_{\beta} \av{ \partial_{\beta}  Q^{(0)}_{\alpha} (\tilde{X}^{(0)}_{\gamma},\trueanom)} \\ 
    &+ \av{ Y^{(1)}_{\beta} \partial_{\beta}  Q^{(0)}_{\alpha} (\tilde{X}^{(0)}_{\gamma},\trueanom)} \\
    & + \av{Q^{(1)}_{\alpha} (\tilde{X}_{\gamma}^{(0)},\trueanom)}.
\end{split}
\end{align}
The secular evolution through 1PA is then governed by~\cref{eqn:MSA_LO_adiabatic,eqn:MSA_NLO_adiabatic_full}, which must be solved sequentially. First, solving the 0PA system, ~\cref{eqn:MSA_LO_adiabatic}, provides the ``background'' solution  $\tilde{X}_{\alpha}^{(0)}$. Evaluating \cref{eqn:Y_osc} on it, then, yields the oscillatory piece $Y^{(1)}_{\alpha}$. Finally, solving the 1PA system \cref{eqn:MSA_NLO_adiabatic_full} yields $\tilde{X}_{\alpha}^{(1)}$.

In general, both the 1PA oscillatory ($Y_{\alpha}^{(1)}$) and secular ($\tilde{X}^{(1)}_{\alpha}$) pieces of $\xn{1}{\alpha}$ contribute to the right-hand side (RHS) of~\cref{eqn:MSA_NLO_adiabatic_full}.
Directly below their Eq.~(12.238), Poisson and Will~\cite{PoissonWill2014} state instead that the contribution involving the secular piece,
$\tilde{X}^{(1)}_{\beta}\av{\partial_{\beta}Q^{(0)}_{\alpha}(\tilde{X}^{(0)}_{\gamma},\trueanom)}$,
vanishes.
However, the averaged derivative in this contribution can be expressed as
\begin{equation} \label{eqn:Jacobian}
\left.
\left\langle
\frac{\partial Q_\alpha^{(0)}(X_{\gamma},\vartheta)}
{\partial X_\beta}
\right\rangle
\right|_{X_{\gamma}=\tilde{X}_{\gamma}^{(0)}}
=
\frac{\partial \left\langle Q_\alpha^{(0)} \right\rangle}
{\partial \tilde{X}^{(0)}_\beta},
\end{equation}
because the secular variables are held fixed while averaging over the fast variables.
Thus, this average is nothing but the Jacobian matrix of the orbit-averaged 0PA sourcing functions, evaluated on the 0PA solution.
This Jacobian matrix need not vanish when the averaged sourcing functions depend on the orbital elements, as we show explicitly for radiation reaction in Sec.~\ref{sec:PA_RR}.
Nonetheless, resumming the secular dynamics following~\cite{Mora:2003wt,Will:2016,Will:2019lfe} absorbs this contribution into the leading orbit-averaged sourcing functions evaluated on the resummed secular variables.
For radiation reaction, the resummed equations recover the form of the final secular evolution equations given in~\cite{PoissonWill2014} [their Eqs.~(12.250)], as we will show in Sec.~\ref{subsec:resum_IC}.

\subsection{Equivalence with resummed MSA and initial conditions} \label{subsec:resum_IC}
An alternative MSA ansatz comes from the formalism of~\cite{Mora:2003wt,Will:2016,Will:2019lfe}, in which the orbital elements read
\begin{align}
X_{\alpha} = \tilde{X}_{\alpha}(\zeta) + \epsilon Y^{(1)}_{\alpha} (\zeta,\vartheta) +\mathcal{O}(\epsilon^2), \label{eqn:resum_ansatz}
\end{align}
so that only the oscillatory piece is expanded in $\epsilon$.
In other words, with~\cref{eqn:resum_ansatz}, one applies a near-identity averaging transformation~\cite{KevorkianCole1996} to the orbital elements by going from $X_{\alpha}$ to $\tilde{X}_{\alpha}$, which is the approach used for evolving EMRIs 
~\cite{VanDeMeent:2018cgn,Lynch:2021ogr} and EOB dynamics~\cite{Lynch:2026ibo,Lynch:2026eyw,Gamba:2026ado}. 

Below, we show that Below, we show that the approach of [37, 43] is equiva- approach of~\cite{Mora:2003wt,Will:2016} is equivalent to that of~\cite{PoissonWill2014}.
To do so, we first resum the secular 1PA evolution in the following way.
Let $\tilde{X}_{\alpha} = \tilde{X}^{(0)}_{\alpha} + \epsilon \tilde{X}^{(1)}_{\alpha} + \mathcal{O}(\epsilon^2)$, which results in the resummed secular variation:
\begin{align}
 \ddtheta{\tilde{X}_{\alpha}} &= \ddtheta{\tilde{X}^{(0)}_{\alpha}} + \epsilon \ddtheta{\tilde{X}^{(1)}_{\alpha}} + \bigO{\epsilon^2} \label{eqn:resum_LHS},
\end{align}
along with the following relations:
\begin{align}
\begin{split}\label{eqn:resum_simplify}
Q^{(0)}_{\alpha} (\tilde{X}_{\gamma},\trueanom) &= Q^{(0)}_{\alpha} (\tilde{X}_{\gamma}^{(0)},\trueanom) \\
 &+ \epsilon \tilde{X}^{(1)}_{\beta}   \partial_{\beta}  Q^{(0)}_{\alpha} (\tilde{X}^{(0)}_{\gamma},\trueanom) + \bigO{\epsilon^2}, \\ 
 \epsilon \partial_{\beta}  Q^{(0)}_{\alpha} (\tilde{X}_{\gamma},\trueanom) &= \epsilon \partial_{\beta}  Q^{(0)}_{\alpha} (\tilde{X}^{(0)}_{\gamma},\trueanom) + \bigO{\epsilon^2},\\
 \epsilon Q_{\alpha}^{(1)}(\tilde{X}_{\gamma},\trueanom) &= \epsilon Q_{\alpha}^{(1)}(\tilde{X}^{(0)}_{\gamma},\trueanom) + \bigO{\epsilon^2}.
\end{split}
\end{align}    
Furthermore, the oscillatory piece $Y^{(1)}_{\alpha} (\zeta,\vartheta)$ is still given by~\cref{eqn:Y_osc}, but with the sourcing functions evaluated at $\tilde{X}_{\beta}$, which is equivalent to evaluating at $\tilde{X}_{\beta}^{(0)}$, up to 2PA corrections.
Adding~\cref{eqn:MSA_LO_adiabatic} and $\epsilon$ times~\cref{eqn:MSA_NLO_adiabatic_full}, and using~\cref{eqn:resum_simplify} to re-express the right-hand side in terms of $\tilde{X}_{\beta}$, we obtain the resummed secular evolution of~\cref{eqn:resum_LHS}:
\begin{align}
\begin{split}\label{eqn:PWresum}  
    \ddtheta{\tilde{X}_{\alpha}} &= \ddtheta{\tilde{X}^{(0)}_{\alpha}} + \epsilon \ddtheta{\tilde{X}^{(1)}_{\alpha}} + \bigO{\epsilon^2} \\
    &= \av{Q^{(0)}_{\alpha} (\tilde{X}_{\gamma},\trueanom)}  \\
    &+ \epsilon  \av{Y^{(1)}_{\beta} \partial_{\beta}  Q^{(0)}_{\alpha} (\tilde{X}_{\gamma},\trueanom)}  \\
    & + \epsilon \av{Q_{\alpha}^{(1)}(\tilde{X}_{\gamma},\trueanom)} + \bigO{\epsilon^2},   
\end{split}
\end{align}
which is the same secular evolution obtained using the formalism of~\cite{Mora:2003wt,Will:2016}. This, therefore, establishes equivalence with the formalism of~\cite{PoissonWill2014}. When using the resummed secular variables, $d \tilde{X}_{\alpha}/ d\zeta$ are explicitly only functions of $\tilde{X}_{\beta}$.  Thus, one needs to only solve a single ODE system, given to 1PA order by~\cref{eqn:PWresum}. In contrast, when using the formulation of~\cite{PoissonWill2014}, one has to solve a system of ODEs, one system \emph{at each order} (given to 1PA by~\cref{eqn:MSA_LO_adiabatic,eqn:MSA_NLO_adiabatic_full}).

What are the initial conditions in the two MSA formulations? 
For clarity, we address this by focusing on only one of the orbital elements, the semi-latus rectum $p$; the same arguments hold for the other orbital elements.
Suppose that $\vartheta = \vartheta_0$ is the initial value of the true anomaly, with $p_0 = p(\vartheta_0)$ being the initial value of the osculating semi-latus rectum.
In the formulation of~\cite{PoissonWill2014}, let the initial values of the 0PA semi-latus rectum be $ \tilde{p}^{(0)}_0$ and its 1PA correction be $ \epsilon \tilde{p}^{(1)}_0 + \epsilon Y^{(1)}_{p,0}$; thus, $p_0 = \tilde{p}^{(0)}_0 + \epsilon \tilde{p}^{(1)}_0 + \epsilon Y^{(1)}_{p,0}$.
A conventional way to split this initial condition (see~\cite{Pound:2007ti,Pound:2007th} for related discussion) is to have $\tilde{p}^{(0)}_0 = p_0$, and $\tilde{p}^{(1)}_0 + Y^{(1)}_{p,0} = 0$. What is essential here is that the 1PA oscillatory part of the orbital element be average-free.

\begin{figure}[htb]
    \centering
    \includegraphics[width=\linewidth]{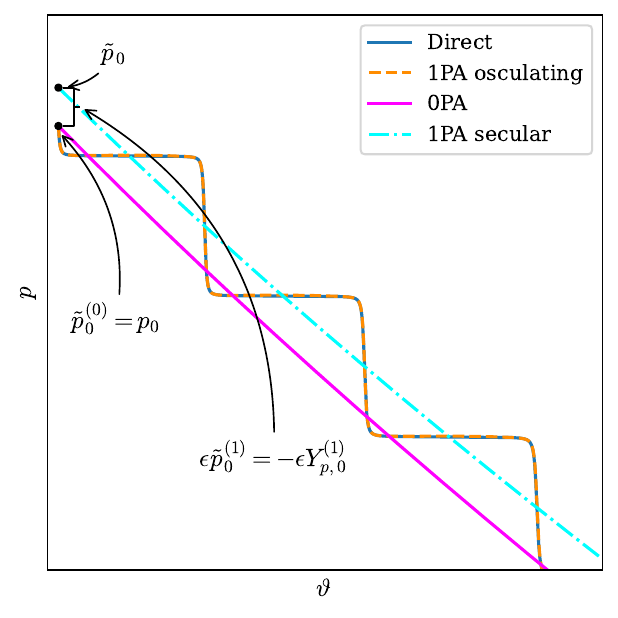}
    \caption{Sketch of the semi-latus rectum's evolution with respect to the true anomaly, illustrating how the initial conditions differ in the 0PA and 1PA approximations. The blue solid line corresponds to the direct (non-averaged) numerical evolution. The 0PA evolution (magenta solid line) is initialized at the osculating value $p_0$. The 1PA secular evolution (cyan dash-dot line) is initialized with $\tilde{p}_0 = p_0 - \epsilon Y^{(1)}_{p,0}$, which differs from the 0PA initial condition by precisely $\epsilon \tilde{p}_0^{(1)}$. The osculating 1PA solution (orange dashed line) tracks the direct solution up to 2PA corrections.}
    \label{fig:p_cartoon}
\end{figure}

From here, we can work out what the initial conditions in the resummed formulation of~\cite{Mora:2003wt,Will:2016} ought to be, so that one obtains consistent solutions between the two formulations. We obtain $\tilde{p}_0 = p_0  -  \epsilon Y^{(1)}_{p,0}$, where $\tilde{p}_0$ is the initial value of the resummed secular semi-latus rectum, with $\tilde{p}_0 = \tilde{p}^{(0)}_0 + \epsilon \tilde{p}^{(1)}_0 $. Note that when working to 0PA order, this just gives us $\tilde{p}_0 = p_0 $.
We illustrate this using a cartoon sketch in Fig.~\ref{fig:p_cartoon}. Thus, in the resummed formulation of~\cite{Mora:2003wt,Will:2016}, when working to 1PA order, one has to apply a 1PA shift to the initial condition for the secular orbital element. This shift is informed by the initial 1PA oscillation of the orbital element. We apply this to all the orbital elements, to obtain $\tilde{X}_{\alpha,0} = X_{\alpha,0} - \epsilon Y^{(1)}_{\alpha,0} $. In other words, the 0PA evolution starts from a snapshot of the initial osculating orbital elements and knows nothing of their initial oscillations, whereas the 1PA evolution accounts for it.

\section{Post-adiabatic effects due to radiation reaction} \label{sec:PA_RR}

This section studies how different PA effects induced by radiation reaction affect the orbital evolution. In Sec.~\ref{sec:qual_PA}, we define the PA corrections under study and qualitatively sketch their contribution to the orbital dynamics. In Sec.~\ref{subsec:rad-reac-balance}, we review the $2.5$PN radiation-reaction force in the ``generalized gauge'', following~\cite{Iyer:1995rn,Arun:2009mc}. In Sec.~\ref{subsec:adiabatic}, we derive the adiabatic evolution, recovering the results of Peters in~\cite{Peters:1964zz}. Finally, in Secs.~\ref{sec:1PA_osc} and~\ref{sec:1PA_sec}, we compute the 1PA oscillatory and secular evolution, respectively. 

\subsection{Qualitative analysis of PA corrections due to radiation reaction}\label{sec:qual_PA}
It is important to clarify what we mean by PA corrections due to radiation reaction within PN theory.
The EoM can be split into a conservative and dissipative sector, with conservative PN effects entering at $1$PN order relative to the Newtonian motion, and the dissipative radiation-reaction effects entering at relative $2.5$PN order. In order to compute the orbital evolution under both conservative and dissipative PN effects, one can treat both the conservative and dissipative PN accelerations as perturbations on the Keplerian dynamics and perform MSA by treating $v \ll 1$ as the formal expansion parameter~\cite{Mora:2003wt,Will:2016,Tucker:2021mvo}. 

PA corrections can arise from both the conservative and dissipative sectors when the motion is described using Keplerian orbital elements~\cite{Mora:2003wt,Will:2016}. Conservative PN effects, beginning at 1PN order, cause the Keplerian energy and angular momentum to oscillate, producing conservative 1PA oscillations in the orbital elements~\cite{Lincoln_Will_1990,PoissonWill2014}. On the other hand, the quasi-Keplerian parametrization~\cite{Damour_Deruelle} incorporates these conservative effects into the underlying orbital description, so its orbital elements do \textit{not} exhibit these conservative 1PA oscillations.
In this quasi-Keplerian parametrization, 1PA corrections are instead sourced only by radiation reaction, and thus, are purely dissipative~\cite{Damour:2004bz,Moore:2016qxz,Boetzel:2019nfw}. In this paper, we focus on the leading PN contributions to the dissipative 1PA terms, which arise from the 2.5PN radiation-reaction acceleration. Higher-order PN corrections to this acceleration produce sub-leading PN corrections to the dissipative 1PA terms. Restricting our calculation to the leading PN contribution allows us to use the Keplerian parametrization and obtain an analytical picture of what controls the accuracy of the 0PA and 1PA approximations.

Prior to carrying out detailed MSA calculations, it is useful to estimate how well separated the orbital and radiation-reaction timescales are. The radiation-reaction timescale scales as $\tau_{\rm rr} \sim (5/64)\, \eta^{-1} v_p^{-5}\, \tau_{\rm orb}$, where $\tau_{\rm orb} \sim \sqrt{a^3/m}$ is the orbital timescale and $v_p \sim \sqrt{m/p}$ is a characteristic Newtonian, pericenter velocity scale\footnote{The Newtonian vis-viva relation gives the pericenter speed $v_{\rm peri,inst}=(1+e)\sqrt{m/p}$, with $p=a(1-e^2)$ (see, e.g.,~\cite{Moore:2019xkm}).
Thus, $v_p=\sqrt{m/p}$ captures its characteristic scaling but omits the factor $(1+e)$.
This relation also holds for instantaneous Keplerian osculating elements.
In Sec.~\ref{sec:num_exp}, we instead use the characteristic velocity $v_{\rm peri}\equiv\sqrt{m/r_{\rm peri}}=v_p\sqrt{1+e}$ to set the initial and stopping conditions.
Here, we use $v_p$ only to characterize orbital scales, so its distinction from the actual pericenter speed does not affect the discussion.}, so that the ratio of timescales is $\epsilon \sim (64/5)\, \eta\, v_p^5$; recall here that  $\eta = q/(1+q)^2$ is the symmetric mass ratio, with $q = m_2/m_1 \leq 1$ the mass ratio. The 1PA dynamics (secular or oscillatory) contribute at $\mathcal{O}(\epsilon)$ relative to the 0PA dynamics.
Therefore, the 1PA corrections to the orbital elements depend linearly on $\eta$, and, for a given semi-major axis, the separation of timescales degrades at larger eccentricity, owing to the smaller $p = a(1-e^2)$ and hence larger pericenter velocity. For example, with $\eta \sim 1/4$ (comparable masses) and $a \sim 100\,m$, we estimate $\epsilon \sim 3 \times 10^{-5}$ for $e \sim 0.1$, whereas increasing the eccentricity to $e \sim 0.9$ gives $\epsilon \sim 2 \times 10^{-3}$, nearly two orders of magnitude larger. Requiring the timescales to be separated by at least a factor $\epsilon_{\rm thresh}^{-1}$, i.e., $\epsilon < \epsilon_{\rm thresh}$, translates into an upper bound on $e$, for given $a$, $m$, and $\eta$:
\begin{align}
e^2 \lesssim 1 - \frac{m}{a}\left(\frac{64}{5}\,\eta\,\epsilon_{\rm thresh}^{-1}\right)^{2/5}.
\end{align}
For instance, $\epsilon_{\rm thresh} \sim 10^{-3}$, with $\eta \sim 1/4$ and $a \sim 100\,m$, gives $e \lesssim 0.86$. The 0PA approximation relies on this separation of timescales, which, in PN theory, depends on the eccentricity. The separation becomes less clean at large eccentricity, as well as during the late inspiral, where $v_p$ increases. Such an estimate helps build intuition for when the PA corrections become relevant, but $\epsilon$ only characterizes the separation of timescales: it does not tell us how the PA corrections accumulate in the orbital phase, nor their detailed structure, for which we need the calculations below, together with an understanding of how the errors scale for each orbital element.

\subsection{Radiation-reaction force in a generalized gauge} \label{subsec:rad-reac-balance}
Radiation-reaction effects enter at leading $2.5$PN order in the EoM. At $2.5$PN order, there is gauge freedom in the near-zone radiation-reaction potentials, and, thus, the radiation-reaction force is gauge-dependent~\cite{Iyer:1995rn,PoissonWill2014}. As demonstrated in~\cite{Iyer:1995rn}, the resulting far-zone fluxes do not identically match the mechanical work done on the binary, with the difference being the gauge-dependent oscillatory Schott terms. In~\cite{Iyer:1995rn}, the authors found that the flux balance laws, can be satisfied when including the Schott terms, with a two-parameter residual gauge freedom at $2.5$PN order.
The resulting generalized radiation-reaction acceleration, which satisfies the balance laws can be expressed as:
\begin{subequations}\label{eqn:gen_rr_acc}
\begin{align}
    \vec{a}_{\mathrm{rr}} &= \dfrac{8}{5} \eta \dfrac{m^2}{r^3} \left[ \dot{r} \vec{n} A_{2.5\mathrm{PN}}- \vec{v} B_{2.5\mathrm{PN}} \right] ,\\
    A_{2.5\mathrm{PN}} &= A_1 v^2 +  A_2 \dfrac{m}{r} + A_3 \dot{r}^2,\\
    B_{2.5\mathrm{PN}} &= B_1 v^2 +  B_2 \dfrac{m}{r} + B_3 \dot{r}^2, \\
    \vec{v} &= \dot{r} \vec{n} + r\dot{\phi} \vec{\lambda},
\end{align}    
\end{subequations}
where the coefficients $A_i$ and $B_i$ are given by   
\begin{equation}
\begin{aligned}
A_1 &= 3+3\beta_g, &
A_2 &= \frac{23}{3}+2\alpha_g-3\beta_g, \quad
A_3 = -5\beta_g, \\
B_1 &= 2+\alpha_g, &
B_2 &= 2-\alpha_g, \quad
B_3 = -3-3\alpha_g, 
\end{aligned}
\end{equation}
with $(\alpha_g,\beta_g)$ the Iyer-Will gauge parametrization. The same coefficients can be represented in the Poisson-Will parameterization~\cite{PoissonWill2014} in terms of $(a_g, b_g)$ using the relations  
\begin{align}
\alpha_g = (35/6) a_g - (5/2) b_g + 4 ,\quad  \beta_g = 5 a_g - (15/4) b_g + 5.
\end{align}
In~\cref{tab:gauge_correspondence}, we tabulate the gauge parameter values for a few commonly used gauge choices. This gauge freedom arises from 2.5PN transformations of the relative separation $\vec{r}$, as shown in~\cite{Iyer:1995rn}, which we will discuss in more detail in \cref{sec:waveform}.
\begin{table}[!t]
\renewcommand{\arraystretch}{1.3}
\centering
\begin{tabular}{lcc}
\hline
\textbf{Gauge} & $(a_g, b_g)$ & $(\alpha_g, \beta_g)$ \\
\hline
Standard harmonic~\cite{Lincoln_Will_1990,Iyer:1995rn} & $(-2/3,\; 4/9)$ & $(-1,\; 0)$ \\
\hline
ADM~\cite{Jaranowski:1997,Konigsdorffer:2003ue} & $(-2/5,\; 0)$ & $(5/3,\; 3)$ \\
\hline
Burke--Thorne~\cite{Iyer:1995rn,Galley:2016zee} & $(0,\; 0)$ & $(4,\; 5)$ \\
\hline
EOB~\cite{Gamboa:2024hli, Gamboa:2024imd} & $(-\frac{2}{3}, \frac{98}{45})$ & $(-\frac{16}{3}, -\frac{13}{2})$ \\
\hline
\end{tabular}
\caption{Gauge parameter correspondence in terms of the $(a_g,b_g)$ Poisson-Will parametrization and the $(\alpha_g,\beta_g)$ Iyer-Will parametrization. 
}
\label{tab:gauge_correspondence}
\end{table}
With $\vec{a}_{rr}$ in hand, we can use the osculating orbit and MSA formalisms of \cref{sec:orbital_MSA} to obtain the 0PA and 1PA evolutions. The details of this implementation are broken down as follows. Using~\cref{eqn:acc_decomp}, we obtain the components of the perturbing acceleration, which are then used to obtain the evolution equations to $\mathcal{O}(\epsilon^2)$, as given in~\cref{eqn:ODE_main}. The sourcing functions $Q_{\alpha}^{(0)}$ and $Q_{\alpha}^{(1)}$ are provided in a supplemental Mathematica file for completeness. With the sourcing functions in hand, we use~\cref{eqn:PWresum} and~\cref{eqn:Y_osc} to compute the 1PA accurate evolution.

\allowdisplaybreaks[4]

\subsection{Adiabatic (0PA) evolution} \label{subsec:adiabatic}
The adiabatic or 0PA evolution of $a,e$, and $\varpi$, expressed using the resummed secular orbital elements (see Sec.~\ref{subsec:resum_IC}) is given by
\begin{subequations}\label{eqn:adiabatic_evol}
\begin{align}
\dfrac{d\tilde{a}}{d \zeta}
&= -\frac{2\left(96 + 292 \tilde{e}^2 + 37 \tilde{e}^4\right)m^{5/2}\eta}
{15\,\tilde{a}^{3/2}(1-\tilde{e}^2)^{7/2}}, \\
\dfrac{d\tilde{e}}{d \zeta} &=
-\frac{\tilde{e}\left(304 + 121\tilde{e}^2\right)m^{5/2}\eta}
{15\,\tilde{a}^{5/2}(1-\tilde{e}^2)^{5/2}},\\
\dfrac{d \tilde{\varpi}}{d \zeta} &= 0,
\end{align}
\end{subequations}
with the adiabatic time evolution given by
\begin{align}
\dfrac{d\tilde{\mft}}{d\zeta} &= \left( \frac{\tilde{a}^3}{m} \right)^{1/2}, \label{eqn:adiabatic_evol_t}
\end{align}
truncated at $\bigO{\epsilon}$. At 0PA order, one can obtain an exact analytic solution for $\tilde{a}$ as a function of $\tilde{e}$, which is one of the main results in~\cite{Peters:1964zz}. This is possible because the sourcing functions only depend on $\tilde{a}$ and $\tilde{e}$ and not on the other orbital elements, after integrating out the short timescale $\vartheta$.

Furthermore, this decoupling at 0PA means that there is a 0PA conserved quantity of the motion, given by~\cite{Peters:1964zz}
\begin{align}
\mathcal{C} =  \tilde{a} (1 - \tilde{e}^2)  \tilde{e}^{-12/19} \left(1 + \frac{121}{304} \tilde{e}^2 \right)^{-870/2299},\label{eqn:adiabatic_const}
\end{align}
which is determined entirely by the initial conditions $a_0$ and $e_0$. We can understand the physical meaning of $\mathcal{C}$ by inspecting its high-eccentricity limit $1-\tilde{e} \ll 1$, which gives $\mathcal{C} \sim \tilde{p}$, up to numerical factors.  In other words, the conservation of the adiabatic constant $\mathcal{C}$ translates to the approximate conservation of the semi-latus rectum for burst sources~\cite{ColemanMiller:2021lky}. 

\subsection{1PA evolution}
\subsubsection{1PA oscillatory corrections}\label{sec:1PA_osc}
From~\cref{eqn:Y_osc}, we obtain the oscillatory 1PA corrections to $a,e$, and $\varpi$ as
\begin{subequations} \label{eqn:1PA_osc_sol}
\begin{align}
Y^{(1)}_a &= \dfrac{\eta m^{5/2}\tilde{e}}{\tilde{a}^{3/2}(1-\tilde{e}^2)^{7/2}}  \sum_{n=1}^{5} f^{(n \vartheta)}_{a} (\tilde{e},a_g,b_g)   \sin (n \vartheta)  , \\
Y^{(1)}_e &= \dfrac{\eta m^{5/2}}{\tilde{a}^{5/2}(1-\tilde{e}^2)^{5/2}}  \sum_{n=1}^{5}f^{(n \vartheta)}_{e} (\tilde{e},a_g,b_g) \sin (n \vartheta) ,  \\
Y^{(1)}_{\varpi} &= \dfrac{\eta m^{5/2}}{\tilde{e} \tilde{a}^{5/2}(1-\tilde{e}^2)^{5/2}}  \sum_{n=1}^{5}f^{(n \vartheta)}_{\varpi} (\tilde{e},a_g,b_g)  \cos (n \vartheta) ,
\end{align}
\end{subequations}
where the function coefficients $f^{(n \vartheta)}_{a,e,\varpi} (\tilde{e},a_g,b_g)$ (see the supplemental Mathematica file for the full expressions) have a polynomial dependence on the 0PA eccentricity $\tilde{e}$ and also depend on the gauge parameters $(a_g, b_g)$. We will address this gauge-dependence later in Sec.~\ref{sec:waveform}. The oscillatory 1PA correction $Y^{(1)}_{\mft}$ can be similarly computed, and we obtain the implicit expression:
\begin{align}
Y^{(1)}_{\mft} = (\ell-\trueanom) \sqrt{\frac{a^3}{m}}, \label{eqn:t_osc}
\end{align}
after using~\cref{eqn:kepler_solution}.
One can easily verify that $Y^{(1)}_{\mft} $ is purely oscillatory due to the fact that both $l$ and $\trueanom$ complete $2 \pi$ radians along the Keplerian osculating orbit. From here, we have that $\mathfrak{t} = \tilde{\mathfrak{t}} + \epsilon Y^{(1)}_{\mathfrak{t}}$, which implies that $t = \tilde{t} + Y^{(0)}_t$, with $Y^{(0)}_t = Y^{(1)}_{\mathfrak{t}}$. This simply shows that the MSA procedure has resulted in the leading decomposition of the time evolution into a secular and oscillatory piece, thereby reproducing the Keplerian solution along each osculating orbit. The oscillatory contributions from the NLO sourcing functions can be similarly computed, but we do not list them here.

The sourcing functions satisfy $\partial_{\varpi}Q_{\beta}^{(0)} = 0$ and $\partial_{\mft} Q_{\alpha} = 0$. This means that $Y^{(1)}_{\varpi}$ and $Y^{(1)}_{\mft}$ will not contribute to the secular evolution of the 1PA orbital elements, while $Y^{(1)}_{a}$ and $Y^{(1)}_{e}$ will. Furthermore, to evaluate~\cref{eqn:1PA_osc_sol} as a function of time to this order, we would need to substitute the 0PA orbital elements as functions of time and simply solve the Kepler equation (cf.~\cref{eqn:Kepler_eqn}) to invert the true anomaly in terms of time.
We explain how to do this in more detail in Sec.~\ref{sec:setup}.

\subsubsection{1PA secular evolution}\label{sec:1PA_sec}
Using the resummed secular orbital elements, we obtain the secular evolution up to 1PA order to be
\begin{subequations}\label{eqn:1PA_sec_resum}
\begin{align}
\dfrac{d \tilde{a}}{d\zeta} &= -\dfrac{2 (96 + 292 \tilde{e}^2 + 37 \tilde{e}^4) m^{5/2} \eta}{
 15 \tilde{a}^{3/2} (1 - \tilde{e}^2)^{7/2}}  + \mathcal{O}(\epsilon^2) , \\
 \dfrac{d \tilde{e}}{d\zeta} &= -\dfrac{\tilde{e} (304+ 121 \tilde{e}^2) m^{5/2} \eta}{15 \tilde{a}^{5/2} (1-\tilde{e}^2)^{5/2}} + \mathcal{O}(\epsilon^2), \\
\dfrac{d \tilde{\varpi}}{d\zeta} &= \epsilon \mathcal{G}(\tilde{e},\tilde{a},a_g,b_g) + \mathcal{O}(\epsilon^2) \label{eqn:1PA_sec_resum_peri} , \\
\dfrac{d \tilde{\mft}}{d\zeta} &= \left( \dfrac{\tilde{a}^3}{m} \right)^{1/2} + \mathcal{O}(\epsilon^2),
\end{align}
\end{subequations}
where the $\mathcal{O}(\epsilon)$ contributions to the evolution equations for $\tilde{a}$, $\tilde{e}$, and $\tilde{\mft}$ vanish, but they do not for $\tilde{\varpi}$, whose secular evolution is explicitly governed by $\mathcal{G}$, which we provide in the supplemental Mathematica file.

Although the only secular 1PA correction in the evolution equations is for $\tilde{\varpi}$, there are 1PA corrections to the solutions for $\tilde{e}$ and $\tilde{a}$, due to the 1PA accurate initial conditions.
In fact, since the 1PA evolution equations for $\tilde{e}$ and $\tilde{a}$ are the same as in the 0PA case, we can compute $\tilde{a}$ as a function of $\tilde{e}$ in terms of a 1PA adiabatic constant, which has the same functional form as~\cref{eqn:adiabatic_const}, but evaluated using the 1PA initial conditions.
Thus, the 1PA secular dynamics for $\tilde{a}$ and $\tilde{e}$ have the same trajectories as in the 0PA case, with contours given by~\cref{eqn:adiabatic_const}. When expanding to $\mathcal{O}(\epsilon)$ order, the 1PA adiabatic constant is
\begin{align}
\mathcal{C}_{\rm 1PA} = \mathcal{C}_{\rm 0PA} & \left[ 1 - \epsilon \dfrac{Y^{(1)}_{a,0}}{a_0} \right. \nn
&+ \epsilon \left. \dfrac{2 (96 + 292 e_0^2 + 37 e_0^4) Y^{(1)}_{e,0}}{e_0 (1 - e_0^2) (304 + 121 e_0^2)} +\mathcal{O}(\epsilon^2) \right],
\end{align}
where $\mathcal{C}_{\rm 0PA} $ is simply evaluated using~\cref{eqn:adiabatic_const} with $\tilde{a}=a_0$ and $\tilde{e} = e_0$. 

We now explicitly demonstrate that the 1PA secular evolution obtained using the formulation of~\cite{PoissonWill2014} is consistent with~\cref{eqn:1PA_sec_resum}.
The secular evolution for $\tilde{a}^{(1)},\tilde{e}^{(1)}$, $\tilde{\varpi}^{(1)}$, and $\mft^{(1)}$ is given by
\begin{subequations}\label{eqn:1PA_sec_standard}
\begin{align}
\dfrac{d \tilde{a}^{(1)}}{d\zeta} &= \tilde{a}^{(1)} \mathcal{F}^a_a (\tilde{a}^{(0)},\tilde{e}^{(0)}) + \tilde{e}^{(1)} \mathcal{F}^a_e (\tilde{a}^{(0)},\tilde{e}^{(0)}),\\
\dfrac{d \tilde{e}^{(1)}}{d\zeta} &= \tilde{e}^{(1)} \mathcal{F}^e_e (\tilde{a}^{(0)},\tilde{e}^{(0)}) + \tilde{a}^{(1)} \mathcal{F}^e_a (\tilde{a}^{(0)},\tilde{e}^{(0)}), \\
\dfrac{d \tilde{\varpi}^{(1)}}{d\zeta} &= \mathcal{G}(\tilde{e}^{(0)},\tilde{a}^{(0)},a_g,b_g),\\
\dfrac{d\tilde{\mft}^{(1)}}{d\zeta} &=  \tilde{a}^{(1)} \mathcal{F}_a^{\mathfrak{t}} (\tilde{a}^{(0)},\tilde{e}^{(0)}), \label{eqn:1PA_sec_t_standard}
\end{align}
\end{subequations} 
where $\mathcal{F}_{\beta}^{\alpha} = \partial \langle Q_{\alpha}^{(0)} \rangle / \partial \tilde{X}_{\beta}^{(0)}$ is the Jacobian matrix defined in~\cref{eqn:Jacobian}, evaluated on the 0PA secular solution for $X_{\alpha} = \{a,e,\varpi,\mathfrak{t}\}$. Its components depend only on $\tilde{a}^{(0)}$ and $\tilde{e}^{(0)}$; explicit expressions are provided in the supplemental Mathematica file. The same components follow by expanding the sourcing terms in~\cref{eqn:1PA_sec_resum} using $\tilde{X}_{\alpha} = \tilde{X}^{(0)}_{\alpha} + \epsilon\,\tilde{X}^{(1)}_{\alpha}$. For example, expanding the leading secular time-evolution source gives $\mathcal{F}_a^{\mathfrak{t}} = \tfrac{3}{2}\sqrt{\tilde{a}^{(0)}/m}$. For the pericenter evolution, the Jacobian terms vanish because $\langle Q^{(0)}_{\varpi}\rangle$ is identically zero. 
Instead, $\mathcal{G}$, which is the same sourcing function as in~\cref{eqn:1PA_sec_resum_peri}, has two contributions: the orbit average of the NLO sourcing function, and the orbit-averaged coupling of the 1PA oscillations with the gradient of the LO sourcing function.

\Cref{eqn:1PA_sec_standard} also clarify how the secular 1PA corrections depend on the initial conditions.
For the eccentricity and semi-major axis, the secular 1PA corrections vanish when $Y^{(1)}_{a,0}=Y^{(1)}_{e,0}=0$, because $\tilde{a}^{(1)}$ and $\tilde{e}^{(1)}$ satisfy coupled linear homogeneous ODEs with vanishing initial conditions.
With this initial condition matching, the secular 1PA correction to $\mathfrak{t}$ is nonzero only when $Y^{(1)}_{a,0}\neq 0$, consistent with~\cite{Loutrel:2017fgu,Loutrel:2018ssg}.
By contrast, the inhomogeneous source $\mathcal{G}$ drives a secular 1PA pericenter evolution, even when the initial pericenter correction vanishes. Together with the initial condition matching, these results explicitly verify the equivalence between the MSA formalisms.

An important feature of the resummed secular 1PA evolution (cf.~\cref{eqn:1PA_sec_resum}) is that the equations describing the orbit-averaged energy and angular momentum balance laws are the same at 1PA order as they are at 0PA order. Specifically, the 1PA secular rate of change of the orbital energy and the angular momentum, neglecting 2PA corrections, is still given by
\begin{align}
\begin{aligned}\label{eqn:balance_law}
\frac{d \tilde{E}}{d\tilde{t}} &= - \frac{32}{5} \epsilon \eta^2 \frac{m^5}{\tilde{a}^5 (1-\tilde{e}^2)^{7/2}} \left ( 1+ \frac{73}{24}\tilde{e}^2 + \frac{37}{96}\tilde{e}^4 \right),\\
\frac{d \tilde{L}}{d\tilde{t}}  &= - \frac{32}{5} \epsilon \eta^2  \frac{m^{9/2}}{\tilde{a}^{7/2} (1-\tilde{e}^2)^2} \left( 1 + \frac{7}{8} \tilde{e}^2 \right),
\end{aligned}
\end{align}
which is what one finds when averaging the radiative fluxes at infinity, as originally derived by Peters and Mathews~\cite{Peters:1963ux,Peters:1964zz}. However, note that~\cref{eqn:balance_law} does not mean there are no secular 1PA corrections to the resummed energy $\tilde{E}$ and angular momentum $\tilde{L}$. In fact, as we discussed earlier, such corrections arise from a 1PA shift in the initial conditions for the resummed secular orbital elements. Thus, in general, when solving for $\tilde{E}$ and $\tilde{L}$, there will be secular 1PA corrections. One can also see this from the MSA formulation of~\cite{PoissonWill2014} by re-expanding $\tilde{E}$ and $\tilde{L}$. For instance, with $\tilde{E} = \tilde{E}^{(0)} + \epsilon \tilde{E}^{(1)}$ , one can obtain the 1PA balance law for $d\tilde{E}^{(1)}/d\zeta$, which will linearly depend on $\tilde{a}^{(1)}$ and $\tilde{e}^{(1)}$ (and likewise for $d\tilde{L}^{(1)}/d\zeta$).

If the initial condition corresponds to a radial turning point, the 1PA oscillations in the energy and angular momentum are zero. In such a case, 0PA is already accurate in describing the orbit-averaged energy and angular momentum, up to 2PA corrections. This can also be understood from the fact that the Schott term contributions vanish when $\dot{r} = 0$. On the other hand, if initialized between radial turning points, the initial oscillations in the energy and angular momentum are non-zero, and thus including 1PA becomes necessary for accurately describing the orbit-averaged energy and angular momentum. This is why the accuracy of the 0PA depends sensitively on the initial true (or mean) anomaly.

\section{Quasi-circular limit \\and comparison to previous works} \label{sec:quasicirc}

We now discuss the quasi-circular limit $e \ll 1$ to obtain further insight into the 0PA and 1PA dynamics, and to compare with previous works. From~\cref{eqn:adiabatic_evol}, we see that the 0PA dynamics are regular in the $\tilde{e} \rightarrow 0$ limit. Therefore, one can perform a systematic post-circular expansion~\cite{Yunes:2009yz} in $\tilde{e} \ll 1$ to obtain closed-form analytic solutions to the 0PA secular evolution at the desired order in $e$, as done in~\cite{Yunes:2009yz,Loutrel:2018ydu,Loutrel:2018ssg,Will:2019lfe}.
To leading order in eccentricity, we find:
\begin{subequations}\label{eqn:small_ecc_sol}
\begin{align}
\tilde{a} &\sim a_0 \left[ 1- 32 \eta \dfrac{m^{5/2}}{a_0^{5/2}}  ( \zeta- \zeta_0) \right]^{2/5}, \\
\tilde{e} & \sim e_0 \left( \dfrac{\tilde{a}}{a_0} \right)^{19/12},
\end{align}
\end{subequations}
where the solution for $e$ just follows from inverting~\cref{eqn:adiabatic_const} in the small-eccentricity limit~\cite{Peters:1964zz}. 

However, for quasi-circular orbits, the pericenter is ill-defined, thus, the quasi-circular limit is singular, and, while this does not have an impact on the 0PA dynamics, it does affect the 1PA dynamics.  This is evident from the fact that the 1PA oscillatory contribution to the pericenter $Y^{(1)}_{\varpi}$ diverges when $\tilde{e} \rightarrow 0$. We can also see this in general from the direct evolution of the osculating orbital elements given by~\cref{eqn:LPE}, where $\dot{\varpi}$ diverges when $e \rightarrow 0$. Furthermore, one is not allowed to perform orbit averaging in the MSA formalism using $\vartheta$ in this limit, because there is a divergent contribution to $\dot{\vartheta}$ that comes from the perturbing acceleration. In fact, when $e \sim \epsilon \ll 1$, the contribution to $\dot{\vartheta}$ from the perturbing acceleration can be comparable to that from the Keplerian acceleration, causing $\vartheta$ to approach a fixed point and $\varpi$ to co-rotate with the orbital phase. Such a dynamical phenomenon was pointed out by~\cite{Lincoln_Will_1990} for the 1PN perturbation, and later emphasized again by~\cite{Loutrel:2018ydu,Loutrel:2018ssg,Will:2019lfe} for the 2.5PN perturbation. In the limit $e \sim \epsilon \ll 1$ when $\vartheta$ is approaching a fixed point, the sourcing functions in~\cref{eqn:ODE_main} are no longer oscillatory with respect to $\vartheta$. Thus, one needs to transform to variables that regularize~\cref{eqn:LPE} and also allow for oscillatory dependence in the sourcing functions, which can then be integrated using the MSA formalism.

A common choice for doing so is to decompose the sourcing terms using the orbital phase $\phi$, and to work in terms of the components of the Runge-Lenz or eccentricity vector $\vec{\mathcal{A}} = \mathcal{A}_x \vec{e}_X + \mathcal{A}_y \vec{e}_Y$, given by~\cite{PoissonWill2014}:
\begin{align}
\vec{\mathcal{A}} = \frac{1}{m} \vec{v} \cross \left (\vec{r} \cross \vec{v} \right) - \vec{n} = e \cos \varpi \vec{e}_X + e \sin \varpi \vec{e}_Y, 
\end{align}
with $e = |\vec{\mathcal{A}}|$ and $\varpi = \tan^{-1}(\mathcal{A}_y/\mathcal{A}_x)$.
One can see that under the transformation from $\{ e, \varpi, \vartheta \}$ to $ \{ \mathcal{A}_x, \mathcal{A}_y, \phi \}$,~\cref{eqn:LPE} becomes regular in the $e \rightarrow 0$ limit. This is why in~\cite{Loutrel:2018ydu,Loutrel:2018ssg,Will:2019lfe} (see also~\cite{Lynch:2026eyw,Gamba:2026ado}), the authors use $ \{ \mathcal{A}_x, \mathcal{A}_y, \phi\}$ to perform MSA for the 2.5PN radiation-reaction perturbation. Additionally, we will work here with $p$ instead of $a$ to simplify the sourcing functions.
In the following, we sketch the main results and implications from this regularized approach, with details provided in Appendix~\ref{sec:quasi_circ}.

In terms of $\{\mathcal{A}_x, \mathcal{A}_y, \phi \}$, there are no explicit NLO sourcing functions in the evolution equations for $dX_{\alpha}/ d\phi$ with $X_{\alpha} = \{p,\, {\mathcal{A}_x},\, {\mathcal{A}_y},\, {\mft} \}$, in contrast with~\cref{eqn:ODE_main}. We now perform MSA with these regularized variables using the resummed formalism by treating $\phi$ as the fast variable and $\Phi=\epsilon \phi$ as the slow variable.
Up to 1PA order, we recover~\cref{eqn:1PA_sec_resum} for the secular dynamics, after transforming back to $\tilde{a},\tilde{e}$, and $\tilde{\varpi}$. On the other hand, the 1PA oscillatory solutions for the components of the Runge-Lenz vector, expanded to leading order in the small eccentricity limit, are given by
\begin{subequations}
\begin{align}
Y^{(1)}_{\mathcal{A}_x} & \sim -\dfrac{64}{5} \dfrac{\eta m^{5/2}}{\tilde{a}^{5/2}} \sin \phi , \\
Y^{(1)}_{\mathcal{A}_y} & \sim \dfrac{64}{5} \dfrac{\eta m^{5/2}}{\tilde{a}^{5/2}} \cos \phi,
\end{align}
\end{subequations}
with the expressions for arbitrary eccentricity given in Appendix~\ref{sec:quasi_circ}, which are consistent with~\cite{Loutrel:2018ydu,Loutrel:2018ssg,Will:2019lfe}. Notably, these 1PA oscillatory solutions for the Runge-Lenz vector components are regular when $e \to 0$, in contrast with~\cref{eqn:1PA_osc_sol}.

Having obtained the regularized 1PA accurate solution, we now comment on the behavior of eccentricity in the $e \sim \epsilon \ll 1$ regime, which can happen during the late inspiral or even during the early inspiral if starting with a very small eccentricity. As noted by previous works~\cite{Lincoln_Will_1990,Loutrel:2018ydu,Loutrel:2018ssg,Will:2019lfe}, the osculating eccentricity grows when $e \sim \epsilon \ll 1$. In other words, the osculating eccentricity never quite vanishes, but instead asymptotes to $e \sim \epsilon$, with $\epsilon$ increasing as the binary inspirals. However, as explained by~\cite{Will:2019lfe}, this is due to the fact that the true anomaly approaches a fixed point, which, in the case of the radiation-reaction perturbation, is an odd multiple of $\pi/2$. Thus, $e \sim \epsilon \ll 1$ indeed corresponds to quasi-circular orbits because the orbital separation $r$ will not have turning points when $\vartheta \rightarrow \vartheta_c$.

We can obtain this increase in the osculating eccentricity from the regularized 1PA solution as shown by~\cite{Loutrel:2018ydu,Loutrel:2018ssg,Will:2019lfe}.
Specifically, following~\cite{Will:2019lfe}, in the quasi-circular limit, we have for the osculating eccentricity:
\begin{align}
e = \left( \tilde{e}^2 + e_c^2 - 2 e_c \tilde{e} \sin [\phi - \tilde{\varpi}] \right)^{1/2}, \label{eqn:1PA_regularized_e}
\end{align}
where
\begin{align}
e_c = \frac{64}{5} \eta \left (\frac{m}{\tilde{a}} \right)^{5/2}, \label{eqn:ecc_critical}
\end{align}
with $\tilde{a} \sim \tilde{p}$ in the small eccentricity limit,
and with $\tilde{e}$ and $\tilde{a}$ given by~\cref{eqn:small_ecc_sol}, but with 1PA accurate initial conditions. In the small eccentricity limit, observe that $e_c \sim \epsilon \sim 1/\tilde{a}^{5/2}$ grows as the binary inspirals and $\tilde{a}$ decreases, in contrast to the secular eccentricity $\tilde{e}$, which decreases as prescribed by~\cref{eqn:small_ecc_sol}. Consequently, when $\tilde{e} \ll e_c$, the osculating eccentricity is instead governed by the increasing $e_c$, which was one of the main results in~\cite{Loutrel:2018ydu,Loutrel:2018ssg,Will:2019lfe}. In the small eccentricity limit, we also have that the osculating pericenter
\begin{align}
\varpi = \tan^{-1} \left[ \frac{\tilde{e} \sin \tilde{\varpi} + e_c \cos \phi}{\tilde{e} \cos \tilde{\varpi} - e_c \sin \phi} \right], 
\end{align}
which gives us $\varpi \sim \phi - k\pi/2$ when $\tilde{e} \ll e_c$, with $k$ being an odd integer, such that $\varpi $ is in the principal domain $(-\pi/2,\pi/2)$. From this, we see that in the quasi-circular limit, $\vartheta \sim \vartheta_c = k\pi/2$.

Thus, the important takeaways regarding the quasi-circular limit are the following:
\begin{itemize}
    \item[(i)] the 1PA secular evolution of $\tilde{e}$ does not grow as the binary inspirals, and 
    \item[(ii)] the growth of the osculating eccentricity during inspiral comes from the true anomaly approaching a fixed point, which coincides with the quasi-circular limit.
\end{itemize}
Another way to obtain the increase in the osculating eccentricity is to directly work with~\cref{eqn:LPE} and expand in small eccentricity and in small perturbations of the true anomaly around $\vartheta_c = k \pi/2$.
When doing so, one can show that the osculating eccentricity must approach $e \sim e_c$, which is consistent with our analysis above. 

\section{Gauge dependence of orbital elements and implications for physical observables}\label{sec:waveform}
As mentioned earlier in Sec.~\ref{subsec:rad-reac-balance}, the instantaneous balance laws admit a two-parameter gauge freedom at 2.5PN order, characterized by $(\alpha_g,\beta_g)$, resulting in a gauge-dependent radiation-reaction acceleration~\cite{Iyer:1995rn}. This freedom reflects the choice of relative separation at 2.5PN order~\cite{Iyer:1995rn}, so coordinate- or trajectory-dependent quantities, such as the Keplerian orbital elements, are also gauge-dependent. Physical observables, such as the radiative fluxes and GW polarizations, remain unchanged when evaluated at the same physical event. Arun et al.~\cite{Arun:2009mc} used the 2.5PN gauge transformation derived in~\cite{Iyer:1995rn} to obtain the corresponding change in the functional form of the radiative fluxes expressed in terms of standard harmonic coordinates. Here we follow their approach, first recovering the acceleration transformation, then deriving the induced transformation of the orbital elements and the corresponding functional transformation of the GW polarizations expressed in terms of standard harmonic orbital elements.

Following~\cite{Iyer:1995rn,Arun:2009mc}, we define the generalized relative separation vector as $\vec{x}_{\rm gen} = \vec{x} + \epsilon \vec{\xi}$, where the gauge transformation $\vec{\xi}$ is given by:
\begin{align}
\vec{\xi} = \frac{8}{15} \frac{m^2 \eta}{r} \left[  -\beta_g \dot{r} \vec{n} + (3+3 \alpha_g - 2\beta_g) \vec{v} \right], \label{eqn:gauge_transform}
\end{align}
with the convention that $\vec{x} = \vec{x}_{\rm SH} $ is the relative separation in standard harmonic coordinates. Note that the $\vec{v}$-dependent term already accounts for the change in the coordinate time~\cite{Iyer:1995rn}.  The resulting generalized relative velocity is then $\vec{v}_{\rm gen} = \vec{v} + \epsilon \dot{\vec{\xi}}$. For the generalized relative acceleration, as shown in~\cite{Iyer:1995rn,Arun:2009mc}, we first have that 
\begin{equation}
\begin{aligned}
\vec{a}_{\rm gen}  & = \vec{a}_K + \epsilon \, \vec{a}_{\rm rr} + \epsilon \, \ddot{\vec{\xi}},
\end{aligned}
\end{equation}
where we recall that $\vec{a}_K$ is the Keplerian acceleration and $\vec{a}_{\rm rr}$ is the radiation-reaction acceleration, both expressed in standard harmonic coordinates. 
We also have from the definition of the generalized acceleration that
\begin{equation}
\begin{aligned}
a^i_{\rm gen} &= a^i_{K, \rm gen}+ \epsilon \, a^i_{ \rm rr, gen} \\
&= a^i_K+ \epsilon \, \xi^j \frac{\partial a^i_K}{\partial x^j} + \epsilon \, a^i_{ \rm rr, gen} ,
\end{aligned}
\end{equation}
which then results in the functional relation between the generalized radiation-reaction acceleration and the standard harmonic one:
\begin{equation}
\begin{aligned}\label{eqn:gen_rr_acc_transform}
a^i_{\rm rr, gen} (\vec{x},\vec{v}) &= a^i_{\rm rr} (\vec{x}, \vec{v}) + \delta a^i,\\
\delta a^i &= \ddot{\xi}^i - \xi^j \frac{\partial a^i_K}{\partial x^j}.
\end{aligned}
\end{equation}
Importantly,~\cref{eqn:gen_rr_acc_transform} is exactly the expression for the radiation-reaction acceleration in~\cref{eqn:gen_rr_acc}, and shows how to consistently transform the acceleration between coordinate systems.

From~\cref{eqn:gauge_transform}, we can compute the functional transformations $\delta r$, $\delta \phi$, $\delta \dot{r}$, and $\delta \dot{\phi}$ in terms of the standard harmonic coordinates. Further, using the coordinate dependent expressions of the reduced energy $E/\mu$, reduced angular momentum $L/\mu$, and Runge-Lenz vector $\vec{\mathcal{A}}$, we obtain their functional transformations, $\delta E / \mu, \delta L/\mu $, and $\delta \vec{\mathcal{A}}$, in terms of the standard harmonic coordinates, the details of which are shown in Appendix~\ref{sec:add_det_gauge}.
We note that these changes in the Keplerian energy and angular momentum are exactly the negative of the changes in the Schott terms between the generalized and standard harmonic gauges, which is what preserves the flux-balance law~\cite{Iyer:1995rn}. 

With $\delta E / \mu, \delta L/\mu $, and $\delta \vec{\mathcal{A}}$ in hand, we then obtain the change in the orbital elements $\epsilon \delta X_{\alpha} \equiv X_{\alpha,\rm gen} - X_\alpha$, where $X_{\alpha, \rm gen}$ and $X_{\alpha}$ are the generalized and standard harmonic orbital elements respectively, and we have inserted $\epsilon$ for book-keeping purposes. In other words, the generalized and standard harmonic orbital elements differ by a near-identity transformation.
Below, we provide the transformation results for each relevant orbital element, with further details in Appendix~\ref{sec:add_det_gauge}:
\par\vspace{0.5\baselineskip}
\begin{widetext}
\begin{subequations}\label{eqn:orbital_elements_gauge_transform}
\begin{align}
\delta a &= -\frac{16}{5}\,\eta\, \frac{ae}{1-e^2}\left(\frac{m}{p}\right)^{5/2}(1+e\cos\vartheta)^2\sin\vartheta\Big[(1+\alpha_g)(1+2e\cos\vartheta+e^2)-\beta_g e^2\sin^2\vartheta\Big], \\
\delta p &= -\frac{16}{5} \eta p e \left( \frac{m}{p} \right)^{5/2} (1+\alpha_g) \sin \vartheta (1+e\cos \vartheta)^2, \\
\delta e &= \frac{8}{5}\,\eta\,e\left(\frac{m}{p}\right)^{5/2}(1+e\cos\vartheta)^2\sin\vartheta\Big[\beta_g e\sin^2\vartheta-2(1+\alpha_g)(e+\cos\vartheta)\Big], \\
\delta\varpi &= \frac{8}{15}\,\eta\,\left(\frac{m}{p}\right)^{5/2}(1+e\cos\vartheta)^2\Big[-6(1+\alpha_g)\sin^2\vartheta+\beta_g(1-2e\cos\vartheta+3e\cos^3\vartheta)\Big], \\
\delta\vartheta &= \frac{8}{5}\,\eta\,\left(\frac{m}{p}\right)^{5/2}(1+e\cos\vartheta)^2\Big[(1+\alpha_g)(3-2\cos^2\vartheta+e\cos\vartheta)-\beta_g(1+e\cos^3\vartheta)\Big].
\end{align}
\end{subequations}
\end{widetext}

From~\cref{eqn:orbital_elements_gauge_transform}, we also obtain the transformation of the time derivatives, $\delta\dot{X}_{\alpha} = (\partial\, \delta X_{\alpha}/\partial \vartheta)\,\dot{\vartheta}$, where $\dot{\vartheta}$ is evaluated using the Keplerian dynamics; contributions from the evolution of the other orbital elements enter at higher order in $\epsilon$. This tells us how the LPE (cf.~\cref{eqn:LPE}) transform, and we verified that $\epsilon \delta \dot{X}_{\alpha}$ agrees with the difference $\dot{X}_{\alpha, \rm gen} - \dot{X}_{\alpha} $, when both $\dot{X}_{\alpha, \rm gen}$ and $ \dot{X}_{\alpha}$ are directly evaluated using the radiation-reaction sourcing terms. In comparison to~\cite{Fumagalli:2025rhc}, we note that our near-identity transformation of the orbital elements is a direct consequence of~\cref{eqn:gauge_transform}, while the transformation of~\cite{Fumagalli:2025rhc} aims to eliminate the dependence on the gauge parameters at the level of the LPE.

Physical observables, such as the GW polarizations $h_{+/ \times }$, are gauge-invariant, which means that they are the same when evaluated at a given physical point in spacetime, irrespective of the coordinates used. However, to evaluate them explicitly, one has to pick a gauge and represent them in terms of the coordinates or in terms of the orbital elements. If $h_{+/ \times } (X_{\alpha})$ represents the GW polarizations in standard harmonic gauge, then gauge invariance means that $h_{+/ \times, \rm gen } (X_{\alpha, \rm gen}) = h_{+/ \times } (X_{\alpha})$.
From this, we can obtain the polarizations in a general gauge, as a function of the standard harmonic orbital elements, similar to how~\cite{Arun:2009mc} obtained the generalized fluxes.

We start from the expressions for $h_{+/\times}(X_{\alpha})$ in standard harmonic gauge.
Following the conventions of~\cite{Loutrel:2018ssg}, we work in a frame in which the orbital angular momentum points along $\vec{e}_Z$, and the GW propagates along $\vec{N} = (\sin\iota\cos\psi, \sin\iota\sin\psi, \cos\iota)$, where $\iota$ is the inclination of the orbital angular momentum relative to the line of sight, and $\psi$ is the azimuthal angle of the line of sight in the orbital plane, referred to as the polarization angle in~\cite{Loutrel:2018ssg}.
The polarization subspace is spanned by $\vec{I} = (\cos\iota\cos\psi, \cos\iota\sin\psi, -\sin\iota)$ and $\vec{\Gamma} = (-\sin\psi, \cos\psi, 0)$, such that $h_+ = \tfrac{1}{2}(I^iI^j - \Gamma^i\Gamma^j)\,h_{ij}$ and $h_\times = \tfrac{1}{2}(I^i\Gamma^j + \Gamma^iI^j)\,h_{ij}$, where, from the quadrupole formula~\cite{PoissonWill2014}, $h_{ij} = (4\eta m/D_L)\left[v_i v_j - (m/r)\, n_i n_j\right]$, with $D_L$ the luminosity distance.
The GW polarizations are then
\begin{widetext}
\begin{subequations}\label{eqn:gw_polarizations}
\begin{align}
h_+ &= -\frac{2 m^2 \eta}{p\, D_L}\left[(1+\cos^2\iota)\left(\cos 2(\phi-\psi) + \frac{5}{4}e\cos(\phi+\varpi-2\psi) + \frac{1}{4}e\cos(3\phi-\varpi-2\psi) + \frac{e^2}{2}\cos 2(\varpi-\psi)\right)\right. \nonumber \\
&\qquad\qquad\qquad \left. + \frac{e}{2}\sin^2\iota\,\bigl(e + \cos(\phi-\varpi)\bigr)\right], \\
h_{\times} &= -\frac{2 m^2 \eta}{p\, D_L}\cos\iota\left[2\sin 2(\phi-\psi) + \frac{5}{2}e\sin(\phi+\varpi-2\psi) + \frac{e}{2}\sin(3\phi-\varpi-2\psi) + e^2\sin 2(\varpi-\psi)\right].
\end{align}
\end{subequations}
\end{widetext}
To first order in $\epsilon$, we then have the functional relationship between the polarizations and the standard harmonic orbital elements:
\begin{align}
h_{+/ \times, \rm gen} (X_{\alpha }) &= h_{+/ \times} (X_{\alpha}) - \epsilon \partial_{\alpha} h_{+/ \times } \delta X^{\alpha}, 
\end{align}
where, in practice, we take $X^{\alpha} = \{ p , e, \varpi, \vartheta \}$.
Thus, the variation $\delta h_{+/\times} = -\partial_{\alpha} h_{+/\times}\, \delta X^{\alpha}$ is precisely what is needed to cancel, to first order in $\epsilon$, the 2.5PN gauge dependence that enters when transforming between the generalized and standard harmonic orbital elements.
This allows for a meaningful comparison, at a given physical event and to first order in $\epsilon$, of waveforms expressed in terms of orbital elements in different coordinates.
In particular, the gauge dependence of the derived 1PA corrections to the orbital elements, such as those entering the PN waveforms of~\cite{Loutrel:2018ydu,Boetzel:2019nfw}, is canceled, to first order in $\epsilon$, by the corresponding change in the functional form of the gravitational-wave polarizations when evaluated at the same physical event.
Although the derived transformation maps the LPE between gauges through $\mathcal{O}(\epsilon)$, transforming the truncated equations generates $\mathcal{O}(\epsilon^2)$ terms that can contribute significantly if they accumulate secularly over the radiation-reaction timescale. Establishing agreement between waveforms from independently truncated evolutions over a full inspiral therefore requires assessing these terms, which we leave to future work.

\section{Accuracy of 0PA and 1PA evolutions: Numerical experiments} \label{sec:num_exp}

We here assess the accuracy of the 0PA and 1PA orbital evolutions, by comparing them against a direct evolution of the orbital elements. The specific questions we address are the following:
\begin{enumerate}
\item How do the 0PA and 1PA approximations affect the various orbital elements, compared to the direct numerical evolution of the EoM? And when is the secular behavior of the direct solution accurately captured by the 0PA and 1PA approximations? How does this depend on the initial conditions?
\item Where in parameter space is the 0PA and 1PA accurate ($\lesssim 10^{-1}$ rad dephasing), for various $e_0$, $p_0/m$, $\vartheta_0$, and $q$? 
\end{enumerate}
For a range of $e_0$, $p_0/ m$ and $\vartheta_0$ and a few different values of $q$, we study the errors due to both 0PA and 1PA approximations, with respect to the direct evolution, and thereby obtain a validity regime for each. To answer the first set of questions, we perform a few detailed case studies. In order to understand the dependence of the errors across parameter space, we vary at most two parameters together, while holding the others fixed.

\subsection{Setup and methods} \label{sec:setup}
To address our questions, we set up the initial conditions and evolutions as follows. For all orbital evolutions, we work in standard harmonic gauge. A reference for the end of the inspiral is the Schwarzschild test-particle last stable orbit (LSO), given by $p_{\rm LSO}/m = 6 + 2e_{\rm LSO}$~\cite{CutlerKennefickPoisson:1994}. Applied to the osculating orbital elements used here, this relation provides only an approximate termination criterion.
However, the PN approximation becomes increasingly inaccurate at high velocities; in the test-particle limit, for instance, the PN energy flux loses accuracy for $v \gtrsim 0.3$, with the precise value depending on the PN order~\cite{Yunes:2008tw}.
For an eccentric binary, the velocity varies along the orbit, and we adopt $v_{\rm peri} \equiv \sqrt{m/r_{\rm peri}} = v_p\sqrt{1+e}$, with $v_p = \sqrt{m/p}$, as a Newtonian characteristic velocity, and require $v_{\rm peri} < 1/3$.
This translates to
\begin{align}
\sqrt{\frac{m(1+e)}{p}} < \frac{1}{3}, \label{eqn:stopping_cond}
\end{align}
which we require to hold for both the initial and stopping conditions of the orbital evolution.
Reference~\cite{Moore:2019xkm} imposes the same bound on their PN parameter $y$, which reduces to $v_p$ at Newtonian order; our condition is thus more conservative, as it accounts for the higher velocities near pericenter. An even more conservative condition would follow from using the instantaneous pericenter velocity, $v_{\rm peri,inst} = (1+e)\sqrt{m/p}$, which exceeds $v_{\rm peri}$ by a further factor of $\sqrt{1+e}$.
When $v_{\rm peri} = 1/3$ at the end of the inspiral, $p/m = 9(1+e)$, which is somewhat more restrictive than reaching the LSO. We note that the threshold of $1/3$ is motivated by quasi-circular results and is not a hard boundary in phase space, unlike the LSO; rather, it serves as a proxy for where the PN approximation may cease to be reliable. These choices determine which systems we set up and study, but do not significantly alter the conclusions of our numerical experiments, as we demonstrate in Appendix~\ref{sec:add_det_num}.

For each orbital time evolution, we use the explicit adaptive $8$th-order Runge-Kutta method (\texttt{DOP853}) of~\cite{HairerNorsettWanner1993}, as implemented in \texttt{scipy}~\cite{Virtanen2020}, and stop the integration when~\cref{eqn:stopping_cond} is violated.  We performed robustness checks of this method, which we discuss in Appendix~\ref{sec:add_det_num}. This ensures that the 0PA and 1PA orbital evolutions stop at the same physical state as the direct evolution. To obtain the direct orbital time-evolution, we numerically integrate the LPEs of~\cref{eqn:LPE}, and we checked that this is consistent with obtaining the orbital elements from a direct integration in terms of $r, \phi, \dot{r}$, and $\dot{\phi}$.
These checks are discussed further in~\cref{sec:add_det_num}.

For the 0PA case, we numerically evolve the secular equations $\dot{\tilde{X}}^{\alpha} = (d\tilde{X}^{\alpha}/d\zeta)/(d\tilde{t}/d\zeta)$ for $X^{\alpha} = \{ p , e, \varpi \}$ in terms of time, as given by~\cref{eqn:adiabatic_evol}.  We also simultaneously evolve the 0PA orbital phase by integrating the instantaneous orbital frequency,
\begin{equation}\label{eqn:dphi_inst}
    \dot{\phi} = \sqrt{\frac{m}{p^3}}(1 + e\cos(\phi-\varpi))^2,
\end{equation}
using the 0PA secular variables $\tilde{X}^{\alpha}$. We set the initial conditions for the 0PA secular variables using $\{\tilde{p}_0 = p_0 , \tilde{e}_0 = e_0, \tilde{\varpi}_0 = \varpi_0 \}$.

For the 1PA secular time-evolution, we proceed as follows. As shown in Sec.~\ref{sec:1PA_sec}, the secular equations for $\tilde{p}$ and $\tilde{e}$ do not change compared to the 0PA case, but there is a non-zero evolution for $\tilde{\varpi}$ unlike its 0PA counterpart. However, evaluating them as functions of $t$ is subtle and not so straightforward, as discussed in~\cite{Pound:2007ti,Pound:2007th}.
This is partly because the averages are performed with respect to phase and not directly with respect to time. Further, the secular variables are evaluated as functions of the secular time $\tilde{t}$, which is related to $t$ through $\tilde{t} = t - Y^{(0)}_t$, with $Y^{(0)}_t$ given by~\cref{eqn:t_osc}.
Taylor expanding around $\tilde{t}$, we obtain:
\begin{align}
\begin{aligned}
\tilde{p}(t) &= \tilde{p}(\tilde{t}) + \epsilon Y^{(0)}_t \frac{d \tilde{p}}{d\tilde{t}}, \\
\tilde{e}(t) &= \tilde{e}(\tilde{t}) + \epsilon Y^{(0)}_t \frac{d \tilde{e}}{d\tilde{t}},\\
\tilde{\varpi}(t) &= \tilde{\varpi}(\tilde{t}) + \epsilon Y^{(0)}_t \frac{d \tilde{\varpi}}{d\tilde{t}},
\end{aligned}
\end{align}
where we inserted the book-keeping parameter $\epsilon$, which comes from the secular time derivatives of the orbital elements. This means that the corrections induced by $Y^{(0)}_t $ are of 1PA order, and need to be included to construct a 1PA accurate osculating solution.
To do so, we first transform to the Runge-Lenz components to avoid singular behavior at small eccentricity, as discussed in Sec.~\ref{sec:quasicirc}.
We then compute the secular time evolution for $\tilde{\mathcal{A}}_x$ and $\tilde{\mathcal{A}}_y$ from $d\tilde{e}/d\tilde{t}$ and $d \tilde{\varpi}/d\tilde{t}$.
Following this, we obtain the 1PA osculating orbital elements from
\begin{align}
\begin{aligned}\label{eqn:1PA_osculating_sol_t}
p(t) &= \tilde{p}(t) - \epsilon Y^{(0)}_t \frac{d \tilde{p}}{d\tilde{t}} + \epsilon Y^{(1)}_p, \\
\mathcal{A}_x(t) &= \tilde{\mathcal{A}}_x (t) - \epsilon Y^{(0)}_t \frac{d \tilde{\mathcal{A}}_x}{d\tilde{t}} + \epsilon Y^{(1)}_{\mathcal{A}_x},\\
\mathcal{A}_y(t) &= \tilde{\mathcal{A}}_y (t) - \epsilon Y^{(0)}_t \frac{d \tilde{\mathcal{A}}_y}{d\tilde{t}} + \epsilon Y^{(1)}_{\mathcal{A}_y},
\end{aligned}
\end{align}
where again we have inserted $\epsilon$ only for book-keeping purposes. We can then compute $e(t)$ and $\varpi(t)$ from~\cref{eqn:1PA_osculating_sol_t}. The initial conditions for the 1PA secular variables are obtained by inverting~\cref{eqn:1PA_osculating_sol_t}, which are then used to integrate the 1PA system.
To concurrently compute the 1PA orbital phase, at each step, we first evaluate the instantaneous orbital frequency $\dot{\phi}$, given by~\cref{eqn:dphi_inst}, using the 1PA osculating orbital elements, given by~\cref{eqn:1PA_osculating_sol_t}, and then integrate it together with the secular 1PA elements.
Since the phase evolves more rapidly than the secular orbital elements, using an adaptive time-step becomes useful, especially for highly eccentric systems. Note that we are justified in evaluating the 1PA oscillatory solution using the 1PA secular elements, as the remainders are of higher order in MSA. By carrying out the calculation in this way, we do not need to recompute the 0PA secular solution when evolving the 1PA system.

\begin{figure}[t]
    \centering
    \includegraphics[width=\linewidth]{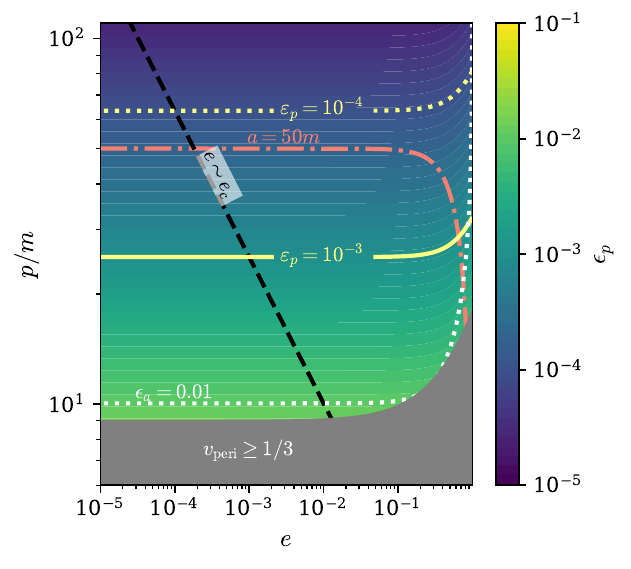}
    \caption{Scaling of orbital to radiation-reaction timescales across the space of $e$ and $p/m$ for $\eta = 1/4$. The gray shaded region is disallowed due to the pericenter velocity violating~\cref{eqn:stopping_cond}. We show contours of constant $\epsilon_p$ (yellow lines), as well as the contour with $\epsilon_a = 10^{-2}$ contour line (red dotted) shown for comparison. The quasi-circular limit corresponds to the region below the black-dashed line where $e \sim e_c$. For reference, we show where $a=50m$ (orange dash-dot line) lies in this plot.
    Observe that the separation of timescales gets weaker at larger $e$ and smaller $p/m$.
    }
    \label{fig:p_e_curve}
\end{figure}

In~\cref{fig:p_e_curve}, we illustrate the region of parameter space $(e, p/m)$ we can study with our numerical experiments. We show contours of constant ratio of orbital to radiation-reaction timescales, which we characterize with the secular variation of the semi-latus rectum\footnote{Alternatively, one can use $\epsilon_e \equiv \tau_{\rm orb}/(e/\dot{e}_{\rm sec})$, which has similar scalings with $\tilde{p}$ and $\tilde{e}$ as $\epsilon_p$.}:
\begin{align}
\epsilon_p \equiv \frac{\tau_{\rm orb}}{p/\dot{p}_{\rm sec}} = \frac{64}{5} \eta \frac{m^{5/2}}{p^{5/2}} \left( 1+\frac{7}{8}e^2 \right).
\end{align}
This measure gives us an eccentricity-dependent correction to $\epsilon$, compared to our Fermi estimate in Sec.~\ref{sec:qual_PA}, which is reflected in the contour lines of Fig.~\ref{fig:p_e_curve}. When $\epsilon_p$ is small, the timescales are well separated, while when $\epsilon_p$ grows toward unity, the MSA is expected to break down. 

At large eccentricity, however, the radiation-reaction timescale is more accurately described by the variation of the semi-major axis
$\epsilon_a = \tau_{\rm orb}/(a/\dot{a}_{\rm sec})$, which is given by
\begin{align}
    \epsilon_a = \frac{64}{5} \eta \frac{m^{5/2}}{p^{5/2} (1-e^2)} \left( 1+\frac{73}{24}e^2 + \frac{37}{96}e^4 \right).
\end{align}
When $1-e \ll 1$, $\epsilon_a$ is larger than $\epsilon_p$ by $\sim (1-e^2)^{-1}$, and thus, it controls whether the radiation-reaction and PA effects are perturbative. We demonstrate this with the contour line corresponding to $\epsilon_a = 10^{-2}$ in~\cref{fig:p_e_curve}, which is larger than $\epsilon_p$ by nearly a factor of $10$ at $e=0.9$. For fixed $p/m$, at larger values of $e$, $\epsilon_a$ can exceed 1, which means that the separation of timescales breaks down. One can study such large eccentricities, provided sufficiently large values of $p/m$~\cite{Will:2016} are considered (see also~\cite{Fumagalli:2025rhc}). In this work, as a conservative measure, we restrict our analysis to $e \leq 0.9$, so that separation of timescales holds throughout parameter space, allowing us to systematically assess the accuracy of the 0PA/1PA approximations. In the remainder of this manuscript, we therefore adopt $\epsilon = \epsilon_p$ as the perturbative parameter. For reference, we also indicate the quasi-circular limit when $e \sim e_c$ (cf.~\cref{eqn:ecc_critical}), and as well as where $a = 50m$ lies in \cref{fig:p_e_curve}. The gray shaded region is disallowed due to~\cref{eqn:stopping_cond} being violated. With~\cref{fig:p_e_curve} as a guideline, we proceed to compare the orbital evolutions.

\subsection{Comparison of 0PA, 1PA and direct evolutions: case studies} \label{sec:comparison_evol}
\subsubsection{Small initial eccentricity}

\begin{figure*}[!t]
    \centering
    \includegraphics[width=\linewidth]{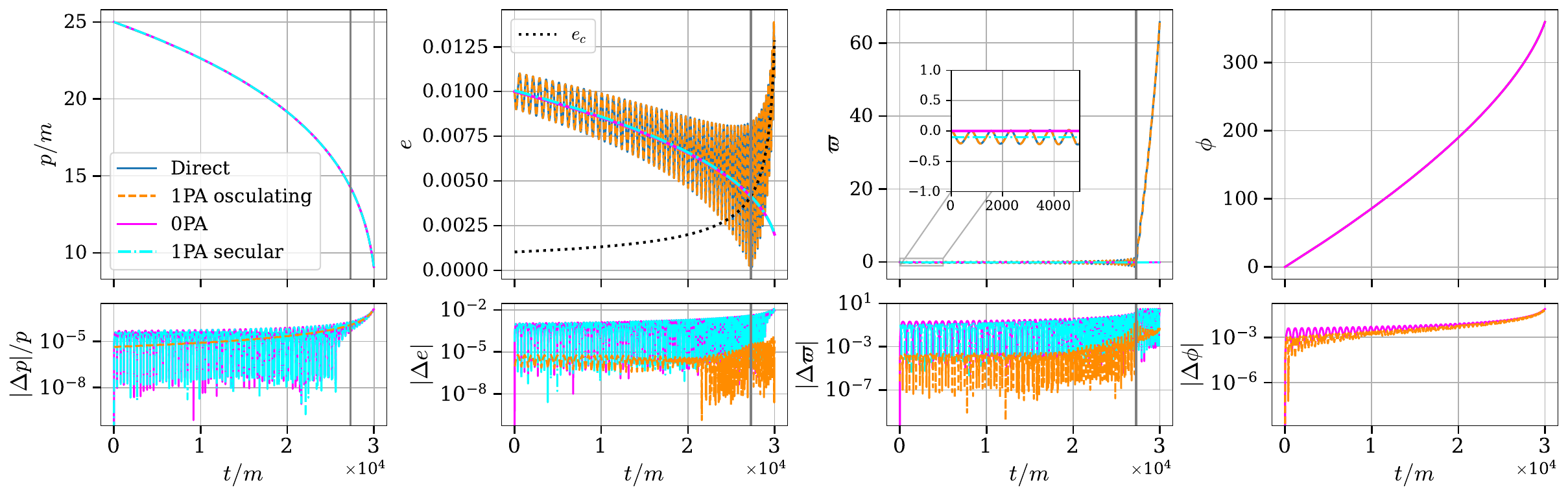}
    \caption{
    Comparison of direct (blue solid), 0PA (magenta solid), and 1PA (osculating in orange dashed, its secular contribution in cyan dash-dot) evolutions for $p$, $e$, $\varpi$, and $\phi$ (top row) along with their residuals against the direct (bottom row) for an equal mass binary with initial conditions $\{p_0/m = 25, e_0 = 0.01, \vartheta_0  = 0, \varpi_0 = 0 \} $. Observe that the osculating 1PA approximation closely matches the direct in all panels. In the second panel of the top row, we also show the contribution from $e_c$ (cf.~\cref{eqn:ecc_critical}) which describes the late-time growth of the osculating eccentricity. The vertical gray line corresponds to when $\tilde{e} \sim e_c$, the quasi-circular limit.}
    \label{fig:e_transition}
\end{figure*}

We first illustrate how the 0PA and 1PA approximations perform against the direct integration for a binary whose initial eccentricity is small. In~\cref{fig:e_transition}, we show an example system with equal masses ($\eta = 1/4$) initialized with $\{p_0/m = 25, e_0 = 0.01, \vartheta_0  = 0, \varpi_0 = 0 \} $ that approaches the quasi-circular limit during the late inspiral. We see that the 1PA osculating approximation closely matches the direct solution for $p$, $e$, $\varpi$, and $\phi$. The secular eccentricities $\tilde{e}$, at 0PA and 1PA order, closely follow each other throughout the inspiral, and reliably describe the secular behavior of the osculating eccentricity during the early inspiral. During the late inspiral, the secular behavior of the osculating eccentricity is instead governed by $e_c$ (black dotted line), as given in~\cref{eqn:ecc_critical}.

As explained in Sec.~\ref{sec:quasicirc}, the transition to the quasi-circular limit happens when the secular 1PA eccentricity $\tilde{e} $ becomes comparable to $e_c$, which for this example occurs at $t/m \sim 27,245$. After this transition, the pericenter co-rotates with the orbital phase, which is clearly seen by the sudden and secular increase of $\varpi$ in the third panel of the top row. This behavior is captured by the 1PA osculating approximation, but not by the secular 0PA or the secular 1PA approximations, highlighting the importance of the oscillatory terms in accurately describing the evolution of the orbital elements. The orbital dephasing (fourth panel of~\cref{fig:e_transition}) between the direct and 0PA evolutions, however, is $\lesssim 0.075$ rad after the inspiral, which is only slightly worse than that of the 1PA osculating approximation, which has a dephasing of $\lesssim 0.072$ rad.

Although including 1PA corrections reduces the error in $e$ and $\varpi$, as can be seen for our example in the bottom row of \cref{fig:e_transition}, it does not significantly decrease the error in $p$, which accumulates after the binary has reached the quasi-circular limit. The error in $p$ dominates the dephasing because $\dot{\phi} \approx \sqrt{m/p^3}$ in this limit. The 1PA approximation neglects 2PA corrections, both in the oscillatory terms that set the initial conditions and in the secular evolution equations. While we do not derive the 2PA terms explicitly, their harmonic structure follows from integrals over $\vartheta$ (or $\phi$) of terms involving $\partial^{\beta} Q^{(0)}_{\alpha}\, Y^{(1)}_{\beta}$ and $\partial^{\beta} Y^{(1)}_{\alpha}\, d\tilde{X}_{\beta}/d\zeta$ (see Appendix~B of~\cite{Will:2016}). Since $Y^{(1)}_p$ is odd in $\vartheta$, these integrands contain parts that are odd in $\vartheta$, so that $Y^{(2)}_p$ contains contributions that are even in $\vartheta$. Consequently, although the 1PA oscillatory contribution to $p$ vanishes at $\vartheta_0 = 0$, the 2PA contribution does not. Moreover, second-order contributions from $Y^{(1)}_{\mathcal{A}_x}$ and $Y^{(1)}_{\mathcal{A}_y}$, as well as from $\partial_{\mathcal{A}_x} Y^{(1)}_p$ and $\partial_{\mathcal{A}_y} Y^{(1)}_p$ will not be suppressed by $e$. Thus, these terms can source a 2PA secular dephasing in the quasi-circular limit, which accumulates during the late inspiral as $\epsilon$ increases with decreasing $p$.

We roughly estimate the 2PA dephasing as $ \approx \epsilon^2 \phi_{\rm direct} \approx 0.06$~rad, with $\epsilon$ and $\phi_{\rm direct}$ evaluated near the end of the inspiral; this explains both the 0PA and 1PA dephasing. Therefore, for this system, initialized at pericenter, the 0PA evolution is already sufficiently accurate, and including 1PA does not improve the dephasing significantly.
To further reduce the dephasing, one would need to go to 2PA order (see~\cite{Lynch:2026ibo} for a recent application of the 2PA approximation to EOB inspirals). In this work, we truncate at 1PA order to focus on the regime of validity of the 0PA approximation.

\subsubsection{Moderate initial eccentricity}

\begin{figure*}[!t]
    \centering
    \includegraphics[width=\linewidth]{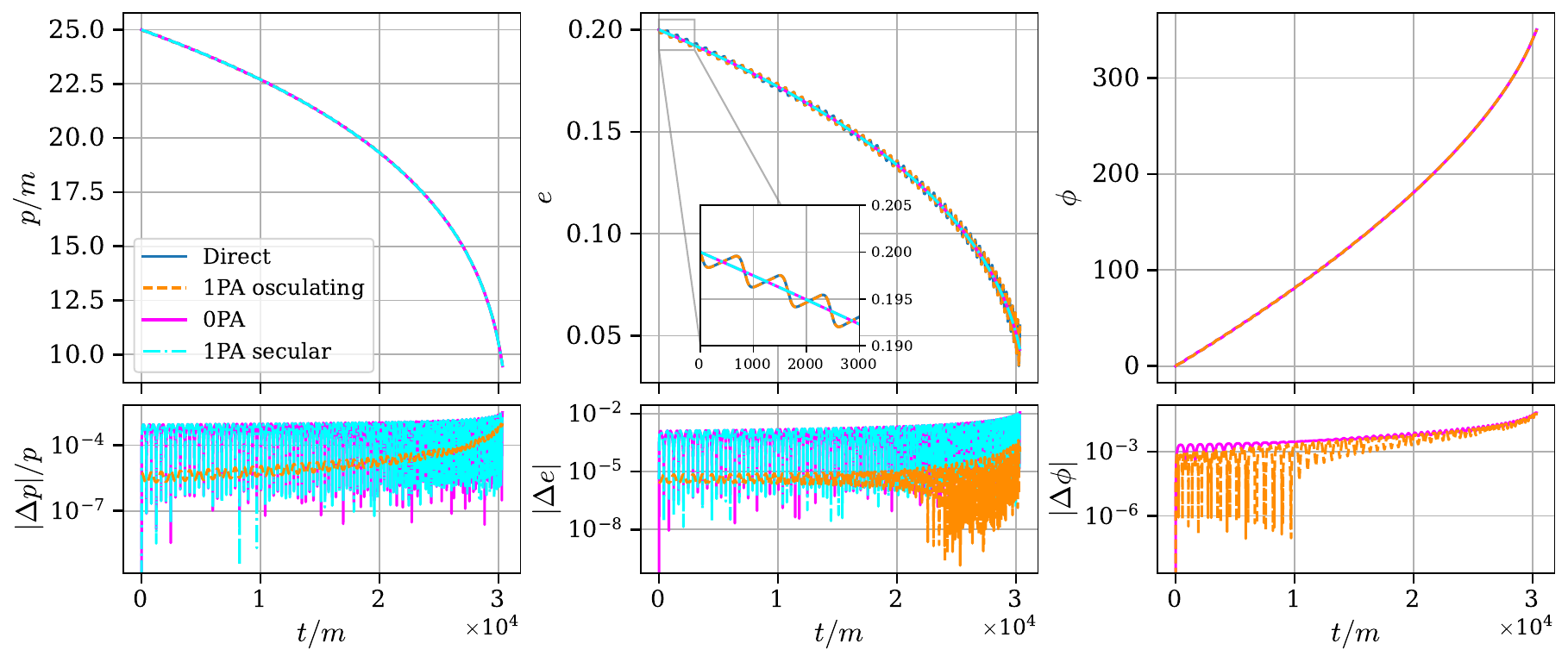}
    \caption{Comparison of direct (blue), 0PA (magenta) and 1PA (orange dashed for osculating and cyan dash-dot for secular) evolutions for system M1 with $\{p_0/m = 25, e_0 = 0.2, \vartheta_0 = 0, \varpi_0 = 0, \eta = 1/4 \} $. In the top row, the inset in the second panel shows the evolution of eccentricity at the start of the inspiral. 
    Observe that both the 0PA and 1PA approximations are accurate relative to the direct solution, with small dephasing over the entire inspiral.}
    \label{fig:e_moderate}
\end{figure*}

\begin{figure*}[!t]
    \centering
    \includegraphics[width=\linewidth]{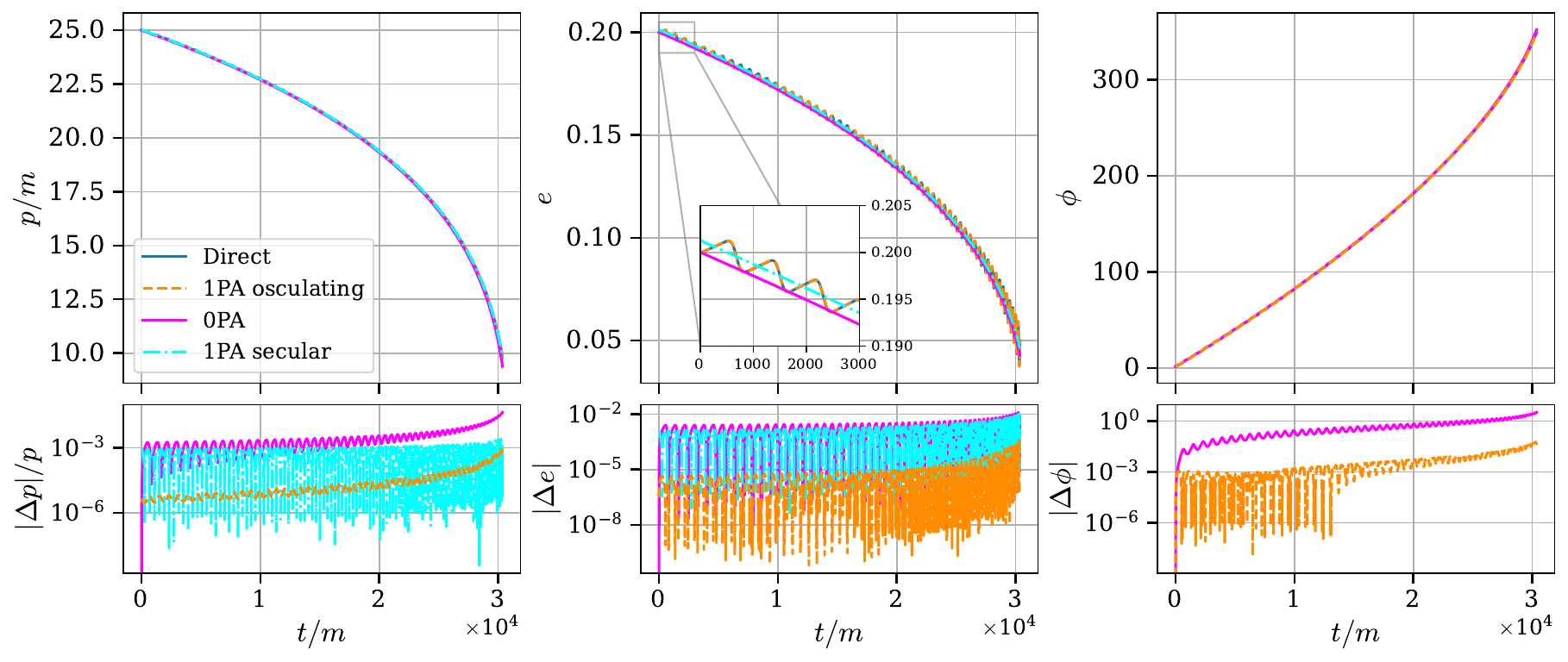}
    \caption{Comparison of direct (blue), 0PA (magenta) and 1PA (orange dashed for osculating and cyan dash-dot for secular) evolutions for system M2 with $\{p_0/m = 25, e_0 = 0.2, \vartheta_0 = \pi/2, \varpi_0 = 0 , \eta = 1/4\} $, with only the initial true anomaly being different from system M1. In the top row, the inset in the second panel shows the evolution of eccentricity at the start of the inspiral. Observe that the 0PA becomes inaccurate with more than 1 rad dephasing relative to the direct solution, while the 1PA remains accurate over the entire inspiral.}
    \label{fig:e_moderate_piby2}
\end{figure*}

We now illustrate the behavior of the orbital elements when the initial eccentricity is moderate.
We initialize an equal mass ($\eta = 1/4$) system with  $\{p_0/m = 25, e_0 = 0.2, \varpi_0 = 0 \} $, but we consider two configurations of the true anomaly: one at pericenter with $\vartheta_0 = 0$ (system M1), and another with $\vartheta_0 = \pi/2$ (system M2). We show the orbital evolutions for systems M1 and M2 in~\cref{fig:e_moderate} and in~\cref{fig:e_moderate_piby2} respectively.
For each case, we plot $p/m$, $e$, and $\phi$ in the top row, with their errors (against the direct solution) in the bottom row. 
As we can see from the bottom row of the third panel in~\cref{fig:e_moderate}, for system M1, both the 0PA and 1PA evolutions perform well, with $\lesssim 0.1$ rad dephasing throughout the inspiral, and with the 1PA model being slightly better than the 0PA model at all times.  As explained earlier, the dephasing will become even smaller if one includes 2PA contributions. The secular evolution of eccentricity under the 0PA and 1PA approximations also coincide and both run through the center of the direct evolution, capturing accurately the secular time evolution.

\begin{figure*}[!t]
    \centering
    \includegraphics[width=\linewidth]{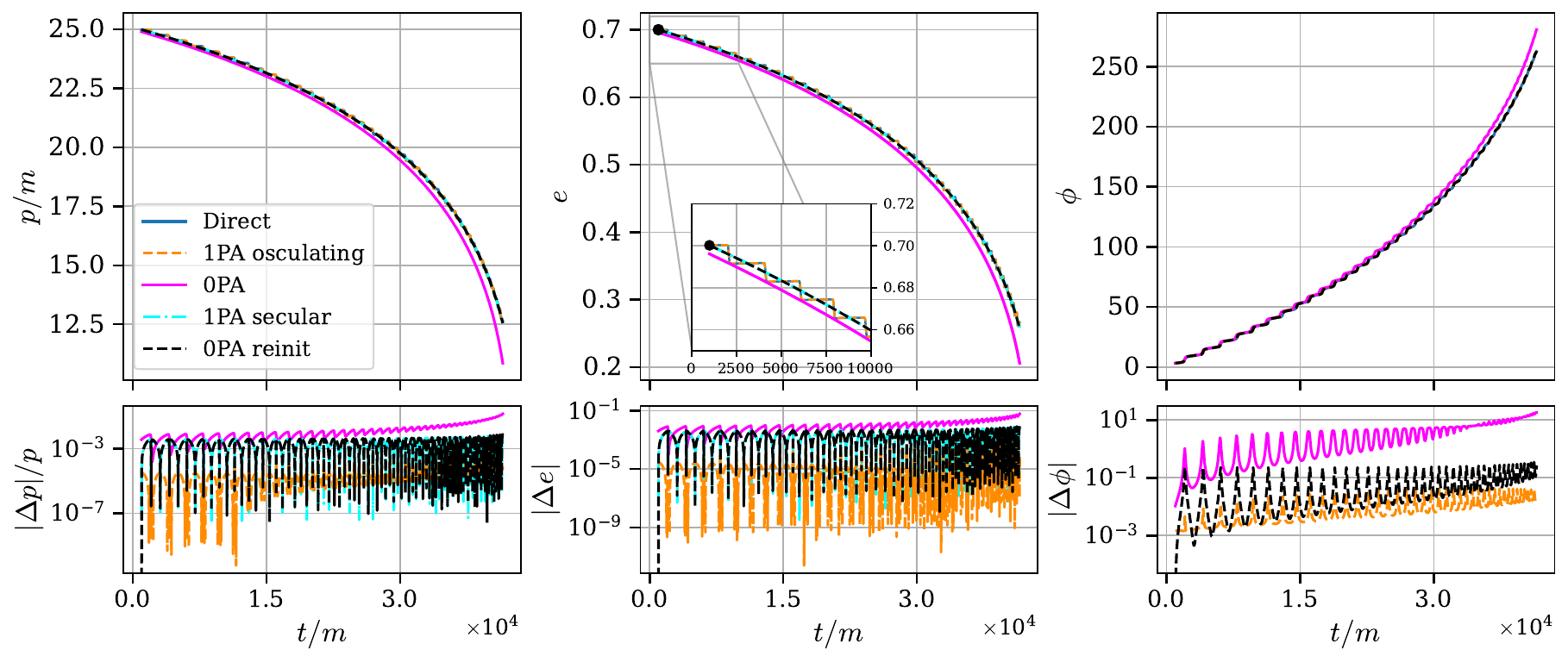}
    \caption{
    Comparison of direct (blue), 0PA (magenta) and 1PA (orange dashed for osculating and cyan dash-dot for secular) evolutions 
    for an equal-mass binary system with $\{p_0/m = 25, e_0 = 0.7, \vartheta_0 = \pi/2, \varpi_0 = 0 \} $. Additionally, the black dashed line in each panel corresponds to the 0PA evolution when reinitialized at the first radial turning point, which happens to be at apocenter. In the top row, the inset in the second panel shows the evolution of eccentricity at the start of the inspiral. Observe that the 1PA approximation is accurate compared to the direct over the inspiral, while the 0PA is only accurate when initialized at the radial turning point. 
    }
    \label{fig:e_large}
\end{figure*}

For system M2 (with the only change being $\vartheta_0 = \pi/2$), however, we see that the 0PA evolution is clearly inaccurate, while the 1PA evolution remains accurate. Specifically, from the third panel of~\cref{fig:e_moderate_piby2}, we see that the 0PA evolution results in a dephasing of $3.21$ rads, compared to the direct evolution, while the 1PA evolution results in $0.05$ rads of dephasing. In the top row of~\cref{fig:e_moderate_piby2}, from the second panel, we see (as is clearly visible in the inset) that the secular 0PA and 1PA evolutions of the eccentricity do not coincide with each other, as there is now a non-zero 1PA correction to the initial condition for $\tilde{p}$ and $\tilde{e}$ (see also~\cite{Pound:2007th,Pound:2007ti}). When initialized at a radial turning point, as for system M1, we have $\dot{r} = 0$, which corresponds to the 1PA oscillations of $e$ and $p$ being initially zero. Furthermore, when $\dot{r} = 0$, the $2.5$PN energy and angular momentum Schott terms vanish, implying that the orbit-averaged energy and angular momentum match their instantaneous values. This is why the 0PA evolution can be accurate relative to the direct evolution when initialized at
a radial turning point. Meanwhile, the 1PA evolution accounts for the Schott term contributions, and, therefore, it can be accurate irrespective of whether the system is initialized at a radial turning point. 
Thus, \emph{the accuracy of the 0PA evolution is sensitive to how and where along the orbit the system is initialized}. A similar sensitivity to the initial true anomaly appears in the inspiral times reported in~\cite{Fumagalli:2025rhc}; the authors compared orbit-averaged and non-adiabatic evolutions through differences in inspiral time, whereas here we examine dephasing as a function of time.

\subsubsection{Large initial eccentricity}
We now consider a system with large initial eccentricity described by the initial conditions $\{p_0/m =25, e_0 =0.7, \varpi_0 = 0, \vartheta_0 = \pi/2 \}$. From~\cref{fig:e_large}, we see that there is a clear and large dephasing of the 0PA evolution relative to the direct one. The dephasing is larger here, as compared to the M2 case, because the eccentricity is larger, which results in larger amplitude 1PA oscillations. Since the system is initialized when the 1PA oscillation in $e$ and $p$ are non-zero (at $\vartheta_0 = \pi/2$), the error incurred by the 0PA evolution is even larger, although $p_0$ is the same as for the M2 system. Meanwhile, the 1PA approximation still performs very well relative to the direct evolution, resulting in a dephasing of $\sim 0.033$ rads, which is two orders of magnitude smaller than the dephasing of the 0PA evolution. We also show here and for the same system, a 0PA evolution initialized at the first radial turning point, which happens to be at apocenter. The resulting evolution remains accurate compared to the direct evolution, even though the initial eccentricity is large, unlike the 0PA case that is initialized at $\vartheta_0 = \pi/2$. In~\cref{sec:scaling_errors}, we explore the region of parameter space where the 0PA evolution is still accurate when initialized at a radial turning point.

\begin{figure}[!t]
    \centering
    \includegraphics[width=\linewidth]{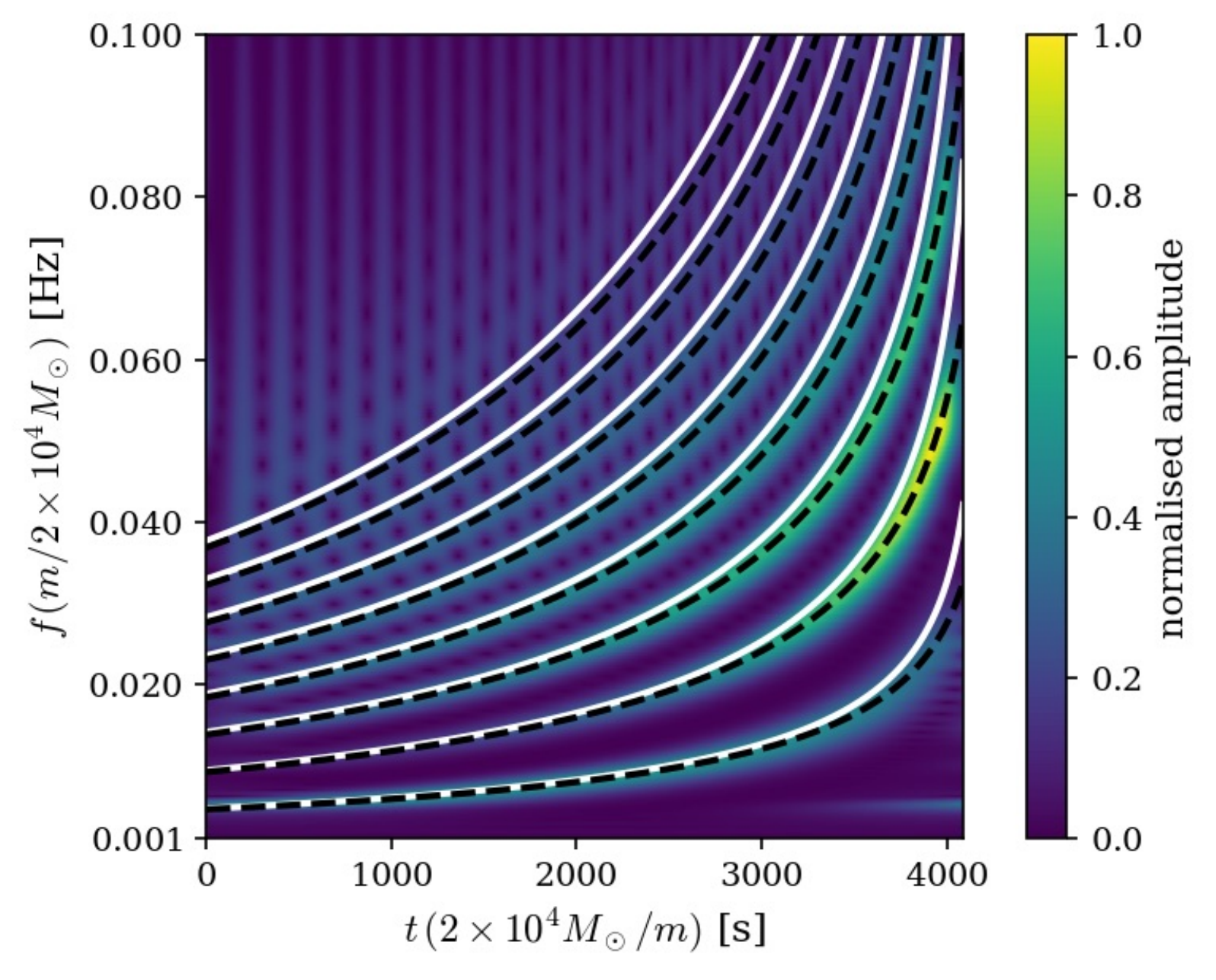}
    \caption{Constant (and normalized) Q-transform of the GW strain computed from the direct evolution of a binary system with $\{p_0/m = 25, e_0 = 0.7 , \vartheta_0 = \pi/2, \varpi_0 = 0\}$, with $m_1 = m_2 = 10^4 M_{\odot}$. The white solid (black dashed) lines are the frequency-time tracks for the first 8 harmonics computed from the 0PA (1PA) secular evolution of the system. The color bar shows the normalized amplitude spectral density.
    Observe that the 1PA tracks each harmonic accurately, while the 0PA does not.
    }
    \label{fig:spectrogram}
\end{figure}

To highlight the observable signatures of such a highly eccentric system, we compute the GW strain $h(t) = h_+(t) F_+ + h_{\times}(t) F_{\times}$ corresponding to the direct evolution, where the GW polarizations are given by~\cref{eqn:gw_polarizations}.
We then compute the (normalized) constant Q-transform (following~\cite{Moore:2019xkm}) of $h(t)$ to show the various harmonics in a frequency-time representation, which we show in~\cref{fig:spectrogram}.
To illustrate the signal from a highly eccentric massive black hole binary in the LISA band, we choose fiducial values of $m_1=m_2=10^4 M_{\odot}$, $D_L=5\,\mathrm{Gpc}$, $\iota=\pi/6$, and fixed antenna factors $F_+=0.65$ and $F_{\times}=0.37$.
Using the LISA sensitivity curve of~\cite{Robson:2018ifk}, with the normalization described in Appendix~\ref{sec:add_det_num}, we obtain a signal-to-noise ratio (SNR) of approximately $25$.
This estimate is illustrative; a complete calculation requires the time-dependent LISA response~\cite{EstabrookWahlquist1975,Cutler:1997ta,Cornish:2002rt,Marsat:2020rtl}.

From~\cref{fig:spectrogram}, we see the various chirping harmonics of the orbital phase, with each harmonic $n$ satisfying~\cite{Yunes:2009yz,Moore:2019xkm} $f_n = n \dot{\tilde{\ell}}/(2\pi)$, with $\dot{\tilde{\ell}}= \sqrt{m/\tilde{a}^3}$ being the mean (orbit-averaged) frequency. The frequency tracks of the first 8 harmonics are shown in~\cref{fig:spectrogram}, where we evaluate $\dot{\tilde{\ell}}$ using the 0PA and 1PA secular evolutions. We see that the 0PA tracks do \textit{not} line up with the chirping harmonics at $f \gtrsim 0.03$Hz, which is where LISA is sensitive.
The 1PA tracks, however, accurately follow the chirping harmonics of the direct evolution all the way across the LISA band. 

To assess the implications of these orbital dephasings for GW matched filtering, we compute the GW mismatch~\cite{Cutler:1994ys} of the 0PA and 1PA waveforms against the direct waveform (see Appendix~\ref{sec:add_det_num} for details).
The 0PA waveform has a mismatch of $\approx 0.67$, whereas the 1PA waveform has a mismatch of $\approx 2\times10^{-5}$.
Although the large 0PA mismatch might suggest that this approximation is unsuitable for matched filtering, the 0PA evolution re-initialized at the first radial turning point (apocenter for this system) achieves a mismatch of $\approx 2\times10^{-5}$, comparable to that of the 1PA waveform. This improvement highlights the importance of initialization and suggests that optimizing over initial conditions could substantially improve the performance of 0PA waveforms. The 1PA approximation systematically accounts for both the initial condition shifts and the dynamical corrections (secular and oscillatory) to the orbital evolution, achieving accuracy without re-initialization. These results motivate incorporating 1PA corrections into waveform models for long-lived, highly eccentric binaries observable with future detectors.

\subsection{Orbital dephasing across parameter space} \label{sec:scaling_errors}

We here study the dependence of the errors due to the 0PA and 1PA approximations by varying $e_0$, $\vartheta_0$, $p_0/m$, and $q$. To do so, we vary two parameters at a time, while holding the others fixed. We explore $p_0/m \in [20,75]$, $e_0 \in [10^{-4},0.9]$, $\vartheta_0 \in [-\pi/2,\pi/2]$, and $q \in [1/10,1]$; for all systems, we fix $\varpi_0 = 0$. The upper bound on $e_0$ is chosen to ensure a clear separation of timescales, as discussed in~\cref{sec:setup} (cf.~\cref{fig:p_e_curve}). The lower bound on $p_0/m$ satisfies~\cref{eqn:stopping_cond} for all $e_0$.  For the grid on $e_0$, we pick a log-spacing for $e_0 \leq 0.1$ and a uniform spacing for $e_0 > 0.1$.  This is to better capture the variation of the dephasing at higher eccentricities. We also ensure that $\vartheta_0 = 0$ is included in the grid. Below, we show the orbital dephasing over the $(e_0,\vartheta_0)$, $(e_0, p_0/m)$, and $(e_0,q)$ grids.

\begin{figure*}[!t]
    \centering
    \includegraphics[width=\linewidth]{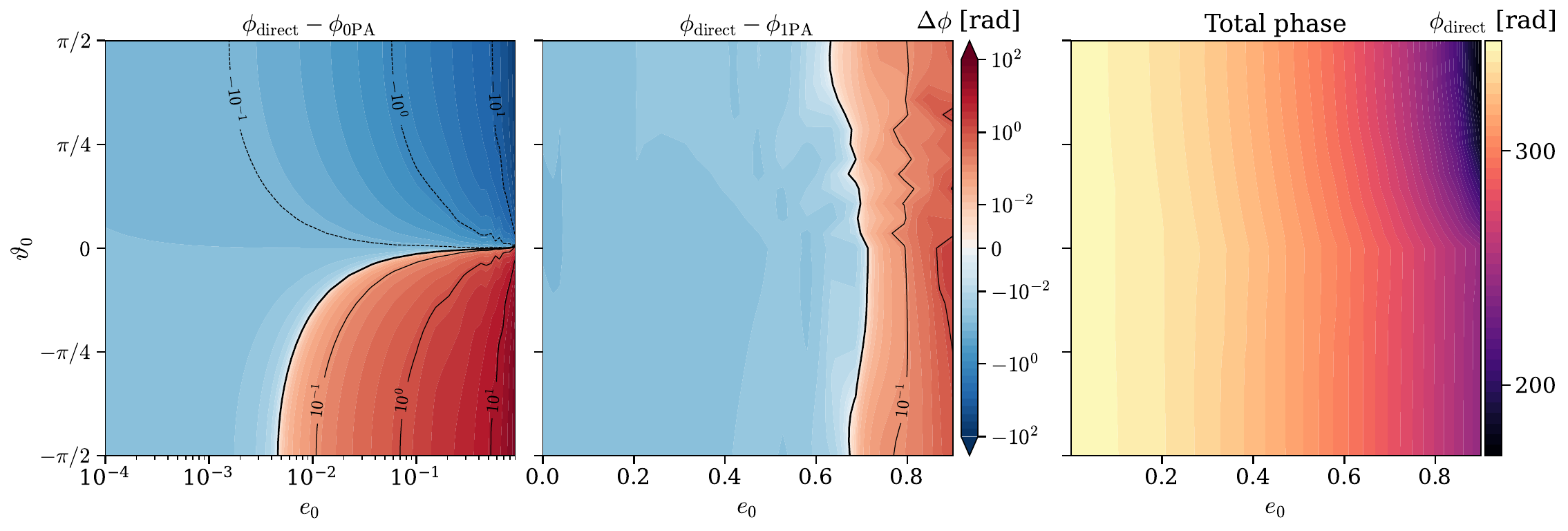}
    \caption{Orbital dephasing as function of initial eccentricity and initial true anomaly, where each system is initialized with $\{ p_0/m = 25, \varpi_0 = 0, \eta = 1/4 \}$. The first two panels show the dephasing (with sign dependence) due to 0PA and 1PA respectively, while the third panel shows the total phase accumulated from the direct integration.
    Note that $e_0$ is plotted on log-scale for the first panel, but on linear scale for the second and third panels.
    Observe that the 0PA dephasing has a strong dependence on $\vartheta_0$, while the 1PA does not.
    }
    \label{fig:dephasing_phi0_e_trueanom0}
\end{figure*}

\subsubsection{Dependence on $e_0$ and $\vartheta_0$}
We first study the dependence on $e_0$ and $\vartheta_0$, by keeping fixed $p_0/m = 25$ and $q=1$.
In~\cref{fig:dephasing_phi0_e_trueanom0}, we show the accumulated orbital dephasing of the 0PA and 1PA evolutions relative to the direct integration, as a function of $e_0$ and $\vartheta_0$. 
The first two panels show the dephasing due to 0PA and 1PA evolutions, respectively, while the third panel shows the total phase corresponding to the direct integration.
The total phase decreases with increasing $e_0$: at fixed $p_0$, a larger eccentricity corresponds to a larger semi-major axis and longer orbital periods, while the enhanced emission at pericenter passages~\cite{Peters:1963ux,Peters:1964zz} shrinks the orbit more rapidly, so that fewer orbits are completed before merger.

Broadly, \cref{fig:dephasing_phi0_e_trueanom0} shows that both the magnitude and the sign of the 0PA and 1PA dephasing depend on $e_0$ and $\vartheta_0$. When initialized far from pericenter, the magnitude of the 0PA dephasing exceeds $1$~rad for $e_0 \gtrsim 0.1$, and remains $\lesssim$~a few $\times\,10^{-1}$~rads only for $e_0 \lesssim 10^{-2}$. The magnitude of the 1PA dephasing, by contrast, reaches $\sim 1$~rad only for $e_0 \sim 0.9$, and remains $\lesssim$~a few $\times\,10^{-1}$~rads for $e_0 \lesssim 0.8$, irrespective of $\vartheta_0$. However, the 1PA evolution offers no significant improvement over the 0PA one for $e_0 \lesssim 10^{-2}$. In this regime, the binary rapidly reaches the quasi-circular limit, where the dephasing is dominated by the error in $p$, which is sourced by the neglected 2PA terms, and thus, it is common to both evolutions (Sec.~\ref{sec:comparison_evol}). Similarly, when initialized at pericenter, the 0PA evolution remains accurate for $e_0 \lesssim 0.8$ and is comparably accurate to the 1PA evolution, since both are limited by the neglected 2PA contributions. The parameter sweep, thus, shows that the conclusions drawn from the case studies in Sec.~\ref{sec:comparison_evol} hold over a wide range of eccentricities.

The 0PA dephasing changes sign along a curve for $\vartheta_0 < 0$ (lower half plane), which starts at $e_0 \approx 5\times10^{-3}$ for $\vartheta_0 = -\pi/2$ and approaches $\vartheta_0 = 0$ as $e_0$ increases.
The 1PA dephasing, in contrast, changes sign at $e_0 \approx 0.65$--$0.7$, with only a weak dependence on $\vartheta_0$. The change in sign can be understood as follows.  The neglected terms start at 1PA order for the 0PA evolution and at 2PA order for the 1PA evolution. The sign of each PA contribution depends on $e_0$ and $\vartheta_0$. For instance, the 1PA oscillatory term $Y^{(1)}_{p}$ is proportional to $e_0$ and it is odd in $\vartheta_0$ (cf.~\cref{eqn:1PA_osc_sol}), so that the neglected 1PA contribution to the 0PA dephasing changes sign with $\vartheta_0$. When the 1PA and 2PA contributions have opposite signs, which for our systems occurs for $\vartheta_0 < 0$, they partially cancel each other during the late inspiral, and the 0PA dephasing changes sign. Similarly, the sign change of the 1PA dephasing can be attributed to the behavior of the neglected 2PA and higher-order contributions with $e_0$. 
In Appendix~\ref{sec:add_det_num}, we illustrate these sign changes through additional case studies.
Such cancellations mean that a higher-order PA approximation need not be more accurate throughout parameter space.
Similar behavior occurs in the PN expansion~\cite{Blanchet:2013haa}, motivating ``optimal'' truncation strategies~\cite{Yunes:2008tw,Faggioli:2024ugn}.

\subsubsection{Dependence on $e_0$ and $p_0/m$}
\begin{figure*}[!t]
    \centering
    \includegraphics[width=\linewidth]{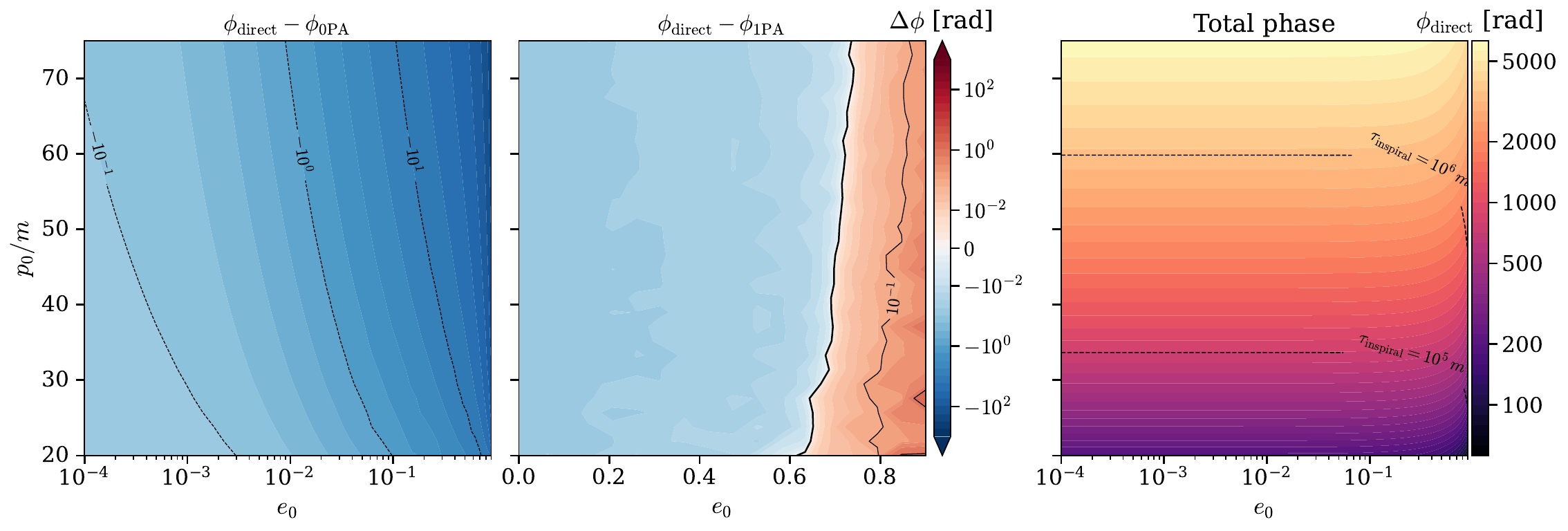}
    \caption{Orbital dephasing as function of initial eccentricity and semi-latus rectum with each system initialized with $\vartheta_0 = \pi/2$. The first two panels show the dephasing due to 0PA and 1PA respectively, while the third panel shows the total phase accumulated from the direct integration.
    In the latter, we show contours of fixed inspiral time $\tau_{\rm inspiral}/m$.
    Note that $e_0$ is plotted on a log-scale in the first and third panels, and on a linear scale in the second panel.
    Observe that the 0PA becomes inaccurate even for $e_0 \gtrsim 0.1$, while the 1PA maintains accuracy up to $e_0 \approx 0.9$.
}
    \label{fig:dephasing_true_anom_piby2}
\end{figure*}

We now study the dependence of the dephasing on the initial eccentricity and semi-latus rectum, for fixed mass ratio $q=1$ and initial true anomaly $\vartheta_0 = \pi/2$. For this choice of initial true anomaly, the magnitude of the 0PA dephasing is largest, as seen in~\cref{fig:dephasing_phi0_e_trueanom0}, so that the parameter sweep over $(e_0, p_0/m)$ provides a pessimistic estimate of the 0PA dephasing.
\Cref{fig:dephasing_true_anom_piby2} shows the 0PA and 1PA dephasing in the first two panels, and the total accumulated phase in the third panel. The contours of constant total phase approximately follow those of constant $\epsilon_p$ (cf.~\cref{fig:p_e_curve}), bending towards larger $p_0/m$ at large $e_0$: since the number of orbits in the inspiral scales roughly as $\epsilon_p^{-1}$, a larger $\epsilon_p$ leads to less accumulated phase. For reference, we also show lines of constant inspiral time $\tau_{\rm inspiral}$ (in units of $m$), computed as the minimum of the end times of the direct, 0PA, and 1PA evolutions; in contrast to the total phase, these bend towards smaller $p_0/m$ at large $e_0$, since at fixed $p_0/m$ a more eccentric binary has a larger semi-major axis and hence longer orbital periods.

At fixed $p_0/m$, the magnitude of the 0PA dephasing increases with $e_0$, as observed in the upper half plane of~\cref{fig:dephasing_phi0_e_trueanom0}. The 0PA dephasing also grows noticeably with $p_0/m$, by a factor of a few between $p_0/m = 20$ and $75$ at fixed $e_0$, since the initial offset is amplified over a longer inspiral; relative to the total accumulated phase, however, the 0PA dephasing decreases with increasing $p_0/m$, and increases with increasing $e_0$. The magnitude of the 0PA dephasing remains $\lesssim$~a few $\times\,10^{-1}$~rad only for $e_0 \lesssim 10^{-3}$--$10^{-2}$, depending on $p_0/m$, while it exceeds $1$~rad for $e_0 \gtrsim 0.03$ across the range of $p_0/m$ considered. The magnitude of the 1PA dephasing, by contrast, remains $\lesssim$~a few $\times\,10^{-1}$~rad for $e_0 \lesssim 0.8$, and reaches $\sim 1$~rad for $e_0 \sim 0.9$. The 0PA dephasing does not change sign, as expected for $\vartheta_0 = \pi/2$, whereas the 1PA dephasing changes sign at $e_0 \approx 0.6$--$0.75$, depending on $p_0/m$; as before, we attribute this sign change to the neglected 2PA and higher-order contributions.

\subsubsection{Dependence on $e_0$ and $q$}

\begin{figure*}[!t]
    \centering
    \includegraphics[width=\linewidth]{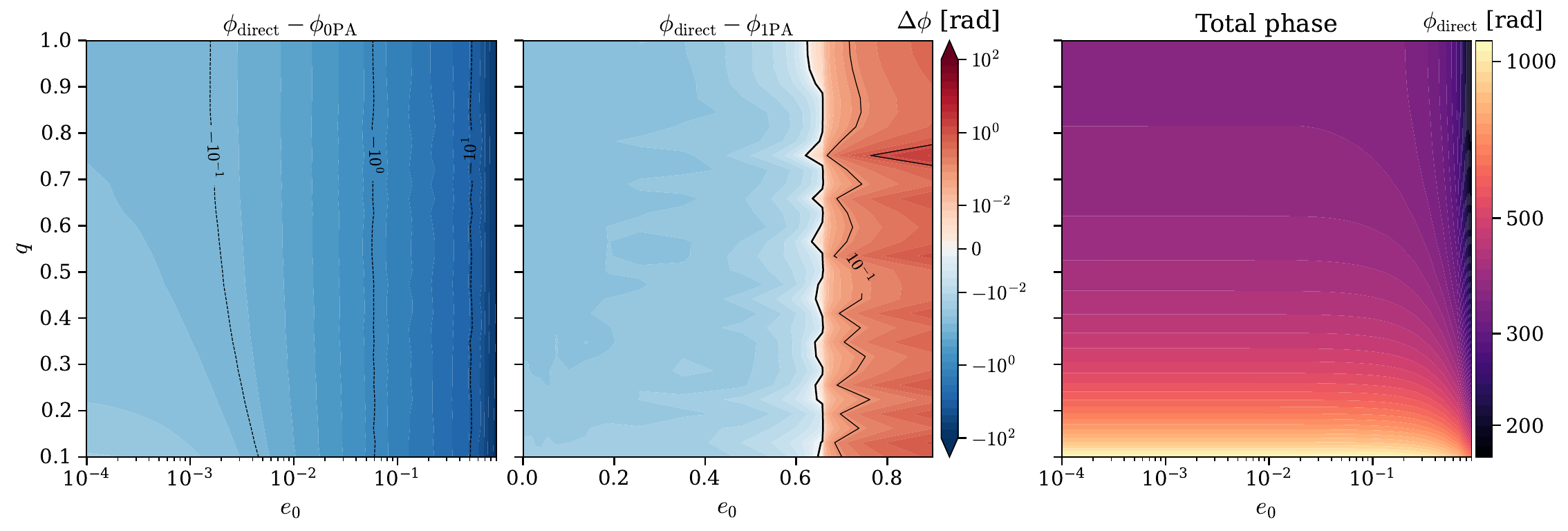}
    \caption{Orbital dephasing as function of initial eccentricity and mass ratio, where each system is initialized with $\{p_0/m = 25,  \varpi_0 = 0, \vartheta_0 = \pi/2 \}$, with the total mass held fixed. The first two panels show the dephasing due to 0PA and 1PA respectively, while the third panel shows the total phase accumulated from the direct integration.
    Note that $e_0$ is plotted on a log-scale in the first and third panels, and on a linear scale in the second panel.
    Observe that at moderate $e_0$, the 0PA dephasing depends weakly on $q$, whereas the 1PA dephasing shows a stronger dependence.
    }
    \label{fig:dephasing_q_e_trueanom_piby2}
\end{figure*}

We now study the dependence of the dephasing on the initial eccentricity and mass ratio, for fixed $p_0 = 25m$ and $\vartheta_0 = \pi/2$. As before, this choice of initial true anomaly provides a pessimistic estimate of the 0PA dephasing. \Cref{fig:dephasing_q_e_trueanom_piby2} shows the 0PA and 1PA dephasing in the first two panels, and the total accumulated phase in the third panel. At fixed $e_0$, the accumulated phase increases as $q$ decreases, since the radiation-reaction force is weaker and thus the binary completes more orbits before the end of the inspiral. At fixed $q$, the accumulated phase decreases with increasing $e_0$, as also seen in~\cref{fig:dephasing_phi0_e_trueanom0,fig:dephasing_true_anom_piby2}.

The features of the 0PA and 1PA dephasing in the $(e_0, q)$ grid are similar to those in the $(e_0, p_0/m)$ grid: the magnitude of the 0PA dephasing increases with $e_0$, and the 1PA dephasing changes sign at $e_0 \approx 0.65$, which we again attribute to the neglected 2PA and higher-order contributions. For $e_0 \gtrsim 10^{-2}$, the 0PA dephasing is nearly independent of $q$. This follows from the scaling of its dominant contribution: the total phase scales as $\phi_{\rm direct} \propto 1/\eta$, while the 1PA corrections scale as $\epsilon \propto \eta$, so that the 0PA dephasing, $\sim \epsilon\, \phi_{\rm direct}$, is independent of $\eta$. At smaller eccentricities, $e_0 \lesssim 5\times10^{-3}$, the magnitude of the 0PA dephasing instead increases with $q$, since the neglected 2PA contributions, which scale as $\sim \epsilon^2 \phi_{\rm direct} \propto \eta$, dominate in this regime. The same scaling applies to the 1PA dephasing at small eccentricities, consistent with its weak increase with $q$. At large eccentricities, $e_0 \gtrsim 0.7$, the 1PA dephasing shows a weak dependence on $q$, which is partly masked by variations between neighboring grid points arising from the sensitivity of the final-time dephasing to the end point of the evolution.

\section{Conclusion} \label{sec:conclusion}
The adiabatic 0PA approximation is typically used to describe the eccentric dynamics of comparable-mass binaries, both for (i) developing waveform models (e.g.,~\cite{Moore:2019xkm,Klein:2021jtd,Arredondo:2024nsl,Morras:2025nlp,Planas:2025feq}) and for (ii) astrophysical formation models or multi-band observations of binaries evolving from large orbital separations to merger~\cite{Sesana:2013dma,Tagawa:2020jnc,Gondan:2020svr,Romero-Shaw:2024klf}. Understanding the regime of validity of the 0PA approximation is therefore an important problem for both waveform and astrophysical modeling. In~\cite{Fumagalli:2025rhc}, the authors partially addressed this question by computing the difference in merger times between direct (non-adiabatic) and 0PA evolutions, and found that the initial true (or mean) anomaly is important, especially for highly eccentric binaries. Separately, in~\cite{Lynch:2026ibo}, the authors developed 1PA and 2PA approximants of an eccentric EOB model, which showed that the PA corrections improve the accuracy across parameter space. Our work focuses on providing an analytical understanding of when and why the 1PA corrections improve upon the 0PA approximation, which provides further insight into the findings of~\cite{Fumagalli:2025rhc} and complements the EOB waveform modeling efforts of~\cite{Lynch:2026ibo,Lynch:2026eyw,Gamba:2026ado}.\\

Towards obtaining such an analytical understanding, we first accounted for three subtleties concerning: (i) the ansatz for the MSA, (ii) the initial conditions for the secular and oscillatory pieces, and (iii) the gauge dependence of the PA corrections. Although the ansatz of~\cite{PoissonWill2014} differs from that of~\cite{Mora:2003wt,Will:2016,Will:2019lfe,Will:2020tri}, we showed explicitly that the two lead to the same secular equations, and determined consistent initial conditions for the secular and oscillatory pieces in each formalism, addressing subtleties (i) and (ii). For subtlety (iii), following~\cite{Arun:2009mc}, we derived the $\mathcal{O}(\epsilon)$ change in the functional form of the gravitational-wave polarizations under 2.5PN gauge transformations of the orbital elements. This change cancels, to $\mathcal{O}(\epsilon)$, the gauge dependence of the 1PA orbital corrections when comparing waveforms at the same physical event. Establishing agreement over a full inspiral between waveforms from independently truncated evolutions in different gauges requires assessing the higher-order terms generated by the gauge transformation.\\

Our main analytical findings regarding the validity of the PA approximations are as follows. The 1PA corrections enter at relative 2.5PN order in the dissipative dynamics, and improve the description of the secular evolution by accounting for the oscillatory behavior of the orbital elements. At 1PA order, the pericenter undergoes a secular precession that is nonzero for all initial conditions. This 1PA precession is not sourced by the orbit-averaged radiative 2.5PN fluxes, but instead arises from the coupling of the 1PA oscillatory terms to the radiation-reaction forcing, and its rate depends on the gauge. Using the resummed formalism, we find that the secular evolution equations for the semi-major axis and eccentricity, and thus for the energy and angular momentum, receive no explicit 1PA corrections, while that for the pericenter does. In general, however, the solutions for the semi-major axis and eccentricity acquire secular 1PA contributions through the initial conditions, namely from the initial values of their 1PA oscillatory terms. When the binary is initialized at a radial turning point, these oscillatory terms, which are odd in the anomaly, vanish initially, so that no such secular 1PA contributions arise for the semi-major axis and eccentricity. This does not hold for the pericenter, whose 1PA oscillatory term is even in the anomaly and generally nonzero at a radial turning point, in addition to its secular 1PA precession. In other words, at pericenter or apocenter, the osculating energy and angular momentum coincide with their orbit-averaged values up to 2PA corrections, so that the 0PA evolution starts from the correct secular initial conditions for these quantities. This is why the initial true (or mean) anomaly is important in determining the accuracy of the 0PA approximation, and, for the same reason, why the 1PA approximation does not offer a significant improvement over 0PA when the system is initialized at a radial turning point.\\

To assess the validity of the 0PA and 1PA approximations, we compared them to direct (non-averaged) solutions over a range of initial eccentricities ($e_0 \in [10^{-4}, 0.9]$), semi-latera recta ($p_0/m \in [20, 75]$), true anomalies ($\vartheta_0 \in [-\pi/2, \pi/2]$), and mass ratios ($q \in [1/10, 1]$).
We confirmed our analytical prediction that the 0PA and 1PA approximations are comparably accurate when initialized at pericenter, with orbital dephasings (relative to the direct solution) of magnitude $\lesssim$~a few $\times\,10^{-1}$~rad for $e_0 \lesssim 0.8$. When initialized away from pericenter ($|\vartheta_0| > \pi/4$), however, the magnitude of the 0PA dephasing exceeds $1$~rad already at moderate eccentricities, $e_0 \gtrsim 0.1$, independently of the mass ratio. The magnitude of the 1PA dephasing, by contrast, depends only weakly on the initial true anomaly, and remains $\lesssim$~a few $\times\,10^{-1}$~rad for $e_0 \lesssim 0.8$ across the parameter space of this study, reaching $\sim 1$~rad only for $e_0 \sim 0.9$. At small eccentricities, $e_0 \lesssim 10^{-2}$, the 0PA approximation is accurate irrespective of the initial true anomaly, and 1PA offers no significant improvement, since both are limited by the neglected 2PA contributions.\\

We also found that the 0PA and 1PA dephasing change sign in parts of the parameter space. The 0PA dephasing changes sign only for $\vartheta_0 < 0$, along a curve in the $(e_0, \vartheta_0)$ plane that starts at $e_0 \approx 5\times10^{-3}$ for $\vartheta_0 = -\pi/2$ and moves to larger $e_0$ as $\vartheta_0$ approaches the pericenter ($\vartheta_0 = 0$). The 1PA dephasing changes sign at $e_0 \approx 0.6$--$0.75$, depending only weakly on $\vartheta_0$, $p_0/m$, and $q$. We attribute these sign changes to competing contributions from successive PA orders: when the 1PA and 2PA contributions have opposite signs, truncating at 0PA can be more accurate than truncating at 1PA, and, likewise, the 1PA approximation can appear more accurate when the neglected higher-order contributions have opposite signs. This behavior is reminiscent of the PN expansion, where successive PN terms can have opposite signs, so that truncating at a lower PN order can be ``accidentally'' more accurate than truncating at a higher one~\cite{Blanchet:2013haa,Yunes:2008tw,Faggioli:2024ugn}.

Our results are limited in the following ways. We did not explicitly compute 2PA corrections in this work, and thus all our calculations have remainders of 2PA order. Nevertheless, we can perform a Fermi estimate of the size of these remainders, from which we find can contribute $\sim \epsilon \phi_{\rm direct} \sim 10^{-2}$ rad dephasing across a wide range of parameter space. In Sec.~\ref{sec:num_exp}, we further argued why these 2PA corrections remain non-zero even in the quasi-circular limit. This agrees with the findings of~\cite{Lynch:2026eyw,Gamba:2026ado}, where 2PA corrections were included in eccentric EOB models in the quasi-circular limit. Another limitation of our work is that we neglected sub-leading PN corrections in computing the 1PA contributions. As explained earlier, this is primarily because our goal is to provide an analytic explanation for the validity of the PA approximation.
The 0PA becomes innacurate mainly because the system is initialized away from a radial turning point which is not sensitive to including PN corrections, provided one is still in the weak-field regime. When including higher PN corrections, the magnitude of the total dephasing (relative to the direct evolution) for the 0PA and 1PA can change, but we expect that their relative dephasing (relative to the direct evolution) to be consistent with our leading PN calculation, as long as one is in the weak-field PN regime. 
For a PN accurate inspiral and waveform model, one should naturally include all relevant PN contributions, which is beyond the scope of our work.

We conclude by stressing that the orbital dephasing of the 0PA (or 1PA) approximation relative to the direct solution does not necessarily translate into systematic errors in parameter estimation or a loss of SNR.
Our dephasing comparisons use fixed initial conditions and system parameters.
The mismatch results illustrate the importance of this choice: re-initializing the 0PA evolution at a radial turning point reduces its mismatch from $\approx 0.67$ to $\approx 2\times10^{-5}$ in our highly eccentric case study.
Assessing the potential SNR loss from using 0PA waveforms requires maximizing the match over the system's parameters to compute the fitting factor~\cite{Cornish:2011ys}.
As waveform systematics become increasingly important with improving detector sensitivity~\cite{Gupta:2024gun,Chandramouli:2024vhw}, it would be instructive in future work to assess parameter biases and SNR loss by analyzing signals from the direct evolution with 0PA and 1PA waveform models.

\section*{Acknowledgments}
R.~S.~C. is thankful to Nicholas Loutrel, Eric Poisson, and Clifford Will for early discussions regarding the MSA formulations and initial conditions, which sparked this project. R.~S.~C. is grateful to Philip Lynch for helpful comments and feedback on early results concerning the numerical experiments. R.~S.~C. also thanks Maarten Van de Meent, Adam Pound, and David Trestini for useful discussions at the Capra 29 meeting. R.~S.~C. and L.~J.~G.~D.~S. are funded by the European Union (ERC grant GWSky/ 101167314). Views and opinions expressed are, however, those of the authors and do not necessarily reflect those of the European Union or the European Research Council Executive Agency. Neither the European Union nor the granting authority can be held responsible for them. 
N.~A.~was supported by the Alfred P.~Sloan Foundation, the UIUC Graduate College, and the UIUC College of Engineering during the writing of this manuscript.
N.~Y.~acknowledges support from Simons Foundation (via Award No.~896696), the
Simons Foundation International (via Grant No.~SFIMPS-BH-00012593-01), and the NSF (via Grant No.~PHY-2512423).

\emph{Software}. 
Analytical calculations were performed using Mathematica, version 14.1 (Wolfram Research, Inc.).
Additionally, the following python packages were used in this work:
\texttt{scipy}~\cite{Virtanen2020},
\texttt{numpy}~\cite{Harris2020},
\texttt{matplotlib}~\cite{Hunter2007},
\texttt{sympy}~\cite{Meurer2017},
\texttt{astropy}~\cite{Astropy2022},
\texttt{jax}~\cite{Bradbury2018}, and
\texttt{diffrax}~\cite{Kidger2021}.

\emph{Data availability}. The code used in this work will be made publicly available on GitHub at \url{https://github.com/rsc496/post-adiabatic-RR}.

\emph{AI statement}. Anthropic's Claude (Opus 5.5) and OpenAI's Codex (GPT-6.1 Sol Ultra) was used for assistance with codes (numerical optimization; converting and checking equations between Mathematica, Python, and LaTeX;  plots), and manuscript editing. The authors are responsible for all content in this manuscript.

\appendix

\section{Regularized 1PA corrections and quasi-circular limit}\label{sec:quasi_circ}
When using $\{ \mathcal{A}_x, \mathcal{A}_y, \phi \}$, the LPE (cf.~\cref{eqn:LPE}) become regular in the $e \rightarrow 0$ limit, and are given by
\begin{subequations}
\label{eq:LPE_reg}
\begin{align}
\frac{dp}{dt} &= 2\sqrt{\frac{p^3}{Gm}}\;
  \frac{\mathcal{S}}{1 + \mathcal{A}_x\cos\phi + \mathcal{A}_y\sin\phi}\,, \\[4pt]
\frac{d\mathcal{A}_x}{dt} &= \sqrt{\frac{p}{Gm}}
  \Big [ \mathcal{R}\sin\phi  \nn
  & \left. + \mathcal{S}\,\frac{\mathcal{A}_x + \left(2 + \mathcal{A}_x\cos\phi + \mathcal{A}_y\sin\phi\right)\cos\phi}
  {1 + \mathcal{A}_x\cos\phi + \mathcal{A}_y\sin\phi}\right], \\[4pt]
\frac{d\mathcal{A}_y}{dt} &= \sqrt{\frac{p}{Gm}}
  \Big [-\mathcal{R}\cos\phi \nn
  &+ \left. \mathcal{S}\,\frac{\mathcal{A}_y + \left(2 + \mathcal{A}_x\cos\phi + \mathcal{A}_y\sin\phi\right)\sin\phi}
  {1 + \mathcal{A}_x\cos\phi + \mathcal{A}_y\sin\phi}\right], \\[4pt]
\frac{d\phi}{dt} &= \sqrt{\frac{Gm}{p^3}}\,
  \left(1 + \mathcal{A}_x\cos\phi + \mathcal{A}_y\sin\phi\right)^2 .
\end{align}
\end{subequations}
Note that the time evolution of the orbital phase is simply the Keplerian result, which is what is expected for an osculating orbit. In this form, when converting derivatives from $d/dt \rightarrow d/d\phi$, there is no need to re-expand the RHS of $d X_{\alpha}/d\phi$ in terms of $\epsilon$, and thus there are no explicit NLO sourcing functions $Q^{(1)}_{\alpha}(X_{\alpha}(\phi),\phi)$, and one only needs $Q^{(0)}_{\alpha}(X_{\alpha}(\phi),\phi)$, which we provide in the supplemental Mathematica file for the case of radiation reaction.
 
We then proceed with the MSA formalism of Sec.~\ref{subsec:expanded-MSA} and Sec.~\ref{subsec:resum_IC}, as we did when using $\{ e, \varpi, \vartheta \}$.  The slow variable is now $\Phi = \epsilon \phi$, and the orbital elements are promoted to $X_{\alpha} (\Phi, \phi)$.  The average is now defined with respect to $\phi$, so we have  $$\av{A} (\Phi) = \frac{1}{2\pi} \int_0^{2 \pi} A (\Phi,\phi) d\phi,$$
which means that oscillatory pieces track oscillations with respect to $\phi$ and not $\vartheta$. Apart from that, the way in which we compute the 1PA oscillatory and secular pieces follow the same steps as in Sec.~\ref{subsec:expanded-MSA} and Sec.~\ref{subsec:resum_IC}. In terms of the resummed secular orbital elements, we have $X_{\alpha} (\Phi, \phi) = \tilde{X}_{\alpha} (\Phi) + \epsilon Y^{(1)}_{\alpha} (\Phi, \phi))$.
For instance, the 1PA oscillatory correction to $p$ is given by
\begin{equation}
Y^{(1)}_p = -\frac{4}{5}\,\eta\,\tilde{p}\left(\frac{Gm}{c^2\tilde{p}}\right)^{5/2}
\tilde{e}\left[\left(36 + 4\tilde{e}^2\right)\sin\vartheta + 5\tilde{e}\sin 2\vartheta\right],
\end{equation}
where we converted back to eccentricity and true anomaly in the standard harmonic gauge. Our $Y^{(1)}_p$ exactly matches Eq.~(13) of~\cite{Will:2019lfe}, and we checked that our $Y^{(1)}_{\mathcal{A}_x}$ and $Y^{(1)}_{\mathcal{A}_y}$ also match theirs.  In our supplemental Mathematica file, we provide the full expressions for these 1PA oscillations in the generalized gauge, together with the resummed 1PA secular evolution equations $d \tilde{X}_{\alpha}/d\Phi$ for the regularized orbital elements.
Further, the 1PA oscillation in $t$ remains the same as~\cref{eqn:t_osc}, and we also find that $d\tilde{\mathfrak{t}}/d \Phi = \sqrt{(\tilde{a}^3/m)} [1+ \mathcal{O}(\epsilon^2)]$.
Thus, when converting to secular time evolution equations, and after reverting to $\tilde{e}$ and $\tilde{\varpi}$, we recover~\cref{eqn:1PA_sec_resum}.

\section{Additional details on gauge transformations}\label{sec:add_det_gauge}

We here provide the intermediate steps in deriving the changes in the orbital elements under the gauge transformation of~\cref{eqn:gauge_transform}.
From $\vec{\xi}$, we first compute $\delta r = \vec{\xi} \cdot \vec{n}$ and $\delta \phi = \vec{\xi} \cdot \vec{\lambda}/r$.
We then compute $\delta \dot{r}$ and $\delta \dot{\phi}$ by taking the time derivatives of $\delta r$ and $\delta \phi $ respectively, and use the Newtonian EoM to order reduce the second order time derivatives.
This gives us
\begin{subequations}\label{eqn:delta_coord}
    \begin{align}
        \delta r &= \frac{8}{5} \eta m^2 (1+\alpha_g - \beta_g) \frac{\dot{r}}{r},\\
        \delta \phi &= \frac{8}{15 } \eta m^2 (3+3\alpha_g -2 \beta_g) \frac{\dot{\phi}}{r},\\
        \delta \dot{r} &= -\frac{8}{5} \eta m^2 (1+\alpha_g - \beta_g) \Big ( \frac{m}{r^3} + \frac{\dot{r}^2}{r^2} -\dot{\phi}^2 \Big ), \\
        \delta \dot{\phi} &= -\frac{8}{5} \eta m^2 (3+3\alpha_g - 2\beta_g) \frac{\dot{r} \dot{\phi}}{r^2}.
    \end{align}
\end{subequations}

The reduced energy $E/\mu$, reduced angular momentum $L / \mu$, and Runge-Lenz vector $\vec{\mathcal{A}} = \mathcal{A}_r \vec{n} + \mathcal{A}_{\phi}\vec{\lambda}$, expressed in terms of the standard harmonic coordinates, reads
\begin{subequations}\label{eqn:conserved_coord}
    \begin{align}
     \frac{E}{\mu} &= \frac{v^2}{2} - \frac{m}{r}, \\
     \frac{L}{\mu} &= r^2 \dot{\phi}, \\
     \mathcal{A}_r & = \frac{L r\dot{\phi}}{\mu m} -1 ,\\
     \mathcal{A}_{\phi} & = - \frac{L \dot{r}}{\mu m}. 
    \end{align}
\end{subequations}
Using~\cref{eqn:delta_coord}, we then compute the changes $\delta E/\mu$, $\delta L/\mu$, $\delta \mathcal{A}_r$, $\delta \mathcal{A}_{\phi}$, which gives
\begin{subequations}\label{eqn:delta_cons_coord}
    \begin{align}
        \frac{\delta E}{\mu} & = -\frac{8}{5} \eta m^2 \dot{r} \Big ( (1+\alpha_g -\beta_g) \frac{\dot{r}^2}{r^2} + (1+\alpha_g) \dot{\phi}^2 \Big ),\\
        \frac{\delta L}{\mu} & = -\frac{8}{5}m^2 \eta (1+\alpha_g) \dot{r} \dot{\phi},\\
        \delta \mathcal{A}_r &= -\frac{8}{15} m \eta (6+6\alpha_g - \beta_g) r \dot{r} \dot{\phi}^2,\\
         \delta \mathcal{A}_{\phi} &= -\frac{8}{15} m \eta \dot{\phi} \Big [ \beta_g \left( \frac{m}{r} - r^2 \dot{\phi}^2 \right) + 3 (-2 - 2\alpha_g + \beta_g) \dot{r}^2 \Big ].
        \end{align}
\end{subequations}
To compute the changes $\delta X_{\alpha}$ in the orbital elements, we vary the Keplerian relations $X_{\alpha}=X_{\alpha}(E/\mu,L/\mu,\vec{\mathcal{A}})$~\cite{PoissonWill2014} with respect to the reduced energy and angular momentum and the Runge--Lenz vector. This gives
\begin{subequations}
    \begin{align}
        \delta a &= \frac{2 a^2}{m \mu} \delta E,\\
        \delta p &= 2 \frac{L}{\mu^2 m} \delta L,\\
        \delta e &= \delta \mathcal{A}_r \cos \vartheta - \delta \mathcal{A}_{\phi} \sin \vartheta,\\
        \delta \varpi &= \frac{1}{e} \left( \delta \mathcal{A}_r \sin \vartheta + \delta \mathcal{A}_{\phi} \cos \vartheta \right),\\
        \delta \vartheta &= \delta \phi - \delta \varpi,
    \end{align}
\end{subequations}
which results in~\cref{eqn:orbital_elements_gauge_transform} after using~\cref{eqn:delta_coord,eqn:delta_cons_coord}, and evaluating $r$, $\phi$, $\dot{r}$, and $\dot{\phi}$ using the Keplerian osculating orbit relations.

\section{Additional details on numerical experiments}\label{sec:add_det_num}
We first discuss numerical errors and sanity checks, then describe the SNR and mismatch calculations, and finally present additional case studies supplementing Sec.~\ref{sec:num_exp}.

 \begin{figure*}[!t]
     \centering
     \includegraphics[width=\linewidth]{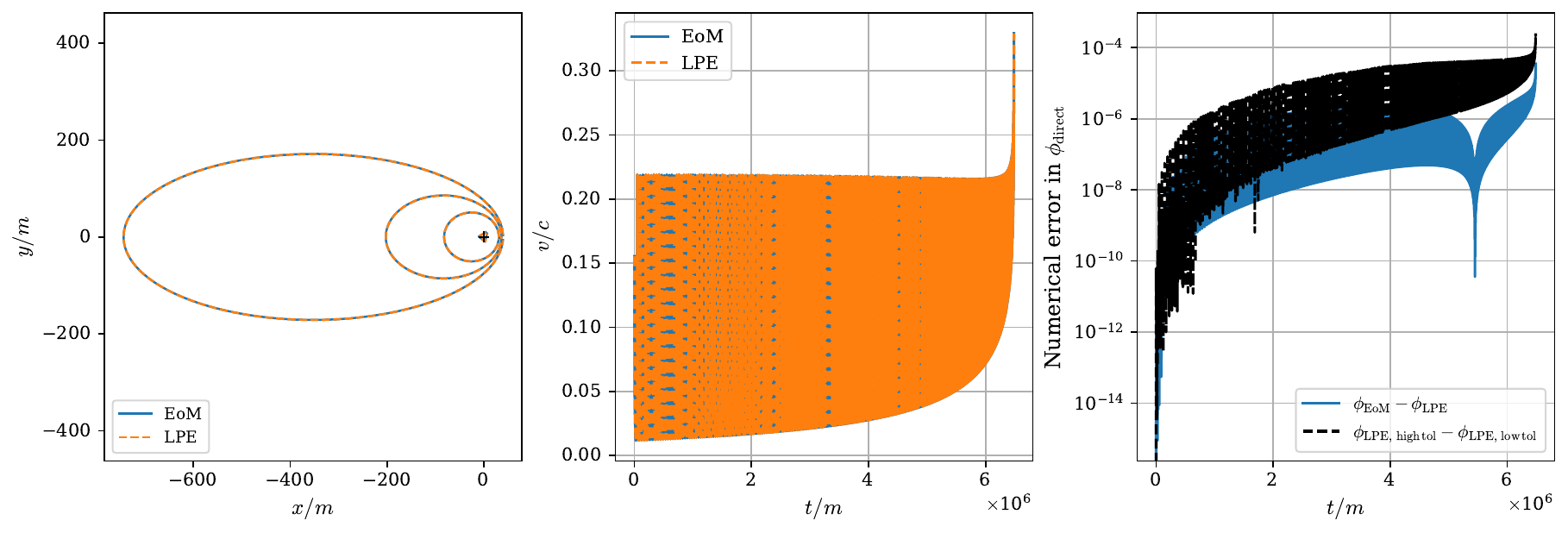}
     \caption{Numerical error analysis in the direct integration of an equal mass system with $\{ p_0/m =75, e_0 = 0.9 , \vartheta_0 = \pi/2, \varpi_0 = 0\}$. The first panel shows a few snapshots of the osculating orbit computed from the trajectories corresponding to EoM (blue) and the LPE (orange dashed). The middle panel shows the velocity as function of time for the entire inspiral from both the EoM (blue) and LPE (orange-dashed). The third panel shows the numerical errors in the phase computed from the difference between the EoM and LPE solutions in blue, and between the LPE solutions at two different tolerances in black dashed lines (see \cref{app:numerical}). Observe that these numerical errors are much smaller than the errors due to the PA approximations.}
     \label{fig:numerical_comparison}
 \end{figure*}
 
\begin{figure*}[!t]
    \centering
    \includegraphics[width=\linewidth]{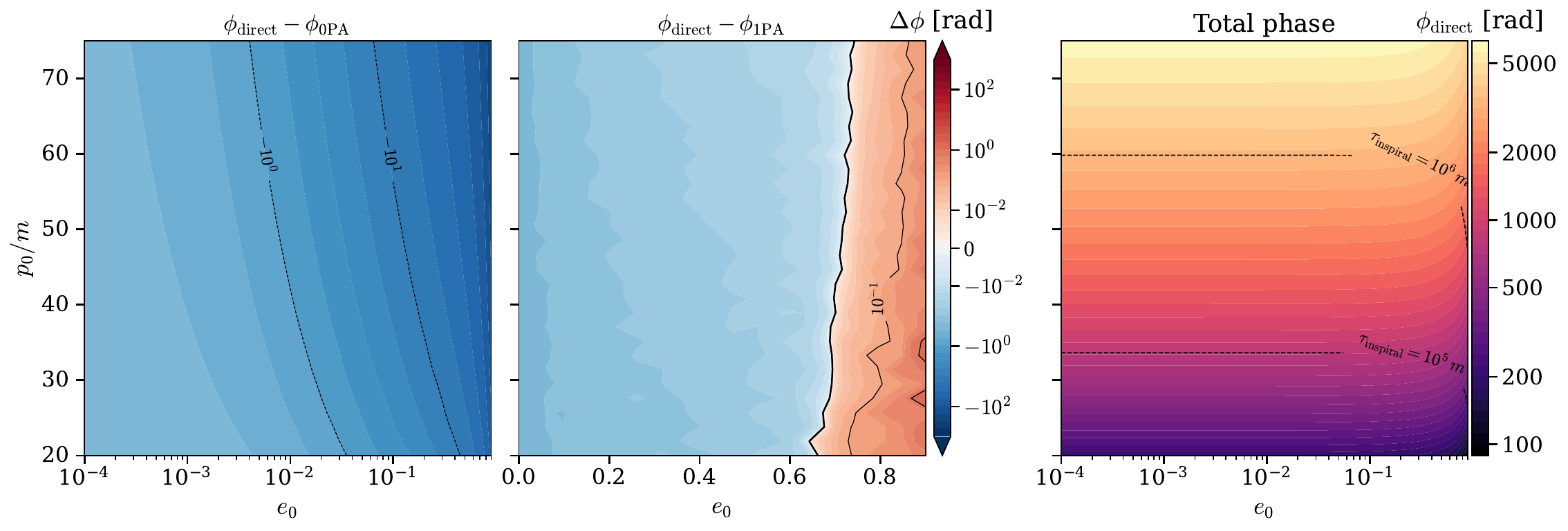}
    \caption{Orbital dephasing as function of initial eccentricity and semi-latus rectum with each system initialized with $\vartheta_0 = \pi/2$, but with each evolution terminated at the LSO. 
    The first two panels show the dephasing due to 0PA and 1PA respectively, while the third panel shows the total phase accumulated from the direct integration.
    In the latter, we show contours of fixed inspiral time $\tau_{\rm inspiral}/m$.
    Note that $e_0$ is plotted on linear scale in the first panel panels, and on log-scale in the other two panels. 
    Observe that the results are robust to stopping condition; using the LSO instead of~\cref{eqn:stopping_cond}, as used in~\cref{fig:dephasing_true_anom_piby2}, does not significantly affect the results.
    }
    \label{fig:dephasing_true_anom_piby2_LSO}
\end{figure*}

\subsection{Numerical errors and sanity checks} \label{app:numerical}
To quantify the numerical errors in the orbital integration, we performed the following checks. We computed the direct (non-averaged) evolution in two independent ways: by integrating the LPE for the osculating orbital elements (cf.~\cref{eqn:LPE}) then reconstructing the trajectory from them, and by integrating the EoM for $\{r, \phi, \dot{r}, \dot{\phi}\}$ directly and extracting the osculating orbital elements from the resulting trajectory. Both integrations use \texttt{DOP853} with \texttt{atol=1e-14} and \texttt{rtol=1e-12}. We then computed the difference between their phases, $\phi_{\rm EoM} - \phi_{\rm LPE}$, as an estimate of the numerical error in $\phi_{\rm direct}$.
Since the two formulations are related by an exact map, this difference arises from the accumulation of integration errors, which differ between the two sets of variables. We also integrated the LPE with two sets of tolerances, \texttt{atol=1e-14}, \texttt{rtol=1e-12} (``high tol'') and \texttt{atol=1e-12}, \texttt{rtol=1e-10} (``low tol''), and computed $\phi_{\rm LPE,\,high\,tol} - \phi_{\rm LPE,\,low\,tol}$ as a second estimate of the numerical error.

In~\cref{fig:numerical_comparison}, we compare these methods and their resulting errors for an equal-mass system with $\{p_0/m = 75,\, e_0 = 0.9,\, \vartheta_0 = \pi/2,\, \varpi_0 = 0\}$.
We chose the largest semi-latus rectum and eccentricity in our grid (see Sec.~\ref{sec:setup}), since a larger $p_0$ leads to a longer inspiral, over which numerical errors accumulate, while a larger $e_0$ leads to more rapid variations near pericenter.
We therefore expect this system to be among the most demanding in our grid, such that its numerical errors provide a conservative estimate.
We find that the resulting numerical errors are several orders of magnitude smaller than the dephasing due to the PA approximations.
Integrators designed for long-term evolutions of non-conservative systems, such as slimplectic~\cite{Tsang_2015} and Magnusian~\cite{DaSilva:2026pdk} schemes, have also been developed; for the purposes of this work, however, \texttt{DOP853} with \texttt{atol=1e-14} and \texttt{rtol=1e-12} is sufficient.

Further, we tested the robustness of our results to the choice of stopping condition.
Instead of using~\cref{eqn:stopping_cond}, we terminated the inspiral at the LSO, which essentially allows for more phase to be accumulated.
In~\cref{fig:dephasing_true_anom_piby2_LSO}, we show the orbital dephasing across $e_0$ and $p_0$ for $\vartheta_0 = \pi/2$. 
We find that the results are qualitatively similar to that of~\cref{fig:dephasing_true_anom_piby2}, with slightly larger magnitude of dephasing across the parameter space.
We obtained similar results for the other parameter sweeps, although we do not show them here.
Thus, our choice of the stopping condition does not significantly alter the conclusions regarding the accuracy of the PA approximations.

\subsection{Signal processing and matched filtering}\label{sec:add_snr}
We here provide relevant details for computing the SNR and mismatch, which we presented for a case study in Sec.~\ref{sec:comparison_evol}.
For a detailed overview on signal processing, matched filtering, and Bayesian analysis of GWs, we refer the reader to~\cite{Cutler:1994ys,Flanagan:1997kp,Maggiore:2007ulw}.
First, the Fourier transform of a signal $h(t)$ is given by
\begin{equation}
    \tilde{h}(f) = \int_{-\infty}^{\infty} dt \, h(t) e^{-i 2\pi f t}.
\end{equation}
The inner product of any two signals $h_1 (t)$ and $h_2(t)$ is given by~\cite{Cutler:1994ys}
\begin{align}
    (h_1 \vert h_2) = 4 \mathrm{Re} \int_0^{\infty} \frac{df}{S_n(f)} \tilde{h}_1^*(f) \tilde{h}_2(f),
\end{align}
where $S_n(f)$ is the one-sided noise power-spectral density.
The optimal SNR $\rho$ of a signal $h(t)$ is then $\rho \equiv \sqrt{(h \vert h)}$, which becomes
\begin{align}
    \rho^2 = 4  \int_0^{\infty} \frac{df}{S_n(f)} |\tilde{h}(f)|^2.
\end{align}
The match $\mathfrak{M}$ between two signals $h_1(t)$ and $h_2(t)$ is then defined as
\begin{align}
    \mathfrak{M} = \max_{t_c, \phi_c} \dfrac{(h_1 | h_2)}{\rho_1 \rho_2},
\end{align}
where the inner product $(h_1 | h_2)$ is maximized over global time and phase offsets $t_c$ and $\phi_c$ respectively.
In practice, we use the implicit maximization over an overall phase shift described in~\cite{Moore:2018kvz}, together with a fast Fourier transform (FFT)~\cite{CooleyTukey1965} search and Brent refinement~\cite{Brent1973} over the relative time shift, similar to PyCBC's \texttt{optimized\_match}~\cite{PyCBC}, as explained below.
The quantity to maximize is
\begin{align}
     \max_{t_c, \phi_c} (h_1 | h_2) \equiv \max_{t_c, \phi_c}  4 \mathrm{Re} \int_0^{\infty} \frac{df}{S_n(f)} \tilde{h}_1^*(f) \tilde{h}_2(f) e^{i 2\pi f t_c + i \phi_c}.
\end{align}
Following~\cite{Moore:2018kvz}, we first define $\tilde{G}(f)$ as
\begin{align}
    \tilde{G}(f) = \frac{\tilde{h}_1^* (f) \tilde{h}_2(f)}{S_n(f)},
\end{align}
which implies that the quantity to maximize is
\begin{align}
\max_{t_c, \phi_c} (h_1 | h_2) = 4 \max_{t_c, \phi_c}  \mathrm{Re} [e^{i \phi_c}G(t_c)], 
\end{align}
where 
\begin{align}
    G(t_c) = \int_0^{\infty} df \, \tilde{G}(f) e^{i 2\pi f t_c}.
\end{align}
The maximization over $\phi_c$ can be done trivially by taking the absolute value:
\begin{align}
\max_{t_c, \phi_c} (h_1 | h_2) = 4 \max_{t_c} |G(t_c)|.
\end{align}
We first estimate $G(t_c)$ on a discrete time grid using an inverse FFT and locate its largest absolute value.
We then refine the corresponding time shift using bounded Brent minimization, as implemented in \texttt{scipy}, evaluating $G(t_c)$ with Simpson quadrature.
With the refined time shift in hand, we compute the maximized match $\mathfrak{M}$, with the mismatch given by $1-\mathfrak{M}$.

For the large-eccentricity case study in Sec.~\ref{sec:comparison_evol}, we apply a Tukey window with taper parameter $0.05$ to each uniformly sampled waveform before computing its Fourier or Q-transform.
We compute the Fourier transform using \texttt{numpy.fft.rfft}, retaining positive frequencies in the chosen analysis band and multiplying by the sampling interval $\Delta t$ to approximate the continuous transform defined above.
For the Q-transform of the waveform, we apply Gaussian frequency-domain filters at 800 logarithmically spaced central frequencies $f_0$ between $0.001$ and $0.1\,\mathrm{Hz}$, with widths $\sigma_f=f_0/Q$ and $Q=20$.
We plot the resulting amplitudes normalized to their maximum.

To compute the SNR and mismatches, we use the LISA sensitivity curve of~\cite{Robson:2018ifk}.
Our waveforms explicitly include a static, single-channel detector response at fixed source angles, with antenna patterns incorporating the $60^\circ$ arm geometry~\cite{Cutler:1997ta,Cornish:2002rt}.
We therefore multiply their effective strain PSD by $3/10$ to remove the averaged low-frequency response factor.

\subsection{More case studies and parameter sweeps comparing direct and 0PA/1PA evolutions}

\begin{figure*}[!t]
    \centering
    \includegraphics[width=\linewidth]{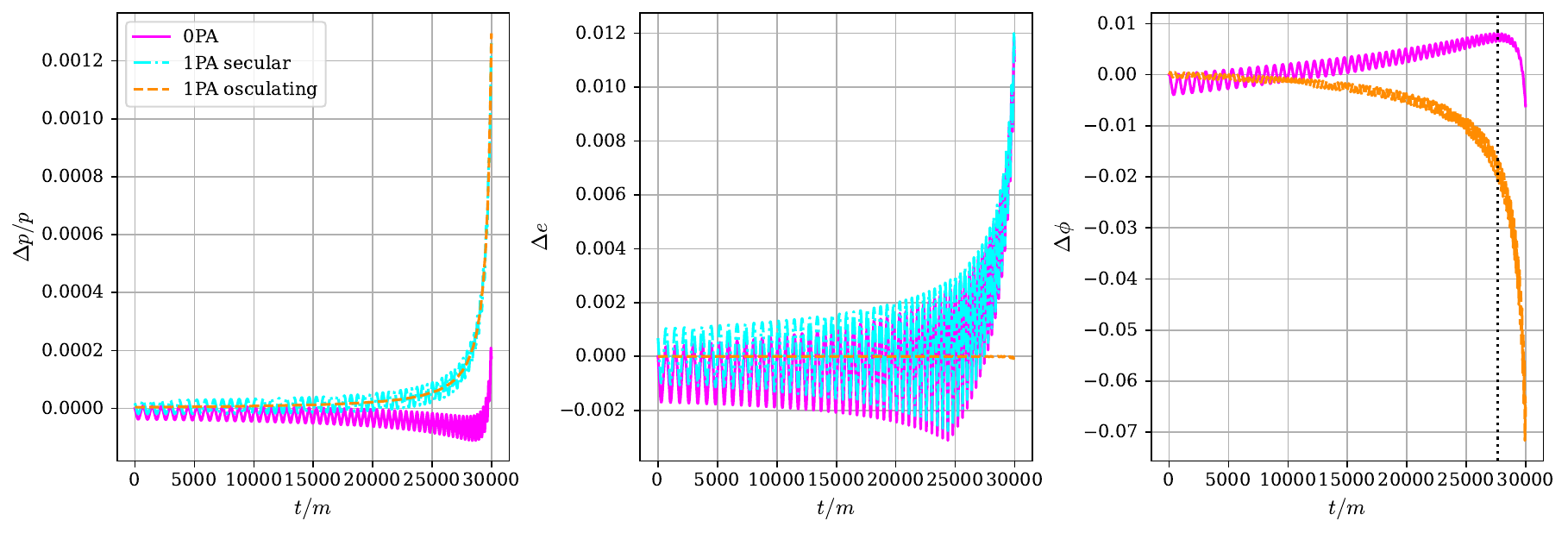}
    \caption{Errors in 0PA (magenta solid), and 1PA (osculating in orange dashed, secular in cyan dash-dot) evolutions (relative to direct) for $p$ (first panel), $e$ (second panel), and $\phi$ (third panel) for an equal mass binary with initial conditions $\{p_0/m = 25, e_0 = 0.006, \vartheta_0  = -\pi/4, \varpi_0 = 0 \} $. The vertical black line in the third panel corresponds to the maximum value of $\phi_{\rm direct} - \phi_{\rm 0PA}$ after which the 0PA dephasing starts to decrease to negative values.
    Observe that the 0PA dephasing undergoes a sign change during the late inspiral.
    }
    \label{fig:e_0PA_signflip}
\end{figure*}

In this section, we provide additional case studies and parameter sweeps to supplement 
the results shown in Sec.~\ref{sec:num_exp}. To highlight the change in sign of the 0PA and 1PA dephasing, we present a case study for each. We pick an equal mass system with $\{p_0/m = 25, e_0 = 0.006, \vartheta_0  = -\pi/4, \varpi_0 = 0 \} $ to highlight the change in sign of the 0PA dephasing, which is shown in~\cref{fig:e_0PA_signflip}. As the binary inspirals, we see the 
0PA dephasing has a secular change towards positive values, while the 1PA dephasing has a secular change towards negative values. While the 1PA dephasing has a monotonic secular change, the 0PA dephasing reaches a maximum when $t/m \approx 27653$, after which it undergoes a secular change towards negative values. Towards the end of the inspiral, the magnitude of dephasing for 0PA is smaller than that of 1PA, although both are negative. This supplements~\cref{fig:dephasing_phi0_e_trueanom0} as it shows how the neglected PA contributions can undergo a sign change during the inspiral. Consequently, this shows that truncating at a lower PA order can result in a smaller total dephasing.
\begin{figure*}[!t]
    \centering
    \includegraphics[width=\linewidth]{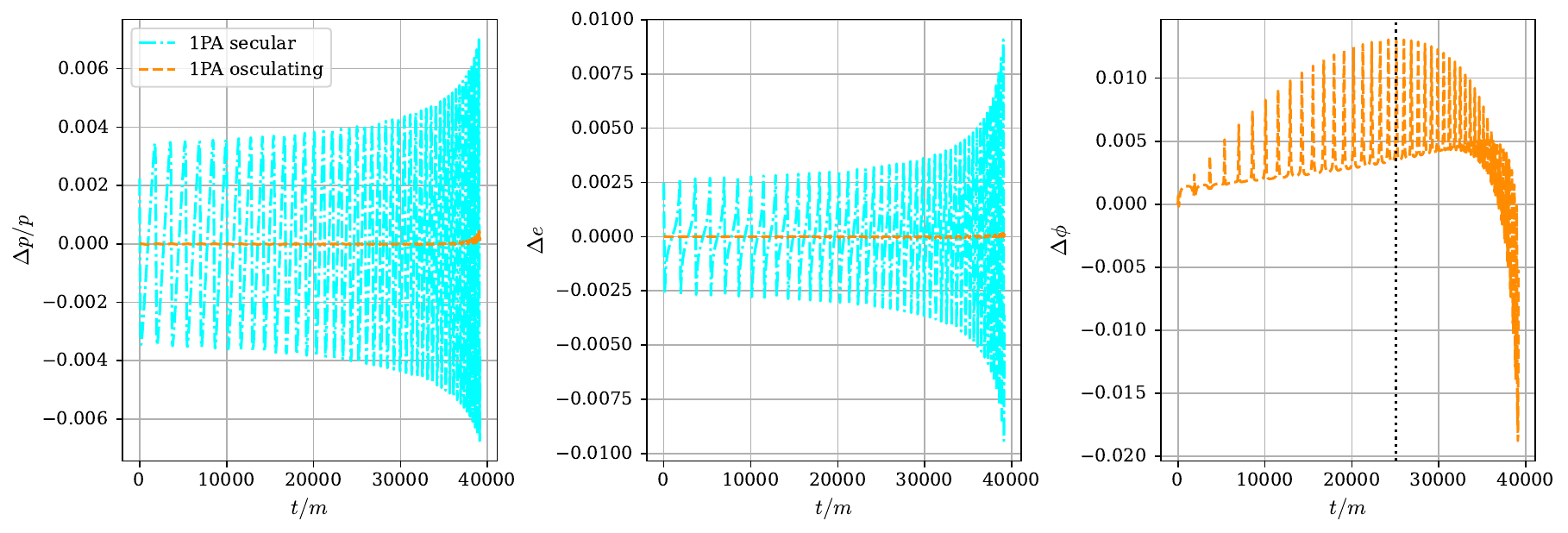}
    \caption{Errors in the 1PA (osculating in orange dashed, secular in cyan dash-dot) evolution (relative to direct) for $p$ (first panel), $e$ (second panel), and $\phi$ (third panel) for an equal mass binary with initial conditions $\{p_0/m = 25, e_0 = 0.67, \vartheta_0  = -\pi/4, \varpi_0 = 0 \} $. The vertical black line in the third panel corresponds to the maximum value of $\phi_{\rm direct} - \phi_{\rm 1PA}$ after which the 1PA dephasing starts to decrease to negative values. The curves for 0PA are left out since the corresponding errors have much larger magnitudes.
    Observe that the 1PA dephasing undergoes a sign change during the late inspiral.
    }
    \label{fig:e_1PA_signflip}
\end{figure*}

Similarly, the 1PA dephasing can also undergo a change in sign, as we shown in~\cref{fig:dephasing_phi0_e_trueanom0}. However, this typically occurs at larger eccentricities, across a range of $\vartheta_0$. We pick an equal mass system with $\{p_0/m = 25, e_0 = 0.67, \vartheta_0  = -\pi/4, \varpi_0 = 0 \} $, with the only change being $e_0$ when compared to the system in~\cref{fig:e_0PA_signflip}.  As shown in~\cref{fig:e_1PA_signflip}, the 1PA dephasing now undergoes a sign change at late times. For this system, the 0PA dephasing has a secular monotonic change, although we do not plot it in~\cref{fig:e_1PA_signflip} because its magnitude is much larger.
\begin{figure*}[!t]
    \centering
    \includegraphics[width=\linewidth]{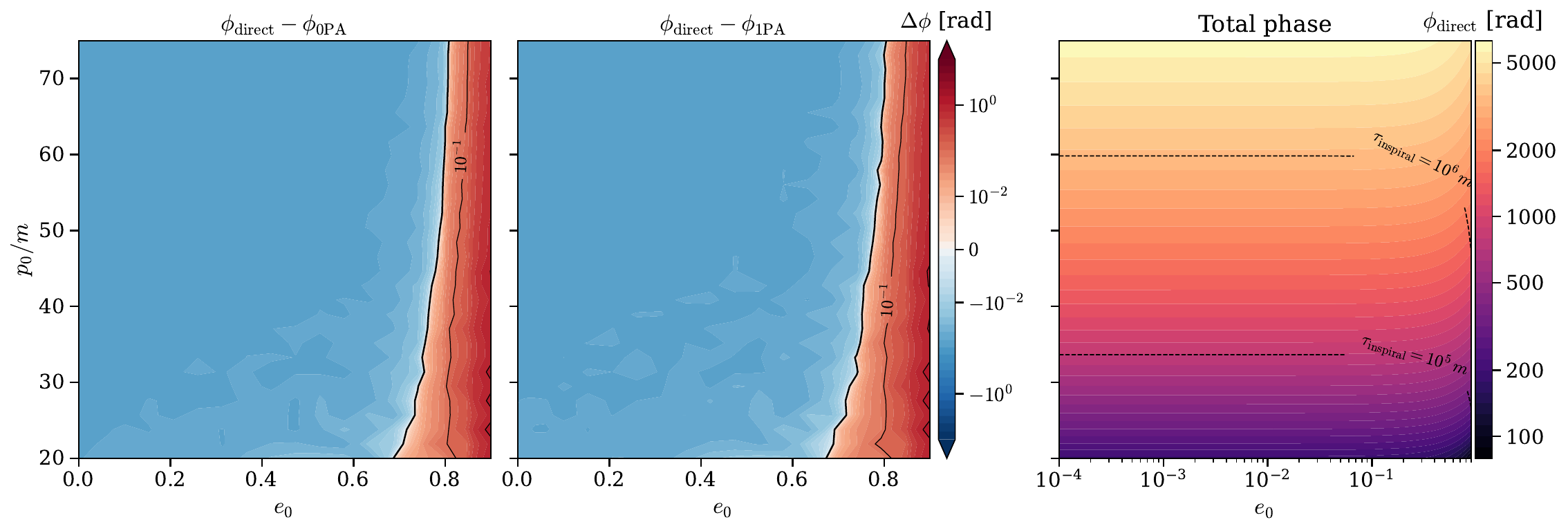}
    \caption{Orbital dephasing as function of initial eccentricity and semi-latus rectum with each system initialized with $\vartheta_0 = 0$ (as opposed to $\vartheta_0 = \pi/2$ in~\cref{fig:dephasing_true_anom_piby2}). The first two panels show the dephasing due to 0PA and 1PA respectively, while the third panel shows the total phase accumulated from the direct integration.
    In the latter, we show contours of fixed inspiral time $\tau_{\rm inspiral}/m$.
    Note that $e_0$ is plotted on linear scale in the first two panels, and on log-scale in the third panel.
    Observe that both the 0PA and 1PA dephasings are comparable to one another, and remain accurate across a large region of parameter space.
}
    \label{fig:dephasing_true_anom_0}
\end{figure*}

In~\cref{fig:dephasing_true_anom_piby2}, we showed how the 0PA and 1PA dephasing vary across $e_0$ and $p_0$ for $\vartheta_0 = \pi/2$, which corresponds to the largest 0PA dephasing, given its dependence on $\vartheta_0$ in~\cref{fig:dephasing_phi0_e_trueanom0}. For completeness, \cref{fig:dephasing_true_anom_0} shows how the 0PA and 1PA dephasing vary across $e_0$ and $p_0$ for $\vartheta_0 = 0$. When initialized at pericenter, the 0PA and 1PA dephasing are comparable across the entire parameter space, with magnitudes $\lesssim 10^{-1}$~rad for $e_0 \lesssim 0.8$.
Both also change sign at the same eccentricity, $e_0 \approx 0.7$--$0.8$, depending on $p_0/m$.
This further shows that, at pericenter, the dephasing in both cases arises from the neglected higher-order PA corrections, starting at 2PA.

\bibliographystyle{apsrev4-2}
\bibliography{ref}

\begin{thebibliography}{89}%
\makeatletter
\providecommand \@ifxundefined [1]{%
 \@ifx{#1\undefined}
}%
\providecommand \@ifnum [1]{%
 \ifnum #1\expandafter \@firstoftwo
 \else \expandafter \@secondoftwo
 \fi
}%
\providecommand \@ifx [1]{%
 \ifx #1\expandafter \@firstoftwo
 \else \expandafter \@secondoftwo
 \fi
}%
\providecommand \natexlab [1]{#1}%
\providecommand \enquote  [1]{``#1''}%
\providecommand \bibnamefont  [1]{#1}%
\providecommand \bibfnamefont [1]{#1}%
\providecommand \citenamefont [1]{#1}%
\providecommand \href@noop [0]{\@secondoftwo}%
\providecommand \href [0]{\begingroup \@sanitize@url \@href}%
\providecommand \@href[1]{\@@startlink{#1}\@@href}%
\providecommand \@@href[1]{\endgroup#1\@@endlink}%
\providecommand \@sanitize@url [0]{\catcode `\\12\catcode `\$12\catcode
  `\&12\catcode `\#12\catcode `\^12\catcode `\_12\catcode `\%12\relax}%
\providecommand \@@startlink[1]{}%
\providecommand \@@endlink[0]{}%
\providecommand \url  [0]{\begingroup\@sanitize@url \@url }%
\providecommand \@url [1]{\endgroup\@href {#1}{\urlprefix }}%
\providecommand \urlprefix  [0]{URL }%
\providecommand \Eprint [0]{\href }%
\providecommand \doibase [0]{https://doi.org/}%
\providecommand \selectlanguage [0]{\@gobble}%
\providecommand \bibinfo  [0]{\@secondoftwo}%
\providecommand \bibfield  [0]{\@secondoftwo}%
\providecommand \translation [1]{[#1]}%
\providecommand \BibitemOpen [0]{}%
\providecommand \bibitemStop [0]{}%
\providecommand \bibitemNoStop [0]{.\EOS\space}%
\providecommand \EOS [0]{\spacefactor3000\relax}%
\providecommand \BibitemShut  [1]{\csname bibitem#1\endcsname}%
\let\auto@bib@innerbib\@empty
%</preamble>
\bibitem [{\citenamefont {Peters}\ and\ \citenamefont
  {Mathews}(1963)}]{Peters:1963ux}%
  \BibitemOpen
  \bibfield  {author} {\bibinfo {author} {\bibfnamefont {P.~C.}\ \bibnamefont
  {Peters}}\ and\ \bibinfo {author} {\bibfnamefont {J.}~\bibnamefont
  {Mathews}},\ }\href {https://doi.org/10.1103/PhysRev.131.435} {\bibfield
  {journal} {\bibinfo  {journal} {Phys. Rev.}\ }\textbf {\bibinfo {volume}
  {131}},\ \bibinfo {pages} {435} (\bibinfo {year} {1963})}\BibitemShut
  {NoStop}%
\bibitem [{\citenamefont {Peters}(1964)}]{Peters:1964zz}%
  \BibitemOpen
  \bibfield  {author} {\bibinfo {author} {\bibfnamefont {P.~C.}\ \bibnamefont
  {Peters}},\ }\href {https://doi.org/10.1103/PhysRev.136.B1224} {\bibfield
  {journal} {\bibinfo  {journal} {Phys. Rev.}\ }\textbf {\bibinfo {volume}
  {136}},\ \bibinfo {pages} {B1224} (\bibinfo {year} {1964})}\BibitemShut
  {NoStop}%
\bibitem [{\citenamefont {Blanchet}(2014)}]{Blanchet:2013haa}%
  \BibitemOpen
  \bibfield  {author} {\bibinfo {author} {\bibfnamefont {L.}~\bibnamefont
  {Blanchet}},\ }\href {https://doi.org/10.12942/lrr-2014-2} {\bibfield
  {journal} {\bibinfo  {journal} {Living Rev. Rel.}\ }\textbf {\bibinfo
  {volume} {17}},\ \bibinfo {pages} {2} (\bibinfo {year} {2014})},\ \Eprint
  {https://arxiv.org/abs/1310.1528} {arXiv:1310.1528 [gr-qc]} \BibitemShut
  {NoStop}%
\bibitem [{\citenamefont {Arun}\ \emph {et~al.}(2009)\citenamefont {Arun},
  \citenamefont {Blanchet}, \citenamefont {Iyer},\ and\ \citenamefont
  {Sinha}}]{Arun:2009mc}%
  \BibitemOpen
  \bibfield  {author} {\bibinfo {author} {\bibfnamefont {K.~G.}\ \bibnamefont
  {Arun}}, \bibinfo {author} {\bibfnamefont {L.}~\bibnamefont {Blanchet}},
  \bibinfo {author} {\bibfnamefont {B.~R.}\ \bibnamefont {Iyer}},\ and\
  \bibinfo {author} {\bibfnamefont {S.}~\bibnamefont {Sinha}},\ }\href
  {https://doi.org/10.1103/PhysRevD.80.124018} {\bibfield  {journal} {\bibinfo
  {journal} {Phys. Rev. D}\ }\textbf {\bibinfo {volume} {80}},\ \bibinfo
  {pages} {124018} (\bibinfo {year} {2009})},\ \Eprint
  {https://arxiv.org/abs/0908.3854} {arXiv:0908.3854 [gr-qc]} \BibitemShut
  {NoStop}%
\bibitem [{\citenamefont {Bender}\ and\ \citenamefont
  {Orszag}(1978)}]{BenderOrszag1978}%
  \BibitemOpen
  \bibfield  {author} {\bibinfo {author} {\bibfnamefont {C.~M.}\ \bibnamefont
  {Bender}}\ and\ \bibinfo {author} {\bibfnamefont {S.~A.}\ \bibnamefont
  {Orszag}},\ }\href@noop {} {\emph {\bibinfo {title} {Advanced Mathematical
  Methods for Scientists and Engineers}}}\ (\bibinfo  {publisher}
  {McGraw--Hill},\ \bibinfo {address} {New York},\ \bibinfo {year}
  {1978})\BibitemShut {NoStop}%
\bibitem [{\citenamefont {Kevorkian}\ and\ \citenamefont
  {Cole}(1996)}]{KevorkianCole1996}%
  \BibitemOpen
  \bibfield  {author} {\bibinfo {author} {\bibfnamefont {J.}~\bibnamefont
  {Kevorkian}}\ and\ \bibinfo {author} {\bibfnamefont {J.~D.}\ \bibnamefont
  {Cole}},\ }\href@noop {} {\emph {\bibinfo {title} {Multiple Scale and
  Singular Perturbation Methods}}},\ \bibinfo {series} {Applied Mathematical
  Sciences}, Vol.\ \bibinfo {volume} {114}\ (\bibinfo  {publisher} {Springer},\
  \bibinfo {address} {New York},\ \bibinfo {year} {1996})\BibitemShut {NoStop}%
\bibitem [{\citenamefont {Chen}\ \emph {et~al.}(1996)\citenamefont {Chen},
  \citenamefont {Goldenfeld},\ and\ \citenamefont {Oono}}]{DRG}%
  \BibitemOpen
  \bibfield  {author} {\bibinfo {author} {\bibfnamefont {L.-Y.}\ \bibnamefont
  {Chen}}, \bibinfo {author} {\bibfnamefont {N.}~\bibnamefont {Goldenfeld}},\
  and\ \bibinfo {author} {\bibfnamefont {Y.}~\bibnamefont {Oono}},\ }\href
  {https://doi.org/10.1103/PhysRevE.54.376} {\bibfield  {journal} {\bibinfo
  {journal} {Phys. Rev. E}\ }\textbf {\bibinfo {volume} {54}},\ \bibinfo
  {pages} {376} (\bibinfo {year} {1996})}\BibitemShut {NoStop}%
\bibitem [{\citenamefont {Galley}\ and\ \citenamefont
  {Rothstein}(2017)}]{Galley:2016zee}%
  \BibitemOpen
  \bibfield  {author} {\bibinfo {author} {\bibfnamefont {C.~R.}\ \bibnamefont
  {Galley}}\ and\ \bibinfo {author} {\bibfnamefont {I.~Z.}\ \bibnamefont
  {Rothstein}},\ }\href {https://doi.org/10.1103/PhysRevD.95.104054} {\bibfield
   {journal} {\bibinfo  {journal} {Phys. Rev. D}\ }\textbf {\bibinfo {volume}
  {95}},\ \bibinfo {pages} {104054} (\bibinfo {year} {2017})},\ \Eprint
  {https://arxiv.org/abs/1609.08268} {arXiv:1609.08268 [gr-qc]} \BibitemShut
  {NoStop}%
\bibitem [{\citenamefont {G{\'a}lvez~Ghersi}\ and\ \citenamefont
  {Stein}(2021)}]{GalvezGhersi:2021sxs}%
  \BibitemOpen
  \bibfield  {author} {\bibinfo {author} {\bibfnamefont {J.~T.}\ \bibnamefont
  {G{\'a}lvez~Ghersi}}\ and\ \bibinfo {author} {\bibfnamefont {L.~C.}\
  \bibnamefont {Stein}},\ }\href {https://doi.org/10.1103/PhysRevE.104.034219}
  {\bibfield  {journal} {\bibinfo  {journal} {Phys. Rev. E}\ }\textbf {\bibinfo
  {volume} {104}},\ \bibinfo {pages} {034219} (\bibinfo {year} {2021})},\
  \Eprint {https://arxiv.org/abs/2106.08410} {arXiv:2106.08410 [hep-th]}
  \BibitemShut {NoStop}%
\bibitem [{\citenamefont {Kim}\ \emph {et~al.}(2026)\citenamefont {Kim},
  \citenamefont {Patil}, \citenamefont {Scheopner},\ and\ \citenamefont
  {Steinhoff}}]{Kim:2025gis}%
  \BibitemOpen
  \bibfield  {author} {\bibinfo {author} {\bibfnamefont {J.-W.}\ \bibnamefont
  {Kim}}, \bibinfo {author} {\bibfnamefont {R.}~\bibnamefont {Patil}}, \bibinfo
  {author} {\bibfnamefont {T.}~\bibnamefont {Scheopner}},\ and\ \bibinfo
  {author} {\bibfnamefont {J.}~\bibnamefont {Steinhoff}},\ }\href
  {https://doi.org/10.1007/JHEP03(2026)241} {\bibfield  {journal} {\bibinfo
  {journal} {JHEP}\ }\textbf {\bibinfo {volume} {03}},\ \bibinfo {pages}
  {241}},\ \Eprint {https://arxiv.org/abs/2511.05649} {arXiv:2511.05649
  [hep-th]} \BibitemShut {NoStop}%
\bibitem [{\citenamefont {Blanco}(2026)}]{Blanco:2026prj}%
  \BibitemOpen
  \bibfield  {author} {\bibinfo {author} {\bibfnamefont {F.~M.}\ \bibnamefont
  {Blanco}},\ }\href@noop {} {\bibfield  {journal} {\bibinfo  {journal}
  {arXiv}\ } (\bibinfo {year} {2026})},\ \Eprint
  {https://arxiv.org/abs/2607.24335} {arXiv:2607.24335 [gr-qc]} \BibitemShut
  {NoStop}%
\bibitem [{\citenamefont {Da~Silva}(2026)}]{DaSilva:2026pdk}%
  \BibitemOpen
  \bibfield  {author} {\bibinfo {author} {\bibfnamefont {L.~J.~G.}\
  \bibnamefont {Da~Silva}},\ }\href@noop {} {\bibfield  {journal} {\bibinfo
  {journal} {arXiv}\ } (\bibinfo {year} {2026})},\ \Eprint
  {https://arxiv.org/abs/2609.03329} {arXiv:2609.03329 [gr-qc]} \BibitemShut
  {NoStop}%
\bibitem [{\citenamefont {Lincoln}\ and\ \citenamefont
  {Will}(1990)}]{Lincoln_Will_1990}%
  \BibitemOpen
  \bibfield  {author} {\bibinfo {author} {\bibfnamefont {C.~W.}\ \bibnamefont
  {Lincoln}}\ and\ \bibinfo {author} {\bibfnamefont {C.~M.}\ \bibnamefont
  {Will}},\ }\href {https://doi.org/10.1103/PhysRevD.42.1123} {\bibfield
  {journal} {\bibinfo  {journal} {Phys. Rev. D}\ }\textbf {\bibinfo {volume}
  {42}},\ \bibinfo {pages} {1123} (\bibinfo {year} {1990})}\BibitemShut
  {NoStop}%
\bibitem [{\citenamefont {Poisson}\ and\ \citenamefont
  {Will}(2014)}]{PoissonWill2014}%
  \BibitemOpen
  \bibfield  {author} {\bibinfo {author} {\bibfnamefont {E.}~\bibnamefont
  {Poisson}}\ and\ \bibinfo {author} {\bibfnamefont {C.~M.}\ \bibnamefont
  {Will}},\ }\href@noop {} {\emph {\bibinfo {title} {Gravity: Newtonian,
  Post-Newtonian, Relativistic}}}\ (\bibinfo  {publisher} {Cambridge University
  Press},\ \bibinfo {address} {Cambridge},\ \bibinfo {year} {2014})\BibitemShut
  {NoStop}%
\bibitem [{\citenamefont {Will}(2019)}]{Will:2019lfe}%
  \BibitemOpen
  \bibfield  {author} {\bibinfo {author} {\bibfnamefont {C.~M.}\ \bibnamefont
  {Will}},\ }\href {https://doi.org/10.1088/1361-6382/ab41d0} {\bibfield
  {journal} {\bibinfo  {journal} {Class. Quant. Grav.}\ }\textbf {\bibinfo
  {volume} {36}},\ \bibinfo {pages} {195013} (\bibinfo {year} {2019})},\
  \Eprint {https://arxiv.org/abs/1906.08064} {arXiv:1906.08064 [gr-qc]}
  \BibitemShut {NoStop}%
\bibitem [{\citenamefont {Galley}(2013)}]{Galley:2012hx}%
  \BibitemOpen
  \bibfield  {author} {\bibinfo {author} {\bibfnamefont {C.~R.}\ \bibnamefont
  {Galley}},\ }\href {https://doi.org/10.1103/PhysRevLett.110.174301}
  {\bibfield  {journal} {\bibinfo  {journal} {Phys. Rev. Lett.}\ }\textbf
  {\bibinfo {volume} {110}},\ \bibinfo {pages} {174301} (\bibinfo {year}
  {2013})},\ \Eprint {https://arxiv.org/abs/1210.2745} {arXiv:1210.2745
  [gr-qc]} \BibitemShut {NoStop}%
\bibitem [{\citenamefont {Tsang}\ \emph {et~al.}(2015)\citenamefont {Tsang},
  \citenamefont {Galley}, \citenamefont {Stein},\ and\ \citenamefont
  {Turner}}]{Tsang_2015}%
  \BibitemOpen
  \bibfield  {author} {\bibinfo {author} {\bibfnamefont {D.}~\bibnamefont
  {Tsang}}, \bibinfo {author} {\bibfnamefont {C.~R.}\ \bibnamefont {Galley}},
  \bibinfo {author} {\bibfnamefont {L.~C.}\ \bibnamefont {Stein}},\ and\
  \bibinfo {author} {\bibfnamefont {A.}~\bibnamefont {Turner}},\ }\href
  {https://doi.org/10.1088/2041-8205/809/1/L9} {\bibfield  {journal} {\bibinfo
  {journal} {The Astrophysical Journal Letters}\ }\textbf {\bibinfo {volume}
  {809}},\ \bibinfo {pages} {L9} (\bibinfo {year} {2015})}\BibitemShut
  {NoStop}%
\bibitem [{\citenamefont {Hinderer}\ and\ \citenamefont
  {Flanagan}(2008)}]{Hinderer:2008dm}%
  \BibitemOpen
  \bibfield  {author} {\bibinfo {author} {\bibfnamefont {T.}~\bibnamefont
  {Hinderer}}\ and\ \bibinfo {author} {\bibfnamefont {E.~E.}\ \bibnamefont
  {Flanagan}},\ }\href {https://doi.org/10.1103/PhysRevD.78.064028} {\bibfield
  {journal} {\bibinfo  {journal} {Phys. Rev. D}\ }\textbf {\bibinfo {volume}
  {78}},\ \bibinfo {pages} {064028} (\bibinfo {year} {2008})},\ \Eprint
  {https://arxiv.org/abs/0805.3337} {arXiv:0805.3337 [gr-qc]} \BibitemShut
  {NoStop}%
\bibitem [{\citenamefont {Damour}\ \emph {et~al.}(2004)\citenamefont {Damour},
  \citenamefont {Gopakumar},\ and\ \citenamefont {Iyer}}]{Damour:2004bz}%
  \BibitemOpen
  \bibfield  {author} {\bibinfo {author} {\bibfnamefont {T.}~\bibnamefont
  {Damour}}, \bibinfo {author} {\bibfnamefont {A.}~\bibnamefont {Gopakumar}},\
  and\ \bibinfo {author} {\bibfnamefont {B.~R.}\ \bibnamefont {Iyer}},\ }\href
  {https://doi.org/10.1103/PhysRevD.70.064028} {\bibfield  {journal} {\bibinfo
  {journal} {Phys. Rev. D}\ }\textbf {\bibinfo {volume} {70}},\ \bibinfo
  {pages} {064028} (\bibinfo {year} {2004})},\ \Eprint
  {https://arxiv.org/abs/gr-qc/0404128} {arXiv:gr-qc/0404128} \BibitemShut
  {NoStop}%
\bibitem [{\citenamefont {Boetzel}\ \emph {et~al.}(2019)\citenamefont
  {Boetzel}, \citenamefont {Mishra}, \citenamefont {Faye}, \citenamefont
  {Gopakumar},\ and\ \citenamefont {Iyer}}]{Boetzel:2019nfw}%
  \BibitemOpen
  \bibfield  {author} {\bibinfo {author} {\bibfnamefont {Y.}~\bibnamefont
  {Boetzel}}, \bibinfo {author} {\bibfnamefont {C.~K.}\ \bibnamefont {Mishra}},
  \bibinfo {author} {\bibfnamefont {G.}~\bibnamefont {Faye}}, \bibinfo {author}
  {\bibfnamefont {A.}~\bibnamefont {Gopakumar}},\ and\ \bibinfo {author}
  {\bibfnamefont {B.~R.}\ \bibnamefont {Iyer}},\ }\href
  {https://doi.org/10.1103/PhysRevD.100.044018} {\bibfield  {journal} {\bibinfo
   {journal} {Phys. Rev. D}\ }\textbf {\bibinfo {volume} {100}},\ \bibinfo
  {pages} {044018} (\bibinfo {year} {2019})},\ \Eprint
  {https://arxiv.org/abs/1904.11814} {arXiv:1904.11814 [gr-qc]} \BibitemShut
  {NoStop}%
\bibitem [{\citenamefont {Loutrel}\ \emph
  {et~al.}(2019{\natexlab{a}})\citenamefont {Loutrel}, \citenamefont
  {Liebersbach}, \citenamefont {Yunes},\ and\ \citenamefont
  {Cornish}}]{Loutrel:2018ssg}%
  \BibitemOpen
  \bibfield  {author} {\bibinfo {author} {\bibfnamefont {N.}~\bibnamefont
  {Loutrel}}, \bibinfo {author} {\bibfnamefont {S.}~\bibnamefont
  {Liebersbach}}, \bibinfo {author} {\bibfnamefont {N.}~\bibnamefont {Yunes}},\
  and\ \bibinfo {author} {\bibfnamefont {N.}~\bibnamefont {Cornish}},\ }\href
  {https://doi.org/10.1088/1361-6382/aaf1ec} {\bibfield  {journal} {\bibinfo
  {journal} {Class. Quant. Grav.}\ }\textbf {\bibinfo {volume} {36}},\ \bibinfo
  {pages} {01} (\bibinfo {year} {2019}{\natexlab{a}})},\ \Eprint
  {https://arxiv.org/abs/1801.09009} {arXiv:1801.09009 [gr-qc]} \BibitemShut
  {NoStop}%
\bibitem [{\citenamefont {Loutrel}\ \emph
  {et~al.}(2019{\natexlab{b}})\citenamefont {Loutrel}, \citenamefont
  {Liebersbach}, \citenamefont {Yunes},\ and\ \citenamefont
  {Cornish}}]{Loutrel:2018ydu}%
  \BibitemOpen
  \bibfield  {author} {\bibinfo {author} {\bibfnamefont {N.}~\bibnamefont
  {Loutrel}}, \bibinfo {author} {\bibfnamefont {S.}~\bibnamefont
  {Liebersbach}}, \bibinfo {author} {\bibfnamefont {N.}~\bibnamefont {Yunes}},\
  and\ \bibinfo {author} {\bibfnamefont {N.}~\bibnamefont {Cornish}},\ }\href
  {https://doi.org/10.1088/1361-6382/aaf2a9} {\bibfield  {journal} {\bibinfo
  {journal} {Class. Quant. Grav.}\ }\textbf {\bibinfo {volume} {36}},\ \bibinfo
  {pages} {025004} (\bibinfo {year} {2019}{\natexlab{b}})},\ \Eprint
  {https://arxiv.org/abs/1810.03521} {arXiv:1810.03521 [gr-qc]} \BibitemShut
  {NoStop}%
\bibitem [{\citenamefont {Fumagalli}\ \emph {et~al.}(2025)\citenamefont
  {Fumagalli}, \citenamefont {Loutrel}, \citenamefont {Gerosa},\ and\
  \citenamefont {Boschini}}]{Fumagalli:2025rhc}%
  \BibitemOpen
  \bibfield  {author} {\bibinfo {author} {\bibfnamefont {G.}~\bibnamefont
  {Fumagalli}}, \bibinfo {author} {\bibfnamefont {N.}~\bibnamefont {Loutrel}},
  \bibinfo {author} {\bibfnamefont {D.}~\bibnamefont {Gerosa}},\ and\ \bibinfo
  {author} {\bibfnamefont {M.}~\bibnamefont {Boschini}},\ }\href
  {https://doi.org/10.1103/znmj-6wvt} {\bibfield  {journal} {\bibinfo
  {journal} {Phys. Rev. D}\ }\textbf {\bibinfo {volume} {112}},\ \bibinfo
  {pages} {024012} (\bibinfo {year} {2025})},\ \Eprint
  {https://arxiv.org/abs/2502.06952} {arXiv:2502.06952 [gr-qc]} \BibitemShut
  {NoStop}%
\bibitem [{\citenamefont {Nee}\ \emph {et~al.}(2025)\citenamefont {Nee} \emph
  {et~al.}}]{Nee:2025zdy}%
  \BibitemOpen
  \bibfield  {author} {\bibinfo {author} {\bibfnamefont {P.~J.}\ \bibnamefont
  {Nee}} \emph {et~al.},\ }\href {https://doi.org/10.1088/1361-6382/ade35d}
  {\bibfield  {journal} {\bibinfo  {journal} {Class. Quant. Grav.}\ }\textbf
  {\bibinfo {volume} {42}},\ \bibinfo {pages} {135011} (\bibinfo {year}
  {2025})},\ \Eprint {https://arxiv.org/abs/2503.05422} {arXiv:2503.05422
  [gr-qc]} \BibitemShut {NoStop}%
\bibitem [{\citenamefont {Lynch}\ \emph
  {et~al.}(2026{\natexlab{a}})\citenamefont {Lynch}, \citenamefont {Buonanno},
  \citenamefont {Gamboa},\ and\ \citenamefont {van~de Meent}}]{Lynch:2026ibo}%
  \BibitemOpen
  \bibfield  {author} {\bibinfo {author} {\bibfnamefont {P.}~\bibnamefont
  {Lynch}}, \bibinfo {author} {\bibfnamefont {A.}~\bibnamefont {Buonanno}},
  \bibinfo {author} {\bibfnamefont {A.}~\bibnamefont {Gamboa}},\ and\ \bibinfo
  {author} {\bibfnamefont {M.}~\bibnamefont {van~de Meent}},\ }\href@noop {}
  {\bibfield  {journal} {\bibinfo  {journal} {arXiv}\ } (\bibinfo {year}
  {2026}{\natexlab{a}})},\ \Eprint {https://arxiv.org/abs/2606.30594}
  {arXiv:2606.30594 [gr-qc]} \BibitemShut {NoStop}%
\bibitem [{\citenamefont {Lynch}\ \emph
  {et~al.}(2026{\natexlab{b}})\citenamefont {Lynch}, \citenamefont {Buonanno},
  \citenamefont {Gamboa},\ and\ \citenamefont {van~de Meent}}]{Lynch:2026eyw}%
  \BibitemOpen
  \bibfield  {author} {\bibinfo {author} {\bibfnamefont {P.}~\bibnamefont
  {Lynch}}, \bibinfo {author} {\bibfnamefont {A.}~\bibnamefont {Buonanno}},
  \bibinfo {author} {\bibfnamefont {A.}~\bibnamefont {Gamboa}},\ and\ \bibinfo
  {author} {\bibfnamefont {M.}~\bibnamefont {van~de Meent}},\ }\href@noop {}
  {\bibfield  {journal} {\bibinfo  {journal} {arXiv}\ } (\bibinfo {year}
  {2026}{\natexlab{b}})},\ \Eprint {https://arxiv.org/abs/2610.29856}
  {arXiv:2610.29856 [gr-qc]} \BibitemShut {NoStop}%
\bibitem [{\citenamefont {Gamba}(2026)}]{Gamba:2026ado}%
  \BibitemOpen
  \bibfield  {author} {\bibinfo {author} {\bibfnamefont {R.}~\bibnamefont
  {Gamba}},\ }\href@noop {} {\bibfield  {journal} {\bibinfo  {journal} {arXiv}\
  } (\bibinfo {year} {2026})},\ \Eprint {https://arxiv.org/abs/2610.30064}
  {arXiv:2610.30064 [gr-qc]} \BibitemShut {NoStop}%
\bibitem [{\citenamefont {Aasi}\ \emph {et~al.}(2015)\citenamefont {Aasi} \emph
  {et~al.}}]{LIGOScientific:2014pky}%
  \BibitemOpen
  \bibfield  {author} {\bibinfo {author} {\bibfnamefont {J.}~\bibnamefont
  {Aasi}} \emph {et~al.} (\bibinfo {collaboration} {LIGO Scientific}),\ }\href
  {https://doi.org/10.1088/0264-9381/32/7/074001} {\bibfield  {journal}
  {\bibinfo  {journal} {Class. Quant. Grav.}\ }\textbf {\bibinfo {volume}
  {32}},\ \bibinfo {pages} {074001} (\bibinfo {year} {2015})},\ \Eprint
  {https://arxiv.org/abs/1411.4547} {arXiv:1411.4547 [gr-qc]} \BibitemShut
  {NoStop}%
\bibitem [{\citenamefont {Acernese}\ \emph {et~al.}(2015)\citenamefont
  {Acernese} \emph {et~al.}}]{VIRGO:2014yos}%
  \BibitemOpen
  \bibfield  {author} {\bibinfo {author} {\bibfnamefont {F.}~\bibnamefont
  {Acernese}} \emph {et~al.} (\bibinfo {collaboration} {VIRGO}),\ }\href
  {https://doi.org/10.1088/0264-9381/32/2/024001} {\bibfield  {journal}
  {\bibinfo  {journal} {Class. Quant. Grav.}\ }\textbf {\bibinfo {volume}
  {32}},\ \bibinfo {pages} {024001} (\bibinfo {year} {2015})},\ \Eprint
  {https://arxiv.org/abs/1408.3978} {arXiv:1408.3978 [gr-qc]} \BibitemShut
  {NoStop}%
\bibitem [{\citenamefont {Akutsu}\ \emph {et~al.}(2021)\citenamefont {Akutsu}
  \emph {et~al.}}]{KAGRA:2020tym}%
  \BibitemOpen
  \bibfield  {author} {\bibinfo {author} {\bibfnamefont {T.}~\bibnamefont
  {Akutsu}} \emph {et~al.} (\bibinfo {collaboration} {KAGRA}),\ }\href
  {https://doi.org/10.1093/ptep/ptaa125} {\bibfield  {journal} {\bibinfo
  {journal} {PTEP}\ }\textbf {\bibinfo {volume} {2021}},\ \bibinfo {pages}
  {05A101} (\bibinfo {year} {2021})},\ \Eprint
  {https://arxiv.org/abs/2005.05574} {arXiv:2005.05574 [physics.ins-det]}
  \BibitemShut {NoStop}%
\bibitem [{\citenamefont {Punturo}\ \emph {et~al.}(2010)\citenamefont {Punturo}
  \emph {et~al.}}]{Punturo:2010zz}%
  \BibitemOpen
  \bibfield  {author} {\bibinfo {author} {\bibfnamefont {M.}~\bibnamefont
  {Punturo}} \emph {et~al.},\ }\href
  {https://doi.org/10.1088/0264-9381/27/19/194002} {\bibfield  {journal}
  {\bibinfo  {journal} {Class. Quant. Grav.}\ }\textbf {\bibinfo {volume}
  {27}},\ \bibinfo {pages} {194002} (\bibinfo {year} {2010})}\BibitemShut
  {NoStop}%
\bibitem [{\citenamefont {Maggiore}\ \emph {et~al.}(2020)\citenamefont
  {Maggiore} \emph {et~al.}}]{Maggiore:2019uih}%
  \BibitemOpen
  \bibfield  {author} {\bibinfo {author} {\bibfnamefont {M.}~\bibnamefont
  {Maggiore}} \emph {et~al.},\ }\href
  {https://doi.org/10.1088/1475-7516/2020/03/050} {\bibfield  {journal}
  {\bibinfo  {journal} {JCAP}\ }\textbf {\bibinfo {volume} {03}},\ \bibinfo
  {pages} {050}},\ \Eprint {https://arxiv.org/abs/1912.02622} {arXiv:1912.02622
  [astro-ph.CO]} \BibitemShut {NoStop}%
\bibitem [{\citenamefont {Evans}\ \emph {et~al.}(2021)\citenamefont {Evans}
  \emph {et~al.}}]{Evans:2021gyd}%
  \BibitemOpen
  \bibfield  {author} {\bibinfo {author} {\bibfnamefont {M.}~\bibnamefont
  {Evans}} \emph {et~al.},\ }\href@noop {} {\bibfield  {journal} {\bibinfo
  {journal} {arXiv}\ } (\bibinfo {year} {2021})},\ \Eprint
  {https://arxiv.org/abs/2109.09882} {arXiv:2109.09882 [astro-ph.IM]}
  \BibitemShut {NoStop}%
\bibitem [{\citenamefont {Amaro-Seoane}\ \emph {et~al.}(2017)\citenamefont
  {Amaro-Seoane} \emph {et~al.}}]{LISA:2017pwj}%
  \BibitemOpen
  \bibfield  {author} {\bibinfo {author} {\bibfnamefont {P.}~\bibnamefont
  {Amaro-Seoane}} \emph {et~al.} (\bibinfo {collaboration} {LISA}),\
  }\href@noop {} {\bibfield  {journal} {\bibinfo  {journal} {arXiv}\ }
  (\bibinfo {year} {2017})},\ \Eprint {https://arxiv.org/abs/1702.00786}
  {arXiv:1702.00786 [astro-ph.IM]} \BibitemShut {NoStop}%
\bibitem [{\citenamefont {Afshordi}\ \emph {et~al.}(2025)\citenamefont
  {Afshordi} \emph {et~al.}}]{LISAConsortiumWaveformWorkingGroup:2023arg}%
  \BibitemOpen
  \bibfield  {author} {\bibinfo {author} {\bibfnamefont {N.}~\bibnamefont
  {Afshordi}} \emph {et~al.} (\bibinfo {collaboration} {LISA Consortium
  Waveform Working Group}),\ }\href
  {https://doi.org/10.1007/s41114-025-00056-1} {\bibfield  {journal} {\bibinfo
  {journal} {Living Rev. Rel.}\ }\textbf {\bibinfo {volume} {28}},\ \bibinfo
  {pages} {9} (\bibinfo {year} {2025})},\ \Eprint
  {https://arxiv.org/abs/2311.01300} {arXiv:2311.01300 [gr-qc]} \BibitemShut
  {NoStop}%
\bibitem [{\citenamefont {Colpi}\ \emph {et~al.}(2024)\citenamefont {Colpi}
  \emph {et~al.}}]{LISA:2024hlh}%
  \BibitemOpen
  \bibfield  {author} {\bibinfo {author} {\bibfnamefont {M.}~\bibnamefont
  {Colpi}} \emph {et~al.} (\bibinfo {collaboration} {LISA}),\ }\href@noop {}
  {\bibfield  {journal} {\bibinfo  {journal} {arXiv}\ } (\bibinfo {year}
  {2024})},\ \Eprint {https://arxiv.org/abs/2402.07571} {arXiv:2402.07571
  [astro-ph.CO]} \BibitemShut {NoStop}%
\bibitem [{\citenamefont {Mora}\ and\ \citenamefont
  {Will}(2004)}]{Mora:2003wt}%
  \BibitemOpen
  \bibfield  {author} {\bibinfo {author} {\bibfnamefont {T.}~\bibnamefont
  {Mora}}\ and\ \bibinfo {author} {\bibfnamefont {C.~M.}\ \bibnamefont
  {Will}},\ }\href {https://doi.org/10.1103/PhysRevD.71.129901} {\bibfield
  {journal} {\bibinfo  {journal} {Phys. Rev. D}\ }\textbf {\bibinfo {volume}
  {69}},\ \bibinfo {pages} {104021} (\bibinfo {year} {2004})},\ \bibinfo {note}
  {[Erratum: Phys.Rev.D 71, 129901 (2005)]},\ \Eprint
  {https://arxiv.org/abs/gr-qc/0312082} {arXiv:gr-qc/0312082} \BibitemShut
  {NoStop}%
\bibitem [{\citenamefont {Van De~Meent}\ and\ \citenamefont
  {Warburton}(2018)}]{VanDeMeent:2018cgn}%
  \BibitemOpen
  \bibfield  {author} {\bibinfo {author} {\bibfnamefont {M.}~\bibnamefont {Van
  De~Meent}}\ and\ \bibinfo {author} {\bibfnamefont {N.}~\bibnamefont
  {Warburton}},\ }\href {https://doi.org/10.1088/1361-6382/aac8ce} {\bibfield
  {journal} {\bibinfo  {journal} {Class. Quant. Grav.}\ }\textbf {\bibinfo
  {volume} {35}},\ \bibinfo {pages} {144003} (\bibinfo {year} {2018})},\
  \Eprint {https://arxiv.org/abs/1802.05281} {arXiv:1802.05281 [gr-qc]}
  \BibitemShut {NoStop}%
\bibitem [{\citenamefont {Lynch}\ \emph {et~al.}(2022)\citenamefont {Lynch},
  \citenamefont {van~de Meent},\ and\ \citenamefont
  {Warburton}}]{Lynch:2021ogr}%
  \BibitemOpen
  \bibfield  {author} {\bibinfo {author} {\bibfnamefont {P.}~\bibnamefont
  {Lynch}}, \bibinfo {author} {\bibfnamefont {M.}~\bibnamefont {van~de
  Meent}},\ and\ \bibinfo {author} {\bibfnamefont {N.}~\bibnamefont
  {Warburton}},\ }\href {https://doi.org/10.1088/1361-6382/ac7507} {\bibfield
  {journal} {\bibinfo  {journal} {Class. Quant. Grav.}\ }\textbf {\bibinfo
  {volume} {39}},\ \bibinfo {pages} {145004} (\bibinfo {year} {2022})},\
  \Eprint {https://arxiv.org/abs/2112.05651} {arXiv:2112.05651 [gr-qc]}
  \BibitemShut {NoStop}%
\bibitem [{\citenamefont {Pound}\ and\ \citenamefont
  {Poisson}(2008{\natexlab{a}})}]{Pound:2007ti}%
  \BibitemOpen
  \bibfield  {author} {\bibinfo {author} {\bibfnamefont {A.}~\bibnamefont
  {Pound}}\ and\ \bibinfo {author} {\bibfnamefont {E.}~\bibnamefont
  {Poisson}},\ }\href {https://doi.org/10.1103/PhysRevD.77.044012} {\bibfield
  {journal} {\bibinfo  {journal} {Phys. Rev. D}\ }\textbf {\bibinfo {volume}
  {77}},\ \bibinfo {pages} {044012} (\bibinfo {year} {2008}{\natexlab{a}})},\
  \Eprint {https://arxiv.org/abs/0708.3037} {arXiv:0708.3037 [gr-qc]}
  \BibitemShut {NoStop}%
\bibitem [{\citenamefont {Pound}\ and\ \citenamefont
  {Poisson}(2008{\natexlab{b}})}]{Pound:2007th}%
  \BibitemOpen
  \bibfield  {author} {\bibinfo {author} {\bibfnamefont {A.}~\bibnamefont
  {Pound}}\ and\ \bibinfo {author} {\bibfnamefont {E.}~\bibnamefont
  {Poisson}},\ }\href {https://doi.org/10.1103/PhysRevD.77.044013} {\bibfield
  {journal} {\bibinfo  {journal} {Phys. Rev. D}\ }\textbf {\bibinfo {volume}
  {77}},\ \bibinfo {pages} {044013} (\bibinfo {year} {2008}{\natexlab{b}})},\
  \Eprint {https://arxiv.org/abs/0708.3033} {arXiv:0708.3033 [gr-qc]}
  \BibitemShut {NoStop}%
\bibitem [{\citenamefont {Iyer}\ and\ \citenamefont
  {Will}(1995)}]{Iyer:1995rn}%
  \BibitemOpen
  \bibfield  {author} {\bibinfo {author} {\bibfnamefont {B.~R.}\ \bibnamefont
  {Iyer}}\ and\ \bibinfo {author} {\bibfnamefont {C.~M.}\ \bibnamefont
  {Will}},\ }\href {https://doi.org/10.1103/PhysRevD.52.6882} {\bibfield
  {journal} {\bibinfo  {journal} {Phys. Rev. D}\ }\textbf {\bibinfo {volume}
  {52}},\ \bibinfo {pages} {6882} (\bibinfo {year} {1995})}\BibitemShut
  {NoStop}%
\bibitem [{\citenamefont {Will}\ and\ \citenamefont
  {Maitra}(2017)}]{Will:2016}%
  \BibitemOpen
  \bibfield  {author} {\bibinfo {author} {\bibfnamefont {C.~M.}\ \bibnamefont
  {Will}}\ and\ \bibinfo {author} {\bibfnamefont {M.}~\bibnamefont {Maitra}},\
  }\href {https://doi.org/10.1103/PhysRevD.95.064003} {\bibfield  {journal}
  {\bibinfo  {journal} {Phys. Rev. D}\ }\textbf {\bibinfo {volume} {95}},\
  \bibinfo {pages} {064003} (\bibinfo {year} {2017})},\ \Eprint
  {https://arxiv.org/abs/1611.06931} {arXiv:1611.06931 [gr-qc]} \BibitemShut
  {NoStop}%
\bibitem [{\citenamefont {Tucker}\ and\ \citenamefont
  {Will}(2021)}]{Tucker:2021mvo}%
  \BibitemOpen
  \bibfield  {author} {\bibinfo {author} {\bibfnamefont {A.}~\bibnamefont
  {Tucker}}\ and\ \bibinfo {author} {\bibfnamefont {C.~M.}\ \bibnamefont
  {Will}},\ }\href {https://doi.org/10.1103/PhysRevD.104.104023} {\bibfield
  {journal} {\bibinfo  {journal} {Phys. Rev. D}\ }\textbf {\bibinfo {volume}
  {104}},\ \bibinfo {pages} {104023} (\bibinfo {year} {2021})},\ \Eprint
  {https://arxiv.org/abs/2108.12210} {arXiv:2108.12210 [gr-qc]} \BibitemShut
  {NoStop}%
\bibitem [{\citenamefont {Damour}\ and\ \citenamefont
  {Deruelle}(1985)}]{Damour_Deruelle}%
  \BibitemOpen
  \bibfield  {author} {\bibinfo {author} {\bibfnamefont {T.}~\bibnamefont
  {Damour}}\ and\ \bibinfo {author} {\bibfnamefont {N.}~\bibnamefont
  {Deruelle}},\ }\href {https://www.numdam.org/item/AIHPA_1985__43_1_107_0/}
  {\bibfield  {journal} {\bibinfo  {journal} {Annales de l'I.H.P. Physique
  th\'eorique}\ }\textbf {\bibinfo {volume} {43}},\ \bibinfo {pages} {107}
  (\bibinfo {year} {1985})}\BibitemShut {NoStop}%
\bibitem [{\citenamefont {Moore}\ \emph {et~al.}(2016)\citenamefont {Moore},
  \citenamefont {Favata}, \citenamefont {Arun},\ and\ \citenamefont
  {Mishra}}]{Moore:2016qxz}%
  \BibitemOpen
  \bibfield  {author} {\bibinfo {author} {\bibfnamefont {B.}~\bibnamefont
  {Moore}}, \bibinfo {author} {\bibfnamefont {M.}~\bibnamefont {Favata}},
  \bibinfo {author} {\bibfnamefont {K.~G.}\ \bibnamefont {Arun}},\ and\
  \bibinfo {author} {\bibfnamefont {C.~K.}\ \bibnamefont {Mishra}},\ }\href
  {https://doi.org/10.1103/PhysRevD.93.124061} {\bibfield  {journal} {\bibinfo
  {journal} {Phys. Rev. D}\ }\textbf {\bibinfo {volume} {93}},\ \bibinfo
  {pages} {124061} (\bibinfo {year} {2016})},\ \Eprint
  {https://arxiv.org/abs/1605.00304} {arXiv:1605.00304 [gr-qc]} \BibitemShut
  {NoStop}%
\bibitem [{\citenamefont {Moore}\ and\ \citenamefont
  {Yunes}(2019)}]{Moore:2019xkm}%
  \BibitemOpen
  \bibfield  {author} {\bibinfo {author} {\bibfnamefont {B.}~\bibnamefont
  {Moore}}\ and\ \bibinfo {author} {\bibfnamefont {N.}~\bibnamefont {Yunes}},\
  }\href {https://doi.org/10.1088/1361-6382/ab3778} {\bibfield  {journal}
  {\bibinfo  {journal} {Class. Quant. Grav.}\ }\textbf {\bibinfo {volume}
  {36}},\ \bibinfo {pages} {185003} (\bibinfo {year} {2019})},\ \Eprint
  {https://arxiv.org/abs/1903.05203} {arXiv:1903.05203 [gr-qc]} \BibitemShut
  {NoStop}%
\bibitem [{\citenamefont {Jaranowski}\ and\ \citenamefont
  {Sch\"afer}(1997)}]{Jaranowski:1997}%
  \BibitemOpen
  \bibfield  {author} {\bibinfo {author} {\bibfnamefont {P.}~\bibnamefont
  {Jaranowski}}\ and\ \bibinfo {author} {\bibfnamefont {G.}~\bibnamefont
  {Sch\"afer}},\ }\href {https://doi.org/10.1103/PhysRevD.55.4712} {\bibfield
  {journal} {\bibinfo  {journal} {Phys. Rev. D}\ }\textbf {\bibinfo {volume}
  {55}},\ \bibinfo {pages} {4712} (\bibinfo {year} {1997})}\BibitemShut
  {NoStop}%
\bibitem [{\citenamefont {Konigsdorffer}\ \emph {et~al.}(2003)\citenamefont
  {Konigsdorffer}, \citenamefont {Faye},\ and\ \citenamefont
  {Schaefer}}]{Konigsdorffer:2003ue}%
  \BibitemOpen
  \bibfield  {author} {\bibinfo {author} {\bibfnamefont {C.}~\bibnamefont
  {Konigsdorffer}}, \bibinfo {author} {\bibfnamefont {G.}~\bibnamefont
  {Faye}},\ and\ \bibinfo {author} {\bibfnamefont {G.}~\bibnamefont
  {Schaefer}},\ }\href {https://doi.org/10.1103/PhysRevD.68.044004} {\bibfield
  {journal} {\bibinfo  {journal} {Phys. Rev. D}\ }\textbf {\bibinfo {volume}
  {68}},\ \bibinfo {pages} {044004} (\bibinfo {year} {2003})},\ \Eprint
  {https://arxiv.org/abs/gr-qc/0305048} {arXiv:gr-qc/0305048} \BibitemShut
  {NoStop}%
\bibitem [{\citenamefont {Gamboa}\ \emph
  {et~al.}(2025{\natexlab{a}})\citenamefont {Gamboa} \emph
  {et~al.}}]{Gamboa:2024hli}%
  \BibitemOpen
  \bibfield  {author} {\bibinfo {author} {\bibfnamefont {A.}~\bibnamefont
  {Gamboa}} \emph {et~al.},\ }\href {https://doi.org/10.1103/jxrc-z298}
  {\bibfield  {journal} {\bibinfo  {journal} {Phys. Rev. D}\ }\textbf {\bibinfo
  {volume} {112}},\ \bibinfo {pages} {044038} (\bibinfo {year}
  {2025}{\natexlab{a}})},\ \Eprint {https://arxiv.org/abs/2412.12823}
  {arXiv:2412.12823 [gr-qc]} \BibitemShut {NoStop}%
\bibitem [{\citenamefont {Gamboa}\ \emph
  {et~al.}(2025{\natexlab{b}})\citenamefont {Gamboa}, \citenamefont {Khalil},\
  and\ \citenamefont {Buonanno}}]{Gamboa:2024imd}%
  \BibitemOpen
  \bibfield  {author} {\bibinfo {author} {\bibfnamefont {A.}~\bibnamefont
  {Gamboa}}, \bibinfo {author} {\bibfnamefont {M.}~\bibnamefont {Khalil}},\
  and\ \bibinfo {author} {\bibfnamefont {A.}~\bibnamefont {Buonanno}},\ }\href
  {https://doi.org/10.1103/rb1c-nx5f} {\bibfield  {journal} {\bibinfo
  {journal} {Phys. Rev. D}\ }\textbf {\bibinfo {volume} {112}},\ \bibinfo
  {pages} {044037} (\bibinfo {year} {2025}{\natexlab{b}})},\ \Eprint
  {https://arxiv.org/abs/2412.12831} {arXiv:2412.12831 [gr-qc]} \BibitemShut
  {NoStop}%
\bibitem [{\citenamefont {Coleman~Miller}\ and\ \citenamefont
  {Yunes}(2021)}]{ColemanMiller:2021lky}%
  \BibitemOpen
  \bibfield  {author} {\bibinfo {author} {\bibfnamefont {M.}~\bibnamefont
  {Coleman~Miller}}\ and\ \bibinfo {author} {\bibfnamefont {N.}~\bibnamefont
  {Yunes}},\ }\href {https://doi.org/10.1088/2514-3433/ac2140} {\emph {\bibinfo
  {title} {{Gravitational Waves in Physics and Astrophysics}}}}\ (\bibinfo
  {publisher} {IOP},\ \bibinfo {year} {2021})\BibitemShut {NoStop}%
\bibitem [{\citenamefont {Loutrel}\ and\ \citenamefont
  {Yunes}(2017)}]{Loutrel:2017fgu}%
  \BibitemOpen
  \bibfield  {author} {\bibinfo {author} {\bibfnamefont {N.}~\bibnamefont
  {Loutrel}}\ and\ \bibinfo {author} {\bibfnamefont {N.}~\bibnamefont
  {Yunes}},\ }\href {https://doi.org/10.1088/1361-6382/aa7449} {\bibfield
  {journal} {\bibinfo  {journal} {Class. Quant. Grav.}\ }\textbf {\bibinfo
  {volume} {34}},\ \bibinfo {pages} {135011} (\bibinfo {year} {2017})},\
  \Eprint {https://arxiv.org/abs/1702.01818} {arXiv:1702.01818 [gr-qc]}
  \BibitemShut {NoStop}%
\bibitem [{\citenamefont {Yunes}\ \emph {et~al.}(2009)\citenamefont {Yunes},
  \citenamefont {Arun}, \citenamefont {Berti},\ and\ \citenamefont
  {Will}}]{Yunes:2009yz}%
  \BibitemOpen
  \bibfield  {author} {\bibinfo {author} {\bibfnamefont {N.}~\bibnamefont
  {Yunes}}, \bibinfo {author} {\bibfnamefont {K.~G.}\ \bibnamefont {Arun}},
  \bibinfo {author} {\bibfnamefont {E.}~\bibnamefont {Berti}},\ and\ \bibinfo
  {author} {\bibfnamefont {C.~M.}\ \bibnamefont {Will}},\ }\href
  {https://doi.org/10.1103/PhysRevD.80.084001} {\bibfield  {journal} {\bibinfo
  {journal} {Phys. Rev. D}\ }\textbf {\bibinfo {volume} {80}},\ \bibinfo
  {pages} {084001} (\bibinfo {year} {2009})},\ \bibinfo {note} {[Erratum:
  Phys.Rev.D 89, 109901 (2014)]},\ \Eprint {https://arxiv.org/abs/0906.0313}
  {arXiv:0906.0313 [gr-qc]} \BibitemShut {NoStop}%
\bibitem [{\citenamefont {Cutler}\ \emph {et~al.}(1994)\citenamefont {Cutler},
  \citenamefont {Kennefick},\ and\ \citenamefont
  {Poisson}}]{CutlerKennefickPoisson:1994}%
  \BibitemOpen
  \bibfield  {author} {\bibinfo {author} {\bibfnamefont {C.}~\bibnamefont
  {Cutler}}, \bibinfo {author} {\bibfnamefont {D.}~\bibnamefont {Kennefick}},\
  and\ \bibinfo {author} {\bibfnamefont {E.}~\bibnamefont {Poisson}},\ }\href
  {https://doi.org/10.1103/PhysRevD.50.3816} {\bibfield  {journal} {\bibinfo
  {journal} {Phys. Rev. D}\ }\textbf {\bibinfo {volume} {50}},\ \bibinfo
  {pages} {3816} (\bibinfo {year} {1994})}\BibitemShut {NoStop}%
\bibitem [{\citenamefont {Yunes}\ and\ \citenamefont
  {Berti}(2008)}]{Yunes:2008tw}%
  \BibitemOpen
  \bibfield  {author} {\bibinfo {author} {\bibfnamefont {N.}~\bibnamefont
  {Yunes}}\ and\ \bibinfo {author} {\bibfnamefont {E.}~\bibnamefont {Berti}},\
  }\href {https://doi.org/10.1103/PhysRevD.77.124006} {\bibfield  {journal}
  {\bibinfo  {journal} {Phys. Rev. D}\ }\textbf {\bibinfo {volume} {77}},\
  \bibinfo {pages} {124006} (\bibinfo {year} {2008})},\ \bibinfo {note}
  {[Erratum: Phys.Rev.D 83, 109901 (2011)]},\ \Eprint
  {https://arxiv.org/abs/0803.1853} {arXiv:0803.1853 [gr-qc]} \BibitemShut
  {NoStop}%
\bibitem [{\citenamefont {Hairer}\ \emph {et~al.}(1993)\citenamefont {Hairer},
  \citenamefont {N{\o}rsett},\ and\ \citenamefont
  {Wanner}}]{HairerNorsettWanner1993}%
  \BibitemOpen
  \bibfield  {author} {\bibinfo {author} {\bibfnamefont {E.}~\bibnamefont
  {Hairer}}, \bibinfo {author} {\bibfnamefont {S.~P.}\ \bibnamefont
  {N{\o}rsett}},\ and\ \bibinfo {author} {\bibfnamefont {G.}~\bibnamefont
  {Wanner}},\ }\href@noop {} {\emph {\bibinfo {title} {Solving Ordinary
  Differential Equations I: Nonstiff Problems}}},\ \bibinfo {edition} {2nd}\
  ed.,\ \bibinfo {series} {Springer Series in Computational Mathematics},
  Vol.~\bibinfo {volume} {8}\ (\bibinfo  {publisher} {Springer},\ \bibinfo
  {address} {Berlin},\ \bibinfo {year} {1993})\BibitemShut {NoStop}%
\bibitem [{\citenamefont {Virtanen}\ \emph {et~al.}(2020)\citenamefont
  {Virtanen}, \citenamefont {Gommers}, \citenamefont {Oliphant}, \citenamefont
  {Haberland}, \citenamefont {Reddy}, \citenamefont {Cournapeau}, \citenamefont
  {Burovski}, \citenamefont {Peterson}, \citenamefont {Weckesser},
  \citenamefont {Bright}, \citenamefont {{van der Walt}}, \citenamefont
  {Brett}, \citenamefont {Wilson}, \citenamefont {Millman}, \citenamefont
  {Mayorov}, \citenamefont {Nelson}, \citenamefont {Jones}, \citenamefont
  {Kern}, \citenamefont {Larson}, \citenamefont {Carey}, \citenamefont {Polat},
  \citenamefont {Feng}, \citenamefont {Moore}, \citenamefont {VanderPlas},
  \citenamefont {Laxalde}, \citenamefont {Perktold}, \citenamefont {Cimrman},
  \citenamefont {Henriksen}, \citenamefont {Quintero}, \citenamefont {Harris},
  \citenamefont {Archibald}, \citenamefont {Ribeiro}, \citenamefont
  {Pedregosa}, \citenamefont {{van Mulbregt}},\ and\ \citenamefont {{SciPy 1.0
  Contributors}}}]{Virtanen2020}%
  \BibitemOpen
  \bibfield  {author} {\bibinfo {author} {\bibfnamefont {P.}~\bibnamefont
  {Virtanen}}, \bibinfo {author} {\bibfnamefont {R.}~\bibnamefont {Gommers}},
  \bibinfo {author} {\bibfnamefont {T.~E.}\ \bibnamefont {Oliphant}}, \bibinfo
  {author} {\bibfnamefont {M.}~\bibnamefont {Haberland}}, \bibinfo {author}
  {\bibfnamefont {T.}~\bibnamefont {Reddy}}, \bibinfo {author} {\bibfnamefont
  {D.}~\bibnamefont {Cournapeau}}, \bibinfo {author} {\bibfnamefont
  {E.}~\bibnamefont {Burovski}}, \bibinfo {author} {\bibfnamefont
  {P.}~\bibnamefont {Peterson}}, \bibinfo {author} {\bibfnamefont
  {W.}~\bibnamefont {Weckesser}}, \bibinfo {author} {\bibfnamefont
  {J.}~\bibnamefont {Bright}}, \bibinfo {author} {\bibfnamefont {S.~J.}\
  \bibnamefont {{van der Walt}}}, \bibinfo {author} {\bibfnamefont
  {M.}~\bibnamefont {Brett}}, \bibinfo {author} {\bibfnamefont
  {J.}~\bibnamefont {Wilson}}, \bibinfo {author} {\bibfnamefont {K.~J.}\
  \bibnamefont {Millman}}, \bibinfo {author} {\bibfnamefont {N.}~\bibnamefont
  {Mayorov}}, \bibinfo {author} {\bibfnamefont {A.~R.~J.}\ \bibnamefont
  {Nelson}}, \bibinfo {author} {\bibfnamefont {E.}~\bibnamefont {Jones}},
  \bibinfo {author} {\bibfnamefont {R.}~\bibnamefont {Kern}}, \bibinfo {author}
  {\bibfnamefont {E.}~\bibnamefont {Larson}}, \bibinfo {author} {\bibfnamefont
  {C.~J.}\ \bibnamefont {Carey}}, \bibinfo {author} {\bibfnamefont
  {{\.I}.}~\bibnamefont {Polat}}, \bibinfo {author} {\bibfnamefont
  {Y.}~\bibnamefont {Feng}}, \bibinfo {author} {\bibfnamefont {E.~W.}\
  \bibnamefont {Moore}}, \bibinfo {author} {\bibfnamefont {J.}~\bibnamefont
  {VanderPlas}}, \bibinfo {author} {\bibfnamefont {D.}~\bibnamefont {Laxalde}},
  \bibinfo {author} {\bibfnamefont {J.}~\bibnamefont {Perktold}}, \bibinfo
  {author} {\bibfnamefont {R.}~\bibnamefont {Cimrman}}, \bibinfo {author}
  {\bibfnamefont {I.}~\bibnamefont {Henriksen}}, \bibinfo {author}
  {\bibfnamefont {E.~A.}\ \bibnamefont {Quintero}}, \bibinfo {author}
  {\bibfnamefont {C.~R.}\ \bibnamefont {Harris}}, \bibinfo {author}
  {\bibfnamefont {A.~M.}\ \bibnamefont {Archibald}}, \bibinfo {author}
  {\bibfnamefont {A.~H.}\ \bibnamefont {Ribeiro}}, \bibinfo {author}
  {\bibfnamefont {F.}~\bibnamefont {Pedregosa}}, \bibinfo {author}
  {\bibfnamefont {P.}~\bibnamefont {{van Mulbregt}}},\ and\ \bibinfo {author}
  {\bibnamefont {{SciPy 1.0 Contributors}}},\ }\href
  {https://doi.org/10.1038/s41592-019-0686-2} {\bibfield  {journal} {\bibinfo
  {journal} {Nature Methods}\ }\textbf {\bibinfo {volume} {17}},\ \bibinfo
  {pages} {261} (\bibinfo {year} {2020})}\BibitemShut {NoStop}%
\bibitem [{\citenamefont {Robson}\ \emph {et~al.}(2019)\citenamefont {Robson},
  \citenamefont {Cornish},\ and\ \citenamefont {Liu}}]{Robson:2018ifk}%
  \BibitemOpen
  \bibfield  {author} {\bibinfo {author} {\bibfnamefont {T.}~\bibnamefont
  {Robson}}, \bibinfo {author} {\bibfnamefont {N.~J.}\ \bibnamefont
  {Cornish}},\ and\ \bibinfo {author} {\bibfnamefont {C.}~\bibnamefont {Liu}},\
  }\href {https://doi.org/10.1088/1361-6382/ab1101} {\bibfield  {journal}
  {\bibinfo  {journal} {Class. Quant. Grav.}\ }\textbf {\bibinfo {volume}
  {36}},\ \bibinfo {pages} {105011} (\bibinfo {year} {2019})},\ \Eprint
  {https://arxiv.org/abs/1803.01944} {arXiv:1803.01944 [astro-ph.HE]}
  \BibitemShut {NoStop}%
\bibitem [{\citenamefont {Estabrook}\ and\ \citenamefont
  {Wahlquist}(1975)}]{EstabrookWahlquist1975}%
  \BibitemOpen
  \bibfield  {author} {\bibinfo {author} {\bibfnamefont {F.~B.}\ \bibnamefont
  {Estabrook}}\ and\ \bibinfo {author} {\bibfnamefont {H.~D.}\ \bibnamefont
  {Wahlquist}},\ }\href {https://doi.org/10.1007/BF00762449} {\bibfield
  {journal} {\bibinfo  {journal} {General Relativity and Gravitation}\ }\textbf
  {\bibinfo {volume} {6}},\ \bibinfo {pages} {439} (\bibinfo {year}
  {1975})}\BibitemShut {NoStop}%
\bibitem [{\citenamefont {Cutler}(1998)}]{Cutler:1997ta}%
  \BibitemOpen
  \bibfield  {author} {\bibinfo {author} {\bibfnamefont {C.}~\bibnamefont
  {Cutler}},\ }\href {https://doi.org/10.1103/PhysRevD.57.7089} {\bibfield
  {journal} {\bibinfo  {journal} {Phys. Rev. D}\ }\textbf {\bibinfo {volume}
  {57}},\ \bibinfo {pages} {7089} (\bibinfo {year} {1998})},\ \Eprint
  {https://arxiv.org/abs/gr-qc/9703068} {arXiv:gr-qc/9703068} \BibitemShut
  {NoStop}%
\bibitem [{\citenamefont {Cornish}\ and\ \citenamefont
  {Rubbo}(2003)}]{Cornish:2002rt}%
  \BibitemOpen
  \bibfield  {author} {\bibinfo {author} {\bibfnamefont {N.~J.}\ \bibnamefont
  {Cornish}}\ and\ \bibinfo {author} {\bibfnamefont {L.~J.}\ \bibnamefont
  {Rubbo}},\ }\href {https://doi.org/10.1103/PhysRevD.67.029905} {\bibfield
  {journal} {\bibinfo  {journal} {Phys. Rev. D}\ }\textbf {\bibinfo {volume}
  {67}},\ \bibinfo {pages} {022001} (\bibinfo {year} {2003})},\ \bibinfo {note}
  {[Erratum: Phys.Rev.D 67, 029905 (2003)]},\ \Eprint
  {https://arxiv.org/abs/gr-qc/0209011} {arXiv:gr-qc/0209011} \BibitemShut
  {NoStop}%
\bibitem [{\citenamefont {Marsat}\ \emph {et~al.}(2021)\citenamefont {Marsat},
  \citenamefont {Baker},\ and\ \citenamefont {Dal~Canton}}]{Marsat:2020rtl}%
  \BibitemOpen
  \bibfield  {author} {\bibinfo {author} {\bibfnamefont {S.}~\bibnamefont
  {Marsat}}, \bibinfo {author} {\bibfnamefont {J.~G.}\ \bibnamefont {Baker}},\
  and\ \bibinfo {author} {\bibfnamefont {T.}~\bibnamefont {Dal~Canton}},\
  }\href {https://doi.org/10.1103/PhysRevD.103.083011} {\bibfield  {journal}
  {\bibinfo  {journal} {Phys. Rev. D}\ }\textbf {\bibinfo {volume} {103}},\
  \bibinfo {pages} {083011} (\bibinfo {year} {2021})},\ \Eprint
  {https://arxiv.org/abs/2003.00357} {arXiv:2003.00357 [gr-qc]} \BibitemShut
  {NoStop}%
\bibitem [{\citenamefont {Cutler}\ and\ \citenamefont
  {Flanagan}(1994)}]{Cutler:1994ys}%
  \BibitemOpen
  \bibfield  {author} {\bibinfo {author} {\bibfnamefont {C.}~\bibnamefont
  {Cutler}}\ and\ \bibinfo {author} {\bibfnamefont {E.~E.}\ \bibnamefont
  {Flanagan}},\ }\href {https://doi.org/10.1103/PhysRevD.49.2658} {\bibfield
  {journal} {\bibinfo  {journal} {Phys. Rev. D}\ }\textbf {\bibinfo {volume}
  {49}},\ \bibinfo {pages} {2658} (\bibinfo {year} {1994})},\ \Eprint
  {https://arxiv.org/abs/gr-qc/9402014} {arXiv:gr-qc/9402014} \BibitemShut
  {NoStop}%
\bibitem [{\citenamefont {Faggioli}\ \emph {et~al.}(2025)\citenamefont
  {Faggioli}, \citenamefont {van~de Meent}, \citenamefont {Buonanno},
  \citenamefont {Gamboa}, \citenamefont {Khalil},\ and\ \citenamefont
  {Khanna}}]{Faggioli:2024ugn}%
  \BibitemOpen
  \bibfield  {author} {\bibinfo {author} {\bibfnamefont {G.}~\bibnamefont
  {Faggioli}}, \bibinfo {author} {\bibfnamefont {M.}~\bibnamefont {van~de
  Meent}}, \bibinfo {author} {\bibfnamefont {A.}~\bibnamefont {Buonanno}},
  \bibinfo {author} {\bibfnamefont {A.}~\bibnamefont {Gamboa}}, \bibinfo
  {author} {\bibfnamefont {M.}~\bibnamefont {Khalil}},\ and\ \bibinfo {author}
  {\bibfnamefont {G.}~\bibnamefont {Khanna}},\ }\href
  {https://doi.org/10.1103/PhysRevD.111.044036} {\bibfield  {journal} {\bibinfo
   {journal} {Phys. Rev. D}\ }\textbf {\bibinfo {volume} {111}},\ \bibinfo
  {pages} {044036} (\bibinfo {year} {2025})},\ \Eprint
  {https://arxiv.org/abs/2405.19006} {arXiv:2405.19006 [gr-qc]} \BibitemShut
  {NoStop}%
\bibitem [{\citenamefont {Klein}(2021)}]{Klein:2021jtd}%
  \BibitemOpen
  \bibfield  {author} {\bibinfo {author} {\bibfnamefont {A.}~\bibnamefont
  {Klein}},\ }\href@noop {} {\bibfield  {journal} {\bibinfo  {journal} {arXiv}\
  } (\bibinfo {year} {2021})},\ \Eprint {https://arxiv.org/abs/2106.10291}
  {arXiv:2106.10291 [gr-qc]} \BibitemShut {NoStop}%
\bibitem [{\citenamefont {Arredondo}\ \emph {et~al.}(2024)\citenamefont
  {Arredondo}, \citenamefont {Klein},\ and\ \citenamefont
  {Yunes}}]{Arredondo:2024nsl}%
  \BibitemOpen
  \bibfield  {author} {\bibinfo {author} {\bibfnamefont {J.~N.}\ \bibnamefont
  {Arredondo}}, \bibinfo {author} {\bibfnamefont {A.}~\bibnamefont {Klein}},\
  and\ \bibinfo {author} {\bibfnamefont {N.}~\bibnamefont {Yunes}},\ }\href
  {https://doi.org/10.1103/PhysRevD.110.044044} {\bibfield  {journal} {\bibinfo
   {journal} {Phys. Rev. D}\ }\textbf {\bibinfo {volume} {110}},\ \bibinfo
  {pages} {044044} (\bibinfo {year} {2024})},\ \Eprint
  {https://arxiv.org/abs/2402.06804} {arXiv:2402.06804 [gr-qc]} \BibitemShut
  {NoStop}%
\bibitem [{\citenamefont {Morras}\ \emph {et~al.}(2025)\citenamefont {Morras},
  \citenamefont {Pratten},\ and\ \citenamefont {Schmidt}}]{Morras:2025nlp}%
  \BibitemOpen
  \bibfield  {author} {\bibinfo {author} {\bibfnamefont {G.}~\bibnamefont
  {Morras}}, \bibinfo {author} {\bibfnamefont {G.}~\bibnamefont {Pratten}},\
  and\ \bibinfo {author} {\bibfnamefont {P.}~\bibnamefont {Schmidt}},\ }\href
  {https://doi.org/10.1103/PhysRevD.111.084052} {\bibfield  {journal} {\bibinfo
   {journal} {Phys. Rev. D}\ }\textbf {\bibinfo {volume} {111}},\ \bibinfo
  {pages} {084052} (\bibinfo {year} {2025})},\ \Eprint
  {https://arxiv.org/abs/2502.03929} {arXiv:2502.03929 [gr-qc]} \BibitemShut
  {NoStop}%
\bibitem [{\citenamefont {Planas}\ \emph {et~al.}(2026)\citenamefont {Planas},
  \citenamefont {Ramos-Buades}, \citenamefont {Garc{\'\i}a-Quir{\'o}s},
  \citenamefont {Estell{\'e}s}, \citenamefont {Husa},\ and\ \citenamefont
  {Haney}}]{Planas:2025feq}%
  \BibitemOpen
  \bibfield  {author} {\bibinfo {author} {\bibfnamefont {M.~d.~L.}\
  \bibnamefont {Planas}}, \bibinfo {author} {\bibfnamefont {A.}~\bibnamefont
  {Ramos-Buades}}, \bibinfo {author} {\bibfnamefont {C.}~\bibnamefont
  {Garc{\'\i}a-Quir{\'o}s}}, \bibinfo {author} {\bibfnamefont {H.}~\bibnamefont
  {Estell{\'e}s}}, \bibinfo {author} {\bibfnamefont {S.}~\bibnamefont {Husa}},\
  and\ \bibinfo {author} {\bibfnamefont {M.}~\bibnamefont {Haney}},\ }\href
  {https://doi.org/10.1103/wz3v-b151} {\bibfield  {journal} {\bibinfo
  {journal} {Phys. Rev. D}\ }\textbf {\bibinfo {volume} {113}},\ \bibinfo
  {pages} {024006} (\bibinfo {year} {2026})},\ \Eprint
  {https://arxiv.org/abs/2503.13062} {arXiv:2503.13062 [gr-qc]} \BibitemShut
  {NoStop}%
\bibitem [{\citenamefont {Sesana}(2013)}]{Sesana:2013dma}%
  \BibitemOpen
  \bibfield  {author} {\bibinfo {author} {\bibfnamefont {A.}~\bibnamefont
  {Sesana}},\ }\href {https://doi.org/10.1088/0264-9381/30/24/244009}
  {\bibfield  {journal} {\bibinfo  {journal} {Class. Quant. Grav.}\ }\textbf
  {\bibinfo {volume} {30}},\ \bibinfo {pages} {244009} (\bibinfo {year}
  {2013})},\ \Eprint {https://arxiv.org/abs/1307.4086} {arXiv:1307.4086
  [astro-ph.CO]} \BibitemShut {NoStop}%
\bibitem [{\citenamefont {Tagawa}\ \emph {et~al.}(2021)\citenamefont {Tagawa},
  \citenamefont {Kocsis}, \citenamefont {Haiman}, \citenamefont {Bartos},
  \citenamefont {Omukai},\ and\ \citenamefont {Samsing}}]{Tagawa:2020jnc}%
  \BibitemOpen
  \bibfield  {author} {\bibinfo {author} {\bibfnamefont {H.}~\bibnamefont
  {Tagawa}}, \bibinfo {author} {\bibfnamefont {B.}~\bibnamefont {Kocsis}},
  \bibinfo {author} {\bibfnamefont {Z.}~\bibnamefont {Haiman}}, \bibinfo
  {author} {\bibfnamefont {I.}~\bibnamefont {Bartos}}, \bibinfo {author}
  {\bibfnamefont {K.}~\bibnamefont {Omukai}},\ and\ \bibinfo {author}
  {\bibfnamefont {J.}~\bibnamefont {Samsing}},\ }\href
  {https://doi.org/10.3847/2041-8213/abd4d3} {\bibfield  {journal} {\bibinfo
  {journal} {Astrophys. J. Lett.}\ }\textbf {\bibinfo {volume} {907}},\
  \bibinfo {pages} {L20} (\bibinfo {year} {2021})},\ \Eprint
  {https://arxiv.org/abs/2010.10526} {arXiv:2010.10526 [astro-ph.HE]}
  \BibitemShut {NoStop}%
\bibitem [{\citenamefont {Gond{\'a}n}\ and\ \citenamefont
  {Kocsis}(2021)}]{Gondan:2020svr}%
  \BibitemOpen
  \bibfield  {author} {\bibinfo {author} {\bibfnamefont {L.}~\bibnamefont
  {Gond{\'a}n}}\ and\ \bibinfo {author} {\bibfnamefont {B.}~\bibnamefont
  {Kocsis}},\ }\href {https://doi.org/10.1093/mnras/stab1722} {\bibfield
  {journal} {\bibinfo  {journal} {Mon. Not. Roy. Astron. Soc.}\ }\textbf
  {\bibinfo {volume} {506}},\ \bibinfo {pages} {1665} (\bibinfo {year}
  {2021})},\ \Eprint {https://arxiv.org/abs/2011.02507} {arXiv:2011.02507
  [astro-ph.HE]} \BibitemShut {NoStop}%
\bibitem [{\citenamefont {Romero-Shaw}\ \emph {et~al.}(2024)\citenamefont
  {Romero-Shaw}, \citenamefont {Goorachurn}, \citenamefont {Siwek},\ and\
  \citenamefont {Moore}}]{Romero-Shaw:2024klf}%
  \BibitemOpen
  \bibfield  {author} {\bibinfo {author} {\bibfnamefont {I.~M.}\ \bibnamefont
  {Romero-Shaw}}, \bibinfo {author} {\bibfnamefont {S.}~\bibnamefont
  {Goorachurn}}, \bibinfo {author} {\bibfnamefont {M.}~\bibnamefont {Siwek}},\
  and\ \bibinfo {author} {\bibfnamefont {C.~J.}\ \bibnamefont {Moore}},\ }\href
  {https://doi.org/10.1093/mnrasl/slae081} {\bibfield  {journal} {\bibinfo
  {journal} {Mon. Not. Roy. Astron. Soc.}\ }\textbf {\bibinfo {volume} {534}},\
  \bibinfo {pages} {L58} (\bibinfo {year} {2024})},\ \Eprint
  {https://arxiv.org/abs/2407.03869} {arXiv:2407.03869 [astro-ph.HE]}
  \BibitemShut {NoStop}%
\bibitem [{\citenamefont {Will}(2021)}]{Will:2020tri}%
  \BibitemOpen
  \bibfield  {author} {\bibinfo {author} {\bibfnamefont {C.~M.}\ \bibnamefont
  {Will}},\ }\href {https://doi.org/10.1103/PhysRevD.103.063003} {\bibfield
  {journal} {\bibinfo  {journal} {Phys. Rev. D}\ }\textbf {\bibinfo {volume}
  {103}},\ \bibinfo {pages} {063003} (\bibinfo {year} {2021})},\ \Eprint
  {https://arxiv.org/abs/2011.13286} {arXiv:2011.13286 [astro-ph.EP]}
  \BibitemShut {NoStop}%
\bibitem [{\citenamefont {Cornish}\ \emph {et~al.}(2011)\citenamefont
  {Cornish}, \citenamefont {Sampson}, \citenamefont {Yunes},\ and\
  \citenamefont {Pretorius}}]{Cornish:2011ys}%
  \BibitemOpen
  \bibfield  {author} {\bibinfo {author} {\bibfnamefont {N.}~\bibnamefont
  {Cornish}}, \bibinfo {author} {\bibfnamefont {L.}~\bibnamefont {Sampson}},
  \bibinfo {author} {\bibfnamefont {N.}~\bibnamefont {Yunes}},\ and\ \bibinfo
  {author} {\bibfnamefont {F.}~\bibnamefont {Pretorius}},\ }\href
  {https://doi.org/10.1103/PhysRevD.84.062003} {\bibfield  {journal} {\bibinfo
  {journal} {Phys. Rev. D}\ }\textbf {\bibinfo {volume} {84}},\ \bibinfo
  {pages} {062003} (\bibinfo {year} {2011})},\ \Eprint
  {https://arxiv.org/abs/1105.2088} {arXiv:1105.2088 [gr-qc]} \BibitemShut
  {NoStop}%
\bibitem [{\citenamefont {Gupta}\ \emph {et~al.}(2024)\citenamefont {Gupta}
  \emph {et~al.}}]{Gupta:2024gun}%
  \BibitemOpen
  \bibfield  {author} {\bibinfo {author} {\bibfnamefont {A.}~\bibnamefont
  {Gupta}} \emph {et~al.},\ }\bibfield  {journal} {\bibinfo  {journal} {arXiv}\
  }\href {https://doi.org/10.21468/SciPostPhysCommRep.5}
  {10.21468/SciPostPhysCommRep.5} (\bibinfo {year} {2024}),\ \Eprint
  {https://arxiv.org/abs/2405.02197} {arXiv:2405.02197 [gr-qc]} \BibitemShut
  {NoStop}%
\bibitem [{\citenamefont {Chandramouli}\ \emph {et~al.}(2025)\citenamefont
  {Chandramouli}, \citenamefont {Prokup}, \citenamefont {Berti},\ and\
  \citenamefont {Yunes}}]{Chandramouli:2024vhw}%
  \BibitemOpen
  \bibfield  {author} {\bibinfo {author} {\bibfnamefont {R.~S.}\ \bibnamefont
  {Chandramouli}}, \bibinfo {author} {\bibfnamefont {K.}~\bibnamefont
  {Prokup}}, \bibinfo {author} {\bibfnamefont {E.}~\bibnamefont {Berti}},\ and\
  \bibinfo {author} {\bibfnamefont {N.}~\bibnamefont {Yunes}},\ }\href
  {https://doi.org/10.1103/PhysRevD.111.044026} {\bibfield  {journal} {\bibinfo
   {journal} {Phys. Rev. D}\ }\textbf {\bibinfo {volume} {111}},\ \bibinfo
  {pages} {044026} (\bibinfo {year} {2025})},\ \Eprint
  {https://arxiv.org/abs/2410.06254} {arXiv:2410.06254 [gr-qc]} \BibitemShut
  {NoStop}%
\bibitem [{\citenamefont {Harris}\ \emph {et~al.}(2020)\citenamefont {Harris},
  \citenamefont {Millman}, \citenamefont {{van der Walt}}, \citenamefont
  {Gommers}, \citenamefont {Virtanen}, \citenamefont {Cournapeau},
  \citenamefont {Wieser}, \citenamefont {Taylor}, \citenamefont {Berg},
  \citenamefont {Smith}, \citenamefont {Kern}, \citenamefont {Picus},
  \citenamefont {Hoyer}, \citenamefont {{van Kerkwijk}}, \citenamefont {Brett},
  \citenamefont {Haldane}, \citenamefont {{Fern{\'a}ndez del R{\'i}o}},
  \citenamefont {Wiebe}, \citenamefont {Peterson}, \citenamefont
  {G{\'e}rard-Marchant}, \citenamefont {Sheppard}, \citenamefont {Reddy},
  \citenamefont {Weckesser}, \citenamefont {Abbasi}, \citenamefont {Gohlke},\
  and\ \citenamefont {Oliphant}}]{Harris2020}%
  \BibitemOpen
  \bibfield  {author} {\bibinfo {author} {\bibfnamefont {C.~R.}\ \bibnamefont
  {Harris}}, \bibinfo {author} {\bibfnamefont {K.~J.}\ \bibnamefont {Millman}},
  \bibinfo {author} {\bibfnamefont {S.~J.}\ \bibnamefont {{van der Walt}}},
  \bibinfo {author} {\bibfnamefont {R.}~\bibnamefont {Gommers}}, \bibinfo
  {author} {\bibfnamefont {P.}~\bibnamefont {Virtanen}}, \bibinfo {author}
  {\bibfnamefont {D.}~\bibnamefont {Cournapeau}}, \bibinfo {author}
  {\bibfnamefont {E.}~\bibnamefont {Wieser}}, \bibinfo {author} {\bibfnamefont
  {J.}~\bibnamefont {Taylor}}, \bibinfo {author} {\bibfnamefont
  {S.}~\bibnamefont {Berg}}, \bibinfo {author} {\bibfnamefont {N.~J.}\
  \bibnamefont {Smith}}, \bibinfo {author} {\bibfnamefont {R.}~\bibnamefont
  {Kern}}, \bibinfo {author} {\bibfnamefont {M.}~\bibnamefont {Picus}},
  \bibinfo {author} {\bibfnamefont {S.}~\bibnamefont {Hoyer}}, \bibinfo
  {author} {\bibfnamefont {M.~H.}\ \bibnamefont {{van Kerkwijk}}}, \bibinfo
  {author} {\bibfnamefont {M.}~\bibnamefont {Brett}}, \bibinfo {author}
  {\bibfnamefont {A.}~\bibnamefont {Haldane}}, \bibinfo {author} {\bibfnamefont
  {J.}~\bibnamefont {{Fern{\'a}ndez del R{\'i}o}}}, \bibinfo {author}
  {\bibfnamefont {M.}~\bibnamefont {Wiebe}}, \bibinfo {author} {\bibfnamefont
  {P.}~\bibnamefont {Peterson}}, \bibinfo {author} {\bibfnamefont
  {P.}~\bibnamefont {G{\'e}rard-Marchant}}, \bibinfo {author} {\bibfnamefont
  {K.}~\bibnamefont {Sheppard}}, \bibinfo {author} {\bibfnamefont
  {T.}~\bibnamefont {Reddy}}, \bibinfo {author} {\bibfnamefont
  {W.}~\bibnamefont {Weckesser}}, \bibinfo {author} {\bibfnamefont
  {H.}~\bibnamefont {Abbasi}}, \bibinfo {author} {\bibfnamefont
  {C.}~\bibnamefont {Gohlke}},\ and\ \bibinfo {author} {\bibfnamefont {T.~E.}\
  \bibnamefont {Oliphant}},\ }\href {https://doi.org/10.1038/s41586-020-2649-2}
  {\bibfield  {journal} {\bibinfo  {journal} {Nature}\ }\textbf {\bibinfo
  {volume} {585}},\ \bibinfo {pages} {357} (\bibinfo {year}
  {2020})}\BibitemShut {NoStop}%
\bibitem [{\citenamefont {Hunter}(2007)}]{Hunter2007}%
  \BibitemOpen
  \bibfield  {author} {\bibinfo {author} {\bibfnamefont {J.~D.}\ \bibnamefont
  {Hunter}},\ }\href {https://doi.org/10.1109/MCSE.2007.55} {\bibfield
  {journal} {\bibinfo  {journal} {Computing in Science \& Engineering}\
  }\textbf {\bibinfo {volume} {9}},\ \bibinfo {pages} {90} (\bibinfo {year}
  {2007})}\BibitemShut {NoStop}%
\bibitem [{\citenamefont {Meurer}\ \emph {et~al.}(2017)\citenamefont {Meurer},
  \citenamefont {Smith}, \citenamefont {Paprocki}, \citenamefont
  {\v{C}ert\'{i}k}, \citenamefont {Kirpichev}, \citenamefont {Rocklin},
  \citenamefont {Kumar}, \citenamefont {Ivanov}, \citenamefont {Moore},
  \citenamefont {Singh}, \citenamefont {Rathnayake}, \citenamefont {Vig},
  \citenamefont {Granger}, \citenamefont {Muller}, \citenamefont {Bonazzi},
  \citenamefont {Gupta}, \citenamefont {Vats}, \citenamefont {Johansson},
  \citenamefont {Pedregosa}, \citenamefont {Curry}, \citenamefont {Terrel},
  \citenamefont {Rou\v{c}ka}, \citenamefont {Saboo}, \citenamefont {Fernando},
  \citenamefont {Kulal}, \citenamefont {Cimrman},\ and\ \citenamefont
  {Scopatz}}]{Meurer2017}%
  \BibitemOpen
  \bibfield  {author} {\bibinfo {author} {\bibfnamefont {A.}~\bibnamefont
  {Meurer}}, \bibinfo {author} {\bibfnamefont {C.~P.}\ \bibnamefont {Smith}},
  \bibinfo {author} {\bibfnamefont {M.}~\bibnamefont {Paprocki}}, \bibinfo
  {author} {\bibfnamefont {O.}~\bibnamefont {\v{C}ert\'{i}k}}, \bibinfo
  {author} {\bibfnamefont {S.~B.}\ \bibnamefont {Kirpichev}}, \bibinfo {author}
  {\bibfnamefont {M.}~\bibnamefont {Rocklin}}, \bibinfo {author} {\bibfnamefont
  {A.}~\bibnamefont {Kumar}}, \bibinfo {author} {\bibfnamefont
  {S.}~\bibnamefont {Ivanov}}, \bibinfo {author} {\bibfnamefont {J.~K.}\
  \bibnamefont {Moore}}, \bibinfo {author} {\bibfnamefont {S.}~\bibnamefont
  {Singh}}, \bibinfo {author} {\bibfnamefont {T.}~\bibnamefont {Rathnayake}},
  \bibinfo {author} {\bibfnamefont {S.}~\bibnamefont {Vig}}, \bibinfo {author}
  {\bibfnamefont {B.~E.}\ \bibnamefont {Granger}}, \bibinfo {author}
  {\bibfnamefont {R.~P.}\ \bibnamefont {Muller}}, \bibinfo {author}
  {\bibfnamefont {F.}~\bibnamefont {Bonazzi}}, \bibinfo {author} {\bibfnamefont
  {H.}~\bibnamefont {Gupta}}, \bibinfo {author} {\bibfnamefont
  {S.}~\bibnamefont {Vats}}, \bibinfo {author} {\bibfnamefont {F.}~\bibnamefont
  {Johansson}}, \bibinfo {author} {\bibfnamefont {F.}~\bibnamefont
  {Pedregosa}}, \bibinfo {author} {\bibfnamefont {M.~J.}\ \bibnamefont
  {Curry}}, \bibinfo {author} {\bibfnamefont {A.~R.}\ \bibnamefont {Terrel}},
  \bibinfo {author} {\bibfnamefont {v.}~\bibnamefont {Rou\v{c}ka}}, \bibinfo
  {author} {\bibfnamefont {A.}~\bibnamefont {Saboo}}, \bibinfo {author}
  {\bibfnamefont {I.}~\bibnamefont {Fernando}}, \bibinfo {author}
  {\bibfnamefont {S.}~\bibnamefont {Kulal}}, \bibinfo {author} {\bibfnamefont
  {R.}~\bibnamefont {Cimrman}},\ and\ \bibinfo {author} {\bibfnamefont
  {A.}~\bibnamefont {Scopatz}},\ }\href {https://doi.org/10.7717/peerj-cs.103}
  {\bibfield  {journal} {\bibinfo  {journal} {PeerJ Computer Science}\ }\textbf
  {\bibinfo {volume} {3}},\ \bibinfo {pages} {e103} (\bibinfo {year}
  {2017})}\BibitemShut {NoStop}%
\bibitem [{\citenamefont {{Astropy Collaboration}}\ \emph
  {et~al.}(2022)\citenamefont {{Astropy Collaboration}}, \citenamefont
  {Price-Whelan}, \citenamefont {Lim}, \citenamefont {Earl} \emph
  {et~al.}}]{Astropy2022}%
  \BibitemOpen
  \bibfield  {author} {\bibinfo {author} {\bibnamefont {{Astropy
  Collaboration}}}, \bibinfo {author} {\bibfnamefont {A.~M.}\ \bibnamefont
  {Price-Whelan}}, \bibinfo {author} {\bibfnamefont {P.~L.}\ \bibnamefont
  {Lim}}, \bibinfo {author} {\bibfnamefont {N.}~\bibnamefont {Earl}}, \emph
  {et~al.},\ }\href {https://doi.org/10.3847/1538-4357/ac7c74} {\bibfield
  {journal} {\bibinfo  {journal} {The Astrophysical Journal}\ }\textbf
  {\bibinfo {volume} {935}},\ \bibinfo {pages} {167} (\bibinfo {year}
  {2022})},\ \Eprint {https://arxiv.org/abs/2206.14220} {arXiv:2206.14220
  [astro-ph.IM]} \BibitemShut {NoStop}%
\bibitem [{\citenamefont {Bradbury}\ \emph {et~al.}(2018)\citenamefont
  {Bradbury}, \citenamefont {Frostig}, \citenamefont {Hawkins}, \citenamefont
  {Johnson}, \citenamefont {Katariya}, \citenamefont {Leary}, \citenamefont
  {Maclaurin}, \citenamefont {Necula}, \citenamefont {Paszke}, \citenamefont
  {VanderPlas}, \citenamefont {Wanderman-Milne},\ and\ \citenamefont
  {Zhang}}]{Bradbury2018}%
  \BibitemOpen
  \bibfield  {author} {\bibinfo {author} {\bibfnamefont {J.}~\bibnamefont
  {Bradbury}}, \bibinfo {author} {\bibfnamefont {R.}~\bibnamefont {Frostig}},
  \bibinfo {author} {\bibfnamefont {P.}~\bibnamefont {Hawkins}}, \bibinfo
  {author} {\bibfnamefont {M.~J.}\ \bibnamefont {Johnson}}, \bibinfo {author}
  {\bibfnamefont {Y.}~\bibnamefont {Katariya}}, \bibinfo {author}
  {\bibfnamefont {C.}~\bibnamefont {Leary}}, \bibinfo {author} {\bibfnamefont
  {D.}~\bibnamefont {Maclaurin}}, \bibinfo {author} {\bibfnamefont
  {G.}~\bibnamefont {Necula}}, \bibinfo {author} {\bibfnamefont
  {A.}~\bibnamefont {Paszke}}, \bibinfo {author} {\bibfnamefont
  {J.}~\bibnamefont {VanderPlas}}, \bibinfo {author} {\bibfnamefont
  {S.}~\bibnamefont {Wanderman-Milne}},\ and\ \bibinfo {author} {\bibfnamefont
  {Q.}~\bibnamefont {Zhang}},\ }\href {https://github.com/jax-ml/jax} {\bibinfo
  {title} {{JAX}: composable transformations of {Python+NumPy} programs}}
  (\bibinfo {year} {2018})\BibitemShut {NoStop}%
\bibitem [{\citenamefont {Kidger}(2021)}]{Kidger2021}%
  \BibitemOpen
  \bibfield  {author} {\bibinfo {author} {\bibfnamefont {P.}~\bibnamefont
  {Kidger}},\ }\emph {\bibinfo {title} {On Neural Differential Equations}},\
  \href@noop {} {Ph.D. thesis},\ \bibinfo  {school} {University of Oxford}
  (\bibinfo {year} {2021}),\ \Eprint {https://arxiv.org/abs/2202.02435}
  {arXiv:2202.02435 [cs.LG]} \BibitemShut {NoStop}%
\bibitem [{\citenamefont {Flanagan}\ and\ \citenamefont
  {Hughes}(1998)}]{Flanagan:1997kp}%
  \BibitemOpen
  \bibfield  {author} {\bibinfo {author} {\bibfnamefont {E.~E.}\ \bibnamefont
  {Flanagan}}\ and\ \bibinfo {author} {\bibfnamefont {S.~A.}\ \bibnamefont
  {Hughes}},\ }\href {https://doi.org/10.1103/PhysRevD.57.4566} {\bibfield
  {journal} {\bibinfo  {journal} {Phys. Rev. D}\ }\textbf {\bibinfo {volume}
  {57}},\ \bibinfo {pages} {4566} (\bibinfo {year} {1998})},\ \Eprint
  {https://arxiv.org/abs/gr-qc/9710129} {arXiv:gr-qc/9710129} \BibitemShut
  {NoStop}%
\bibitem [{\citenamefont {Maggiore}(2007)}]{Maggiore:2007ulw}%
  \BibitemOpen
  \bibfield  {author} {\bibinfo {author} {\bibfnamefont {M.}~\bibnamefont
  {Maggiore}},\ }\href
  {https://doi.org/10.1093/acprof:oso/9780198570745.001.0001} {\emph {\bibinfo
  {title} {{Gravitational Waves. Vol. 1: Theory and Experiments}}}}\ (\bibinfo
  {publisher} {Oxford University Press},\ \bibinfo {year} {2007})\BibitemShut
  {NoStop}%
\bibitem [{\citenamefont {Moore}\ \emph {et~al.}(2018)\citenamefont {Moore},
  \citenamefont {Robson}, \citenamefont {Loutrel},\ and\ \citenamefont
  {Yunes}}]{Moore:2018kvz}%
  \BibitemOpen
  \bibfield  {author} {\bibinfo {author} {\bibfnamefont {B.}~\bibnamefont
  {Moore}}, \bibinfo {author} {\bibfnamefont {T.}~\bibnamefont {Robson}},
  \bibinfo {author} {\bibfnamefont {N.}~\bibnamefont {Loutrel}},\ and\ \bibinfo
  {author} {\bibfnamefont {N.}~\bibnamefont {Yunes}},\ }\href
  {https://doi.org/10.1088/1361-6382/aaea00} {\bibfield  {journal} {\bibinfo
  {journal} {Class. Quant. Grav.}\ }\textbf {\bibinfo {volume} {35}},\ \bibinfo
  {pages} {235006} (\bibinfo {year} {2018})},\ \Eprint
  {https://arxiv.org/abs/1807.07163} {arXiv:1807.07163 [gr-qc]} \BibitemShut
  {NoStop}%
\bibitem [{\citenamefont {Cooley}\ and\ \citenamefont
  {Tukey}(1965)}]{CooleyTukey1965}%
  \BibitemOpen
  \bibfield  {author} {\bibinfo {author} {\bibfnamefont {J.~W.}\ \bibnamefont
  {Cooley}}\ and\ \bibinfo {author} {\bibfnamefont {J.~W.}\ \bibnamefont
  {Tukey}},\ }\href {https://doi.org/10.1090/S0025-5718-1965-0178586-1}
  {\bibfield  {journal} {\bibinfo  {journal} {Mathematics of Computation}\
  }\textbf {\bibinfo {volume} {19}},\ \bibinfo {pages} {297} (\bibinfo {year}
  {1965})}\BibitemShut {NoStop}%
\bibitem [{\citenamefont {Brent}(1973)}]{Brent1973}%
  \BibitemOpen
  \bibfield  {author} {\bibinfo {author} {\bibfnamefont {R.~P.}\ \bibnamefont
  {Brent}},\ }\href@noop {} {\emph {\bibinfo {title} {Algorithms for
  Minimization without Derivatives}}}\ (\bibinfo  {publisher} {Prentice-Hall},\
  \bibinfo {address} {Englewood Cliffs, New Jersey},\ \bibinfo {year}
  {1973})\BibitemShut {NoStop}%
\bibitem [{\citenamefont {Nitz}\ \emph {et~al.}(2024)\citenamefont {Nitz},
  \citenamefont {Harry}, \citenamefont {Brown} \emph {et~al.}}]{PyCBC}%
  \BibitemOpen
  \bibfield  {author} {\bibinfo {author} {\bibfnamefont {A.}~\bibnamefont
  {Nitz}}, \bibinfo {author} {\bibfnamefont {I.}~\bibnamefont {Harry}},
  \bibinfo {author} {\bibfnamefont {D.}~\bibnamefont {Brown}}, \emph {et~al.},\
  }\href {https://doi.org/10.5281/zenodo.10473621} {\bibinfo {title}
  {{gwastro/pycbc: v2.3.3 release of PyCBC}}} (\bibinfo {year}
  {2024})\BibitemShut {NoStop}%
\end{thebibliography}%

\end{document}